\documentclass{article} 
\usepackage{iclr2025_conference,times}

\usepackage{amsmath,amsfonts,bm}

\def\eqref#1{equation~\ref{#1}}

\def\1{\bm{1}}

\DeclareMathAlphabet{\mathsfit}{\encodingdefault}{\sfdefault}{m}{sl}
\SetMathAlphabet{\mathsfit}{bold}{\encodingdefault}{\sfdefault}{bx}{n}

\usepackage{hyperref}
\usepackage{url}

\usepackage[utf8]{inputenc} 
\usepackage[T1]{fontenc}    
\usepackage{hyperref}       
\usepackage{url}            
\usepackage{booktabs}       
\usepackage{amsfonts}       
\usepackage{nicefrac}       
\usepackage{microtype}      
\usepackage{xcolor}         
\usepackage{amsmath}
\usepackage{graphicx} 
\usepackage{caption}  
\usepackage{tikz}

\usepackage{array} 
\usepackage{amsmath} 
\usepackage{amssymb} 
\usepackage{verbatim}

\newcommand{\Tau}{\mathrm{T}}

\usepackage{amsmath}
\usepackage{amsfonts} 
\usepackage{bm}

\usepackage{algorithm}
\usepackage[noend]{algpseudocode}
\usepackage{amsmath} 

\title{\textbf{CREIMBO}: \textbf{C}ross-\textbf{R}egional \textbf{E}nsemble \textbf{I}nteractions in \textbf{M}ulti-view
\textbf{B}rain \textbf{O}bservations}

\author{Noga Mudrik \\
Biomedical Engineering, Kavli NDI, CIS  \\
 The Johns Hopkins University\\
   Baltimore, MD, USA. \\
  \texttt{nmudrik1@jhu.edu} \\
\And
Ryan Ly \\
  Scientific Data Division \\ 
   Lawrence Berkeley National Laboratory \\
   Berkeley, CA, USA. \\
  \texttt{rly@lbl.gov} \\
   \And
  Oliver Rübel \\
  Scientific Data Division \\ 
   Lawrence Berkeley National Laboratory \\
   Berkeley, CA, USA. \\
  \texttt{ORuebel@lbl.gov} \\
\And
 Adam S. Charles \\
  Biomedical Engineering, Kavli NDI, CIS  \\
 The Johns Hopkins University\\
   Baltimore, MD, USA. \\
  \texttt{adamsc@jhu.edu} \\
}

\iclrfinalcopy 
\begin{document}

\maketitle

\begin{abstract}

Modern recordings of neural activity provide diverse observations of neurons across brain areas, behavioral conditions, and subjects; presenting an exciting opportunity to reveal the fundamentals of brain-wide dynamics. Current analysis methods, however, often fail to fully harness the richness of such data, as they provide either uninterpretable representations (e.g., via deep networks) or oversimplify models (e.g., by assuming stationary dynamics or analyzing each session independently). Here, instead of regarding asynchronous neural recordings that lack alignment in neural identity or brain areas as a limitation, we leverage these diverse views into the brain 
to learn a unified model of neural dynamics. 
Specifically, we assume that brain activity is driven by multiple hidden global sub-circuits. These sub-circuits represent global basis interactions between neural ensembles---functional groups of neurons---such that the time-varying decomposition of these sub-circuits defines how the ensembles' interactions evolve over time non-stationarily and non-linearly.
We discover the neural ensembles underlying non-simultaneous observations, along with their non-stationary evolving interactions, with our new model, \textbf{CREIMBO} (Cross-Regional Ensemble Interactions in Multi-view Brain Observations). CREIMBO identifies the hidden composition of per-session neural ensembles through novel graph-driven dictionary learning and models the ensemble dynamics on a low-dimensional manifold spanned by a sparse time-varying composition of the global sub-circuits. Thus, CREIMBO disentangles overlapping temporal neural processes while preserving interpretability due to the use of a shared underlying sub-circuit basis. Moreover, CREIMBO distinguishes session-specific computations from global (session-invariant) ones by identifying session covariates and variations in sub-circuit activations. We demonstrate CREIMBO's ability to recover true components in synthetic data, and uncover meaningful brain dynamics in human high-density electrode recordings, including cross-subject neural mechanisms as well as  inter- vs. intra-region dynamical motifs. {Furthermore, using mouse whole-brain recordings, we show CREIMBO's ability to discover dynamical interactions that capture task and behavioral variables and meaningfully align with the biological importance of the brain areas they represent.}

\end{abstract}


\begin{figure}[t]
    \centering
    \includegraphics[width=0.9\textwidth]{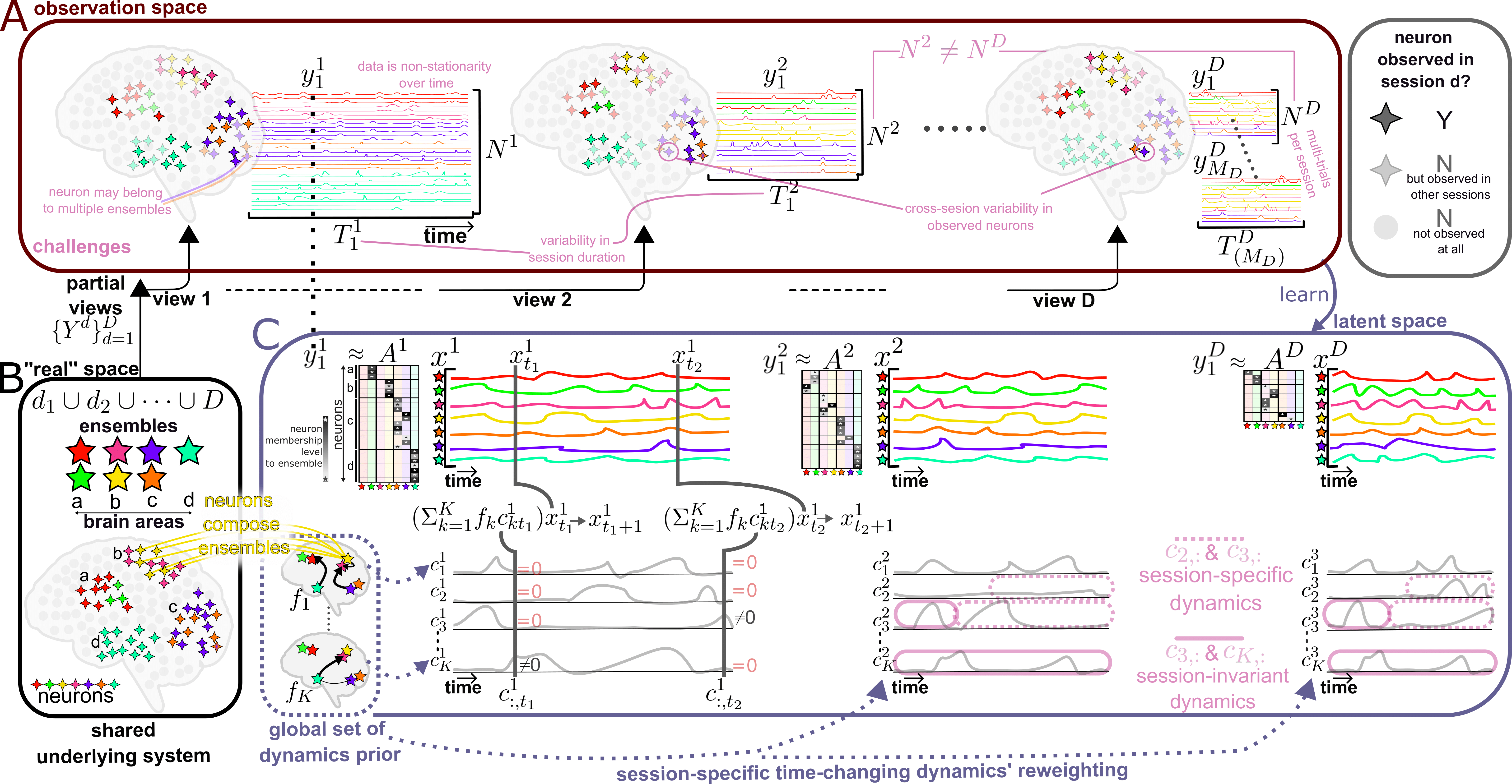}
    \caption{\textbf{CREIMBO's Illustration.} 
    \textbf{A:} Real-world multi-regional neural datasets consist of multiple ($D$) non-simultaneous recording sessions ($\{ {\bm{Y}^d}\}^{D}_{d=1}$) that cannot be matched in terms of individual neurons' identities, quantity, or cross-region distribution, resulting in neurons appearing only in certain sessions (4-pointed stars) or not at all (gray circles). This variability hinder our ability to draw unified conclusions of whole-brain neural computation, while analyzing sessions individually could lead to session-specific biased results. This challenge is further complicated by variability in session (or trial) duration, the non-stationarity of brain dynamics (even within a session), and the presence of multiple trials within each session. 
\textbf{B $\rightarrow$ A:} Instead of considering these asynchronous sessions as a challenge, we frame them as an opportunity to obtain distinct, potentially complementary ``views'' onto the shared underlying brain system, thereby facilitating the learning of a unified brain dynamics model. We assume that brain computations are mediated by multiple neural ensembles---$\bigstar$ groups of same-region neurons with shared functionalities---whose interactions yield meaningful neural representations. 
However, the specific neurons and their membership degrees within each ensemble are unknown, and a neuron may belong to multiple ensembles with varying membership levels based on its diverse functionalities.
    \textbf{C:} 
CREIMBO leverages partial brain views to co-learn ensemble compositions per session ($\{ \bm{A}^d \}_{d=1}^D$), their temporal activity ($\{\bm{x}^d\}_{d=1}^D$), and cross-regional interactions $\bm{x}^d_{t+1} = \bm{F}_t^d \bm{x}^d_t$. It posits that these non-stationary interactions stem from a session-shared dictionary prior of up to $K$ global interactions $\{ \bm{f}_{k} \}_{k=1}^K$, whose sparse time-changing decomposition $\bm{F}^d_t = \sum_{k=1}^K \bm{f}_k \bm{c}_{kt}^d$ shapes the overall ensemble dynamics at each time point $t$ and session $d$.  The interactions' coefficients can help distinguish session-specific interactions from session-invariant interactions in different time periods (pink right, solid vs. dotted in subplot C right). 
With the session-shared ensemble-interactions dictionary and certain assumptions (Sec.~\ref{sec:assumptions}), the model auto-sorts ensembles across sessions by functionality, aligning the $j$-th ensemble of session $d$ with the $j$-th ensemble of session $d'$, despite differences in observed neurons.
\vspace{-0.25cm}
    }
    \label{fig:demofig}
    \vspace{-7pt}
\end{figure}

\section{Introduction}
Identifying the interactions between and within brain areas is fundamental to advancing our understanding of how the brain gives rise to behavior. 
Recent advances in neural recording present an exciting opportunity to study brain-wide interactions by enabling simultaneous recording of neural activity across many brain areas through multiple high-density electrodes. 
Such experiments, repeated over many sessions with different implantation patterns, offer multiple asynchronous recordings of the brain system, with each session encompassing hundreds of distinct neurons across regions. 


Individual recording sessions provide a singular perspective of the brain dynamics, as the reinsertion of recording devices results in the capture of different subsets of neurons in each session. Thus, fitting population-level models often reverts to per-session analysis (i.e., fitting each session independently), which is inherently constrained by the inability to incorporate the full set of recorded activity across all recorded brain areas across all sessions.
Moreover, analyzing the data session-by-session, may hinder our ability to distinguish between computations that are session or subject specific from task related or global session-invariant neural processes. 
Compounding the difficulty of merging different sessions' data are the non-linear and non-stationary nature of neural activity, as well as the presence of noise and variability across trials---both of which require additional model complexity to account for.
Moreover, to extract scientifically meaningful insights, the models must be interpretable, with model parameters directly related to task variables or connections between recorded units.
This gap between the opportunities offered by modern neural data and the limited capabilities of current methods necessitates new approaches to leverage the richness in modern brain data and discover the fundamental neural sub-circuits governing brain activity.
\\
One current approach used to combine multi-session data into unified model is 
deep learning. By training to predict all sessions' data, deep networks implicitly merge the datasets into the fit network weights. 
While these powerful models can combine information across sessions, their uninterpretable 
nature and complexity typically hinders their ability to reveal the fundamental building blocks of neural computation.
Other approaches, including tensor factorization and its variations (e.g.,~\cite{NIPS2015_b3967a0e,harshman1970foundations}) 
may be limited in their ability to merge sessions of varying durations or model the interactions between the identified neural components.

Here, we re-frame cross-session variability in neurons recorded and experimental settings, not as a drawback, but as a valuable advantage, viewing this data as providing multiple, complementary ``views'' into a single, shared brain system (Fig.~\ref{fig:demofig}B $\rightarrow$ A).
 This approach enables a more thorough and holistic discovery of the underlying system designed to extract joint information from the entire dataset.
In particular, we hypothesize that
the population dynamics at each session is driven by a common set of global time-invariant sub-circuits shared across sessions. 
To effectively capture the non-linear and non-stationary dynamics of the brain, we enable these sub-circuits to adjust their activity levels over time and across sessions. This approach allows us to identify evolving neural patterns through the collective activity of the sub-circuits, modeled via their time-varying decomposition.
We assume that the latent dynamics governed by functional neural groups are shared across sessions, even though the specific individual neurons observed may vary. To address this, we propose a per-session transformation from the joint low-dimensional circuit space to each session's observation space, ensuring cross-session alignment of the neural groups' functionalities.


In this work, we lay out this new model, which we term \textbf{C}ross-\textbf{R}egional \textbf{E}nsemble \textbf{I}nteractions in \textbf{M}ulti-view \textbf{B}rain \textbf{O}bservations ({CREIMBO}), and demonstrate its ability to capture latent dynamics in multi-view neural data.  Specifically, our contributions include:
\begin{itemize}
    \item We discover multi-regional brain dynamics through leveraging the richness of modern neural datasets while ensuring interpretability.
    \item We distinguish between intra- and inter-region interactions by identifying sparse, same-area neural ensembles, deploying a specialized structured prior over per-session projections from the circuit latent space to the observed neural space.
    \item We accurately recover ground truth components in synthetic data and discover meaningful cross-regional brain interactions underlying human {and mouse} brain data. 
\end{itemize}

\section{Related Work}
Over the past decade, remarkable advances in neural data acquisition technologies have enabled the measurement of numerous neurons across brain regions, subjects, and experimental settings. This has spurred follow-up work to leverage this extensive data for novel neural discoveries, including merging non-simultaneous sessions and examining cross-regional brain interactions, all while accounting for the complex, non-linear, and non-stationary nature of neural activity.


\textbf{Merging non-simultaneous neural recording sessions.}
Existing work for combining non-simultaneous neuronal population recordings, here termed ``sessions'', 
include
~\cite{turaga2013inferring} who introduced the term ``stitching'' for this problem and modeled neural activity  with a shared latent dynamical systems model 
and~\cite{soudry2013shotgun} 
who proposed to 
stitch asynchronous recordings for inferring functional connectivity by assuming shared cross-session latent input and demonstrated the advantage in observing more neurons for inferring connectivity. 
However, these methods are not designed to capture more than two simultaneously interacting neural sub-circuits across multiple populations, nor do they distinguish between within- and between-area interactions or specifically address temporal non-stationarity.
\cite{bishop2015combining} and~\cite{bishop2014deterministic} proposed integrating the neural activity structure into inference and expanding
stitching to study the communication between two neuronal populations from potentially different brain areas. 
In~\cite{nonnenmacher2017extracting}, the authors proposed 
extracting low-dimensional dynamics from
 multiple sessions  by learning  temporal co-variations across neurons and time.
Both of these methods, through, assume at least a small overlap in the identity of cross-session observed neurons. 
LFADS~\citep{Keshtkaran2021AutoLFADS,sussillo2016lfads} enables stitching 
while addressing the need for limited neural overlap 
by finding a shared non-linear latent dynamical model.
However, its latent dynamics may be difficult to interpret regarding the specific neural sub-circuits that compose it, and due to its foundation on a variational auto-encoder, it may be challenging to tune.

\textbf{Maintaining Interpretability in Non-Stationary Neural Dynamics Models. }
Real-world neural activity follows non-stationary nature due to changing environmental settings. 
Classical dynamical models often
simplified neural dynamics to linear (e.g.,~jPCA~\cite{churchland2012neural}) or non-linear but stationary (e.g.,~\cite{gallego2020long}). 
Models that capture non-stationarity often derive from ``black box'' deep learning models (e.g.,~\cite{schneider2023cebra, sussillo2016lfads, zhu2022deep}, which are powerful but often present limited interpretability with respect to neural interactions. Other models are built on switching piece-wise linear models ~\cite{linderman2016recurrent, linderman2017bayesian, murphy1998switching}, which do not enable the identification of multiple co-active dynamic processes.
dLDS~(decomposed Linear Dynamical Systems, ~\cite{mudrik2024decomposed}) and its multi-step extension~\citep{mudrik2024linocs}), addresses this gap by generalizing the switching assumption to sparse time-changing decomposition of linear dynamical systems, thus enabling 
capturing co-active sub-circuits.
However, while dLDS can be extended to co-process multi-session recordings with cross-session variability in neuron identity, this modification may not yield interpretable latent-to-observed projections and lacks  the structured multi-regional prior over the emission matrix necessary to partition multi-area recordings and study between- vs. within-area interactions.

\textbf{Identification of Functional Neural Ensembles. }
A potential way to find interpretable loading matrix that links low  to high dimensional space  is through the identification of functional neural ensembles---here referred to as groups of neurons with similar functionalities---that determine the axes in the latent space. For instance, the mDLAG model used Automatic Relevance Determination (ARD) to promote population-wise
 sparsity patterns of each latent variable. 
In~\cite{mudrik2023sibblings}, the authors presented an approach to address sparsity in loading matrices while accounting for multi-trial and task-conditions variability. They proposed a graph-driven regularization technique that facilitates the grouping of co-active neurons into the same ensembles and pushing apart neurons that present different functionality, while allowing for structural adjustments in ensemble configurations across different task conditions. They further enable per-trial variability in temporal activity of the ensembles. However, their model does not model the dynamic interactions between ensembles or give a closed form prior for the evolution of the ensembles' temporal traces.
While other methods, e.g., clustering approaches like ~\citep{grossberger2018unsupervised} or Sparse Principal Component Analysis (SPCA)~\cite{zou2006sparse}, enable the recognition of underlying sparse groups, they either do not allow a neuron to belong to multiple groups with varying degrees of membership or do not support controlled adaptations of structural membership across conditions. 


\textbf{Multi-regional brain interactions. }
With the ability to record from multiple brain areas and the recognition of the importance of considering multi-area communication~\citep{pesaran2021multiregional}, recent works have delved into understanding the communication between or within brain areas.
In~\cite{gokcen2022disentangling,gokcen2024uncovering}, the authors proposed models to identify time delays between cross-regional brain interactions, overcoming the challenge of distinguishing indirect, concurrent, and bi-directional interactions in two or more populations. However, their model focuses on analyzing sessions individually and is not intended to uncover the full underlying set of transition matrices, but rather to recognize meaningful time delays that can imply interactions.
Other models addressed multi-regional interactions through a communication subspace~\cite{semedo2019cortical} using dimensionality reduction, or by Generalized Linear Models (GLMs), with either Poisson~\citep{yates2017functional} or Gaussian~\citep{yates2017functional} statistics
to identify functional coupling between areas.
These approaches,  though, are not tailored to capture non-simultaneous cross-sessions variability in neuron identity and subject processing. 
Other recent methods~(\cite{karniol-tambour2024modeling,glaser2020recurrent,li2024onehot}) discussed the potential usage of their model to multi-regional interactions, have proposed studying multi-regional interactions using switching dynamical systems approaches, where the model switches between different states over time to capture distinct periods. However, this approach cannot disentangle interactions among multiple sub-circuits active simultaneously, nor can it identify interactions at multiple temporal resolutions that occur concurrently and distinguish between these or multiple co-processed processes whose encoding occurs gradually rather than via sharp transitions. 
\vspace{-5pt}

\section{Problem Introduction and Approach}
\vspace{-10pt}
\textbf{Problem Intuition:} 
Let $\{\bm{Y}^{d}\}_{d=1}^D$ be a set of estimated neural firing rates (e.g. from neuropixels data) over $D$ asynchronous sessions, indexed by ${d = 1\dots D}$. The neural recordings of each $d$ session, $\bm{Y}^d \in \mathbb{R}^{N^d \times T^d}$, capture $T^d$ time observations of the activity of $N^d$ neurons from up to $J$ distinct brain areas, thus offering a partial view of the brain system. Note that sessions may refer to the same or different subjects, can vary in duration or number of neurons, i.e. $N^d$ and $T^d$ need not be equal to $N^{d'}$ and $T^{d'}$ for different sessions $d,d'$, and can differ in the subset of brain areas (out of $J$) they capture.  
Hence, these cross-session data matrices cannot be aligned into a single data array, e.g., via a tensor---hindering the direct application of existing analyses as tensor factorization.

We assume that these brain observations reflect the hidden activity $\bm{X}^d \in \mathbb{R}^{p \times T^d}$ of $p<< N^d$ functional neural groups that evolve and interact over time. Each of these groups encompasses a sparse set of same-area neurons, while each neuron can belong to more than one group with varying degrees of cross-group membership, encoded in the values of $\bm{X}^d$. 
The low-dimensional latent space in which the groups interact is shared across sessions and projected to each session's observed neurons via an unknown per-session projection \( g^d : \mathbb{R}^p \rightarrow \mathbb{R}^{N^d} \)
 that captures the groups' compositions in the visible neurons, such that \( \bm{Y}^d = g^d(\bm{X}^d) \).
The interactions between these functional groups follow a per-session non-linear and non-statioary  dynamics $\bm{F}_t^d : \mathbb{R}^p \rightarrow \mathbb{R}^{p}$, where $\bm{X}^d_t = \bm{F}^d_t (\bm{x}^d_{t-1})$. As in general non-stationary systems cannot be fit without constraints, we follow the dLDS model and assume that these dynamics can be described by a set $\{\bm{f}_k\}_{k=1}^K$ of $K$ global time- and session-invariant ``sub-circuits''. Here each sub-circuit  ${\bm{f}_k \in \mathbb{R}^{p\times p}}$ represents a basis interaction between these groups that is reused by the system at different time-points throughout its trajectory. 
Each of these latent sub-circuits may capture either ``global'' session-invariant brain interactions and/or session- and subject-specific interactions.
Moreover, these sub-circuits may be active simultaneously or intermittently, yielding a model that can flexibly fit neural trajectories that differ between sessions with a single underlying mechanism. 
The problem CREIMBO addresses is thus identifying the set of \( K \) latent multi-regional sub-circuits \(\{\bm{f}_k \}_{k = 1}^K\), capturing interactions between unknown functional sparse neural ensembles \(\bm{A}\), along with their non-stationary activation levels \(\{ \{\bm{c_{kt}^d}\}_{k=1}^K \}_{d=1}^D \), by leveraging the joint information from asynchronously collected observations \(\{\bm{Y}^d\}_{d=1}^D\).

\textbf{CREIMBO: }
A naive approach to identifying the unknown sub-circuits involves first projecting each session's data into a low-dimensional space and subsequently fitting the dynamics model in the low-dimensional space. However, this two-step approach can be sub-optimal as the dimensionality reduction projections may not prioritize the same dimensions needed to preserve dynamic fidelity. 
An additional advantage of simultaneously identifying the ensembles and dynamics is that in the cross-session model it automatically aligns the cross-session components based on the shared dynamics prior. Such alignment is unlikely given the sensitivity of the subspace ordering in dimensionality reduction methods to each session's distinct statistics. 

\vspace{-3pt}
We therefore present CREIMBO, an algorithm that simultaneously fits the global sub-circuits that underlie the inter-ensemble interactions across multiple sessions.
Our model is predicated on three key assumptions that   
1) ensembles of neurons, rather than individual neurons, form the basic units that interact in brain dynamics, 
2) it is feasible to identify functionally analogous neuronal ensembles across sessions, and 3) the full repertoire of interactions between these ensembles are governed by a global set of dynamical primitives in the form of a basis of linear dynamical systems. 

\vspace{-3pt}
Let $\bm{A}^d$ be a sparse matrix where each column, $\bm{A}^d[:,j]$, encodes the neuronal composition of the $j$-th ensemble in the $d$-th session, such that non-zero values within these columns represent the membership of neurons in the ensemble and their magnitudes reflecting the degree of membership.
Since we aim to capture multi-regional interactions, we design the ensemble matrix $\bm{a} := \bm{A}^d$ as a block diagonal matrix where the $j$-th block,  $\bm{a}^{j} \in \mathbb{R}^{n_j \times p_j}$, contains the neural ensembles for the $j$-th area. We denote $p_j$ as the number of ensembles within that area (note that $p = \sum_{j=1}^J p_j$) and $n_j := N_j^d$ as the number of neurons in that area (i.e., the total number of neurons observed in session $d$ is $N^d = \sum_{j=1}^J n_j$).  
Furthermore, let $\tau := \Tau_m^d$ indicate the number of time-points recorded in each $m$-th trial of session $d$ ($m = 1\dots M_d$), such that the observations in that trial are captured by  $\bm{Y}_m^d \in \mathbb{R}^{N^d\times \Tau_m^d}$.
In each trial $m$, the temporal activity for the $p$ ensembles are denoted as $\bm{x} := \bm{X}_m^d \in \mathbb{R}^{p \times \tau}$, leading to a model where the observations $\bm{y} := \bm{Y}_m^d$ in trial $m$ are assumed to arise from the joint activity of all these ensembles, up do a normally-distributed error $\epsilon \sim \mathcal{N}(0,\sigma)$, i.e., $\bm{y} = \bm{a} \bm{x} + \epsilon.$ {Please refer to Section~\ref{sec:poisson_dev} for Poisson statistics inference in low spiking rate regimes.}
The evolution of the ensembles' activities ($\bm{x}$) reflect the latent interactions between ensembles, which we model with a non-stationary linear dynamical system $\bm{x}_t = \bm{F}_t \bm{x}_{t-1}$ for  $t = 1 \ldots \tau$. The time-changing transition matrix $\bm{F}_t \in \mathbb{R}^{p \times p}$ captures both non-linear and non-stationary  brain activity. 
A key desired property of our model is to capture the overlapping activity of multiple sub-circuits of interacting ensembles. We thus follow dLDS and model the interactions ($\bm{F}_t$) through a time-varying sparse decomposition ${\bm{F}_t = \sum_{k=1}^K \bm{c}_{kt}\bm{f}_k}$, where $\{\bm{f}_{k}\}_{k=1}^K$ are the $K$ global ensemble interactions, and $\bm{c}_{kt}$ are their time-varying coefficients that capture the modulation of each interaction.



Fitting this model requires identifying the  global operators $\{\bm{f}_k\}_{k=1}^K$  and ensembles $\{ \bm{A}^d\}_{d=1}^D$. 
For this, we employ an alternating 
approach and iterate between updating $\{ \bm{f}_k \}_{k=1}^K$ and inferring, 
for each session $d$, 1)~the ensemble compositions $\bm{a} := \bm{A}^d$, 2)~their dynamics $\bm{x}$, and 3) the sub-circuits' temporal coefficients $\bm{c}_t$. These steps are iterated until convergance. 

\textbf{Ensemble update:} We update $\bm{a}$ per-row (neuron) $n$ ($\bm{a}_n$). 
Following the work of~\cite{mudrik2023sibblings}, we infer the sparse ensemble structures with graph-driven re-weighted $\ell_1$ regularization that groups together neurons with shared activity patterns, while pushing apart neurons who do not behave similarly. The graph for each area $j$, $\bm{h}^j \in \mathbb{R}^{n_j \times n_j}$, is calculated based on a data-driven Gaussian kernel that measures the temporal similarity between neurons, such that $\bm{h}^j_{n1, n2} = exp(\frac{\|\bm{y}^j_{n1} - \bm{y}^j_{n2}\|_2^2}{\sigma_h})$ where $\sigma_h$ is a hyperparameter. 
Specifically, for each row $\bm{a}_{n,:}$ and considering only the ensembles in region $j_n$ that neuron $n$ is part of, 
\vspace{-5pt}
\begin{align}
    \widehat{\bm{a}}_{n,:} = \arg \min_{\bm{a}_{n:} }  \| \bm{y}_{n:} - \bm{a}_{n:}\bm{x}_{j_n:} \|_2^2 
            + 
\sum_{j =1}^p \bm{\lambda}^d_{n,j}  |\bm{a}_{n,j}|   
\textrm{ where } 
 \bm{\lambda}_{n,j} = \frac{\beta_1}{\beta_2 + |\widehat{\bm{a}}_{n,j}| +  \beta_3 | \bm{h}_{n:}^j  \widehat{\bm{a}}_{:j} |}. \label{eqn:updateLambda}   
 \vspace{-5pt}
\end{align}
Here, $\bm{x}_{j_n,:}$ is the activity of the ensembles of the $j_n$-th region and $\beta_1$, $\beta_2$, and $\beta_3$ are scalars that control the effect of the graph on the regularization. The graph adjacency  $\bm{h}_{n:}  \widehat{\bm{a}}_{:j}$ weights the sparse membership in the \(j\)-th ensemble based on temporal similarity. Specifically, if $\bm{h}_{n:}  \widehat{\bm{a}}_{:j}$ is high (i.e., the neighbors neuron $n$ on the graph are in the current estimate of the $j$-th ensemble), 
$\bm{\lambda}_{n,j}$  is small, resulting in lower sparsity regularization on neuron $n$ for that ensemble, promoting it to be included. Alternatively, if this correlation is small, the regularization weight $\bm{\lambda}_{n,j}$ is large,  and neuron $n$ will be more likely not to be included in ensemble $j$. 

\textbf{Latent state and dynamics coefficient update:} The ensemble activities $\bm{x}$ and coefficients $\bm{c}$ are updated iterative for each time-point $t$ via the LASSO optimization
\begin{align}
    {\widehat{\bm{x}}_t, \widehat{\bm{c}}_t = \arg \min_{\bm{x}_t, \bm{c}_t} \| \bm{y}_t - \bm{a}\bm{x}_t \|_2^2 
+ \left\| \bm{x}_{t} - \left( \sum_{k=1}^K \bm{f}_k \bm{c}_{k,(t-1)} \right)  \bm{x}_{t-1} \right\|_2^2  + \lambda_c \|\bm{c}_t\|_1}\label{eqn:update_x_c}
\end{align}
where $\lambda_c \in \mathbb{R}$ 
is sparsity-regularization weights on the basis interactions' activations.

\textbf{Dynamical system update:} The global ensembles interactions, $\{ \bm{f}_k \}_{k=1}^K$ are assumed to be sparse (i.e. ensembles interactions are not all to all) and are identified directly by:
\begin{align}
\bm{F}^\textrm{all} = \arg \min_{\bm{F}^\textrm{all}} \| \bm{x}^{+} - 
\bm{F}^{\textrm{all}} (\bm{cx}) \|_2^2 + \lambda_{F}  \|\text{vec}(\bm{F}^\textrm{all})\|_1  + \lambda_{\rho} \sum_{{k_1},{k_2}  , ({k_1} \neq {k_2})  } \rho({\bm{f}}_{k_1}, {\bm{f}}_{k_2}) \label{eq:F_infer}
\end{align}
where $\bm{F}^{\textrm{all}} \in \mathbb{R}^{p \times pK}$ is an horizontal concatenation of all $\bm{f}$s, and  $\bm{x}^{+} \in \mathbb{R}^{p\times \sum_{d \in \textrm{batch} \Tau^d}}$ is the horizontal concatenation of all $\{\{\bm{x}_{t+1}^{d}\}_{t=1}^{\Tau^d}\}_{d \in \text{batch}}$, where ``batch'' means that this operation is in practice taken on a random subset of sessions for computational complexity considerations.  
Above, $\bm{cx}_t^\textrm{batch} \in \mathbb{R}^{Kp\times 1} = [(\bm{c}^{d}_t \otimes [1]_{1\times p}) \circ ([1]_{1\times K} \otimes \bm{x}^{d}_t)^T]^T$, such that $\bm{cx} \in \mathbb{R}^{Kp \times \sum_{d \in \text{batch}} \Tau^d }$ is the horizontal concatenation of all $\{\{ (\bm{cx})_t^d \}_{t=1}^{\Tau^d}\}_{d \in \text{batch}}$. 
The operator \(\text{vec}(\cdot)\) flattens a matrix to a vector and $\lambda_F$ is $\ell_1$  sparsity-promoting weight on $\bm{F}^\textrm{all}$.
$\rho(\cdot)$ represents correlation, such that
$ \sum_{{k_1}, {k_2}  } \rho(\bm{f}_{k_1}, \bm{f}_{k_2})$ is a de-correlation term with weight $\lambda_{\rho}$ used to ensure that distinct $\bm{f}$s are not too similar. 
See Algorithm~\ref{algo:optimization} for a method summary and Section~\ref{sec:model_complexity} for complexity analysis.
\vspace{-5pt}

\begin{figure}[t]
    \centering
    \includegraphics[width=0.9\textwidth]{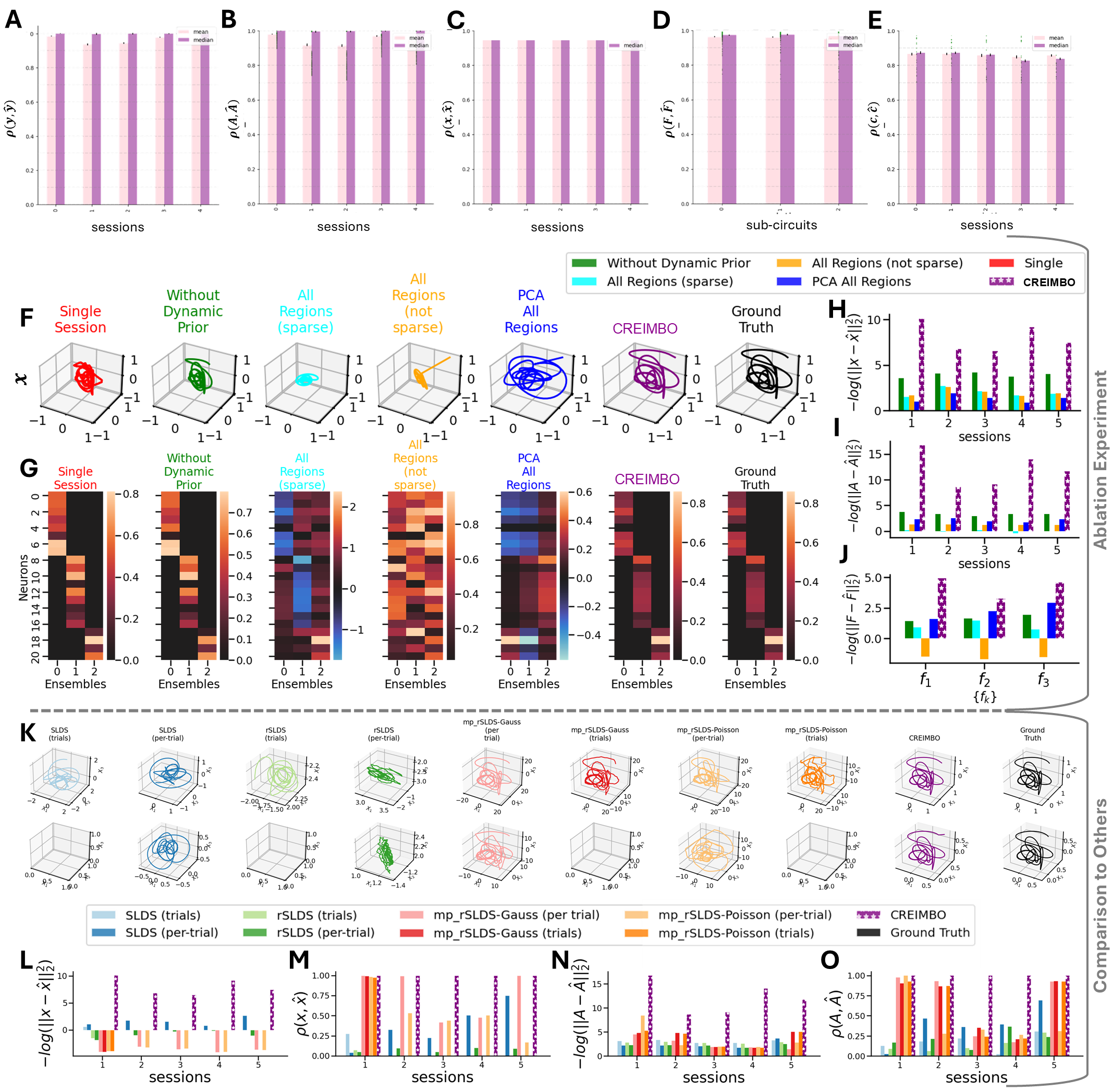}
\caption{
 \textbf{A-E:} Testing CREIMBO over 312 repeats with varying random initializations and random seeds (Sec.~\ref{sec:params_exp_simple}) reveals high correlations between ground truth and fit components for all unknowns, 
$\bm{Y}$, 
$\bm{A}$, 
 $\bm{x}$,
 $\bm{F}$, 
 $\bm{c}$
(subplots \textbf{A}-\textbf{E} respectively).
  \textbf{F-G:}\textbf{Ablation Experiment: } Comparison of the ensemble trajectory ($\bm{x}^d$, subplot F) and the ensemble compositions ($\bm{A}$, subplot G) for the $1$-st session as identified by CREIMBO vs. various ablations (details in Sec.~\ref{sec:baslines}, see Figure~\ref{fig:synth_baselines_comp} for more sessions).
  \textbf{H-J:} Comparing CREIMBO to ablation variants in terms of $-\log(\text{MSE})$ between ground truth and identified dynamics, ensembles, and sub-circuits. {``Single Session'' refers to training CREIMBO on sessions individually. Details in Sec.~\ref{sec:baslines}.} 
  \textbf{K:} \textbf{Baseline Comparisons:} the latent dynamics identified by the baselines (SDLS, rSLDS~\citep{linderman2016recurrent}, multi-population rSLDS~\citep{glaser2020recurrent}, details in Sec.~\ref{sec:baslines}) for the first two sessions (merged sessions baselines provide a single trajectory). 
  \textbf{L-O:} Comparison of the latent dynamics and  identified ensemble matrices compared to the ground truth across the baselines.  
    \vspace{-0.65cm}  } 
  \label{fig:synth_simple_results_concise}
   \vspace{-5pt} 
\end{figure}

\begin{figure}[t]
    \centering
    \includegraphics[width=0.95\textwidth]{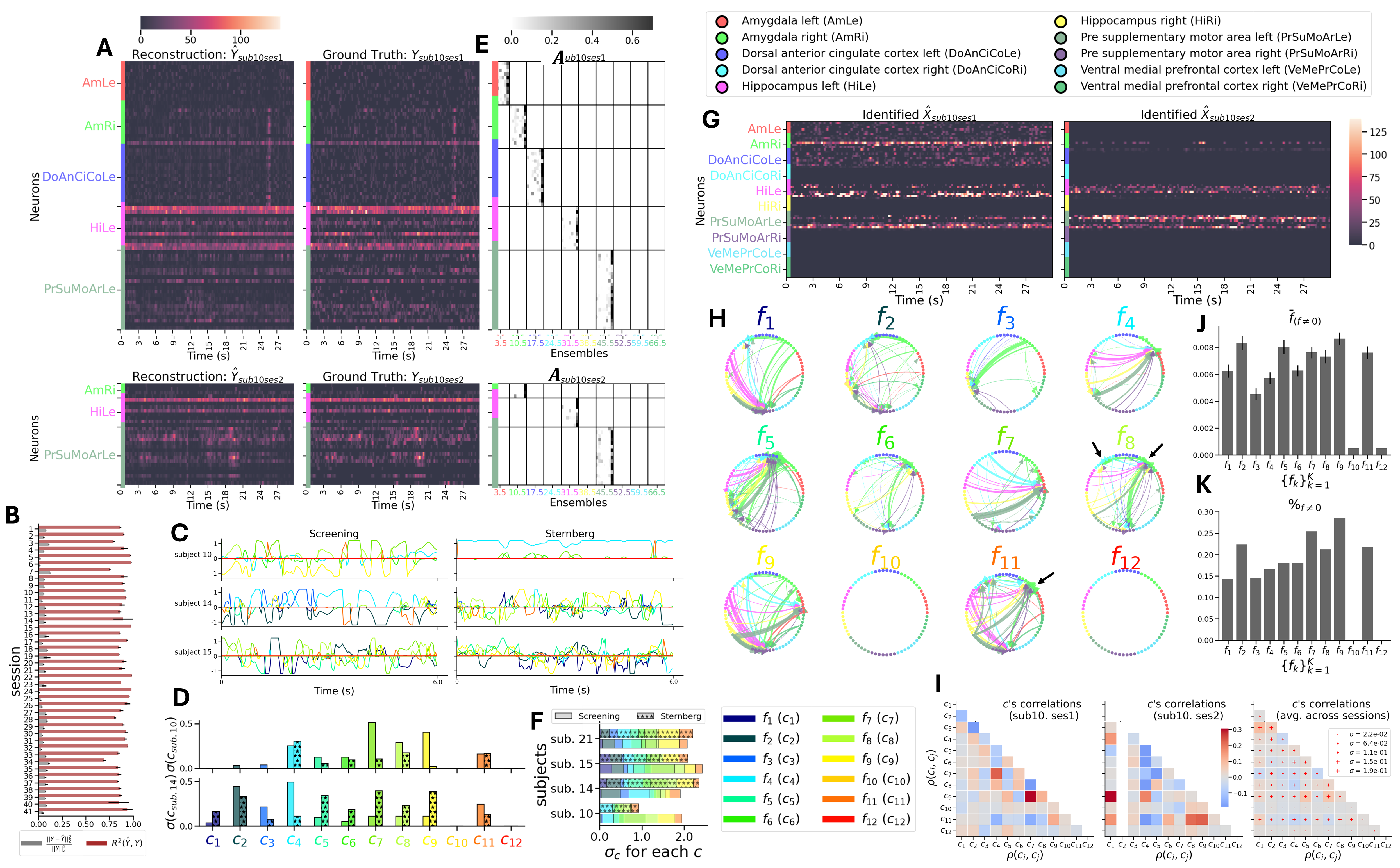}
\vspace{-6pt}
\caption{
CREIMBO identifies cross-regional neural sub-circuits underlying multi-session human brain recordings. 
\textbf{A:} Two exemplary observations ($\bm{Y}$, right) compared to their reconstruction by CREIMBO ($\widehat{\bm{Y}}$, left).
\textbf{B:} Reconstruction performance in terms of $R^2$ and relative error across all sessions.
\textbf{C:} The sub-circuit coefficients for 3 exemplary subjects.
\textbf{D:} The standard deviation (std) of the coefficients over time across the two exemplary sessions.
\textbf{E:} Exemplary identified sparse ensemble matrices ($\{\bm{A}^d \}_{d=1}^D$) for the observations from \textbf{A}.
\textbf{F:} Relative std of the coefficients across four subjects reveals usage of similar sub-circuits.
\textbf{G:} The identified ensembles' activity ($\bm{X}^d$) for the two exemplary sessions.
\textbf{H:} The identified sub-circuits. Width indicates effect magnitude. Each circle corresponds to an ensemble with its color indicating the ensemble's area. Black arrows near $\bm{f}_8$ and $\bm{f}_{11}$ highlight within-region interactions.
\textbf{I:} Pairwise correlations between within-session sub-circuit coefficients (left, middle) and average within-session correlations across all sessions. STD marked by  ``+'' marker size.
\textbf{J:} Mean and STD of connections $f$s.
\textbf{K:} Number of non-zero connections per  $\bm{f}_k$.
\vspace{-0.7cm}}\label{fig:results_D_y_f_full_letters}
\end{figure}

\section{Experiments}

\vspace{-8pt}
\textbf{CREIMBO recovers Ground Truth Components in Synthetic Data}:
We first assess CREIMBO's ability to recover ground-truth components from synthetic data. We generated $K = 3$  sub-circuits represented by a set of 3 rotational matrices (Fig.~\ref{fig:synth_simple_description}B). These sub-circuits capture interactions among $p = 3$ ensembles from $J = 3$ regions ($p_j = 1$ for all $j = 1, \dots, J$). Each region consists of a random number of neurons drawn from a uniform distribution between $4$ to $9$. 
We finally generated the data by simulating 5 sessions with different neuron identities and varying numbers of neurons per region(Fig.~\ref{fig:synth_simple_description}A and D).
CREIMBO was able to recover the ground truth components with high accuracy, as measured by correlation with the ground truth, across $312$ random noisy initialization repeats with random parameters from a range described in~\ref{tab:parameters_synth}. Accurate recover included all parameters: the reconstructed observations $\bm{y}$ (Fig.~\ref{fig:synth_simple_results_concise}A), ensemble composition $\bm{A}$ (Fig.~\ref{fig:synth_simple_results_concise}B) and activity $\bm{x}$  (Fig.~\ref{fig:synth_simple_results_concise}C),   
sub-circuits $\bm{F}$ (Fig.~\ref{fig:synth_simple_results_concise}D),   
and coefficients $\bm{c}$ (Fig.~\ref{fig:synth_simple_results_concise}E). 
Comparing CREIMBO's results to components identified under diverse ablations 
(details in~\ref{sec:baslines}), we found that approaches that either separate the ensembles and dynamics identifications steps, analyze each session separately, or do not consider the localized structure of the ensembles, fail to recover the ground-truth components, i.e., the resulting fit model parameters are less correlated with the ground-truth observations (Fig.~\ref{fig:synth_simple_results_concise}F-J, \ref{fig:synth_baselines_comp}). 
We further evaluated CREIMBO's ability to recover hidden components compared to other methods, (Fig.~\ref{fig:synth_simple_results_concise} K-O, Fig.\ref{fig:slds_compare}. baselines details in Sec.~\ref{sec:baslines}). While the latter shows improved performance, CREIMBO  more accurately recovers the ground truth hidden dynamics and ensembles. Importantly, these models are not designed to identify multiple co-occurring processes or sparse ensembles, in contrast to CREIMBO, and they are not suitable for multi-sessions with varying neural identities, which limits the comparison from the outset.

Next, we tested CREIMBO on a richer synthetic example with $D = 15$ sessions. Each session had a varying number of neurons (a random integer between $14$ and $19$ per region from a uniform distribution) and distinct sub-circuit coefficients (see Fig.~\ref{fig:results_for_2nd_synthetic_experiment} for the generated ground truth components and~\ref{fig:second_synth_area_dist} for regional distributions and masking). The sub-circuits were set as rotational matrices, similar to the previous experiment. Over $204$ random initializations with different parameters (see Sec.~\ref{sec:params_exp_simple}) we again found that CREIMBO robustly identifies the ground truth latent dynamics, ensemble compositions, and sub-circuits. Moreover, CREIMBO accurately reconstructed the observations with high correlation to the ground truth across all initializations (Fig.~\ref{fig:results_for_2nd_synthetic_experiment}E) with low relative error (Fig.~\ref{fig:results_for_2nd_synthetic_experiment}F). 
We further tested CREIMBO on more advanced data that included $D = 40$ sessions with a maximum of $4$ regions and $p_j = 3$ ensembles per region (Fig.~\ref{fig:synth_fig_multi_ens}). We found that CREIMBO manages to recover the components with high correlation with the ground-truth (Fig~\ref{fig:synth_fig_multi_ens_res}). 
\vspace{-6pt}

\\
\\
\textbf{CREIMBO Discovers Multi-Regional Dynamics in Electrophysiology  Data: } 
We then tested CREIMBO on human neural recordings from a high-density electrode array provided by~\cite{kyzar2024dataset}. The data consists of neural activity from overall $p = 10$ brain areas with limited cross-session regional overlap (Fig.~\ref{fig:sars_regions_figure}, \ref{fig:num_neurons_per_area_per_session}),
recorded while subjects performed a screening task (details in~\cite{kyzar2024dataset}). The data encompasses $21$ subjects across $D = 41$ non-simultaneous sessions. For each subject (except Subject~19), the data offers 
1) a ``Screening'' session, 
and 2) a ``Sternberg test'' session  (details in~\cite{kyzar2024dataset}).
We first converted the spike-sorted units to firing rate matrices by convolving the spike trains with a $30$ms Gaussian kernel (Fig.~\ref{fig:example_rate_estimate}), 
and then tested CREIMBO on 
all sessions together with a maximum of $K = 12$ sub-circuits
and as most $p_j = 7$ ensembles per region (full parameter list in Tab.~\ref{tab:parameters_human}). 
First we note that CREIMBO can accurately reconstruct the data with high accuracy across all sessions (Fig.~\ref{fig:results_D_y_f_full_letters}A, B). Interestingly, the ensemble compositions identified by CREIMBO (Fig.~\ref{fig:results_D_y_f_full_letters}E) present sparse patterns with one dense ensemble for most regions that we hypothesize capture the ``background'' mean field activity of that region (right-most column of diagonal blocks in Fig.~\ref{fig:results_D_y_f_full_letters}E). The sparse ensembles we hypothesize capture the more specialized, nuanced functionality.
The ensembles activity ($\bm{X}^d$, Fig.~\ref{fig:results_D_y_f_full_letters}G) reveal some constantly-active ensembles, implying these neuron groups are important for neural processing.

The universal sub-circuits ($\{\bm{f}_k\}_{k=1}^K$) exhibit distinct localized motifs (Fig.~\ref{fig:results_D_y_f_full_letters}H), i.e., most sub-circuits present clear trends of either sourcing or targeting the same area. This is in contrast to other approaches (Sec.~\ref{sec:baslines}), that yield overlapping uninterpretable sub-circuits (Fig.~\ref{fig:f_corrs_baselines_human}) that emphasize only limited circuitry (Fig.~\ref{fig:human_baselines_f}).
Interestingly, sub-circuits $10$, $12$ ended up almost empty (mainly zero-ish values, Fig.~\ref{fig:results_D_y_f_full_letters}H, \ref{fig:fs_heatmaps}). This highlights CREIMBO's ability to automatically nullify redundant sub-circuits---i.e., a form of model selection---via the sparsity regularization. The number of active connections and average connection strength (Fig.~\ref{fig:example_rate_estimate}J, K)
 show modest variability between sub-circuits, highlighting the model's ability to identify distinct sub-circuits with varying interaction resolutions, including those of varying strengths rather than being limited to just k-sparse sub-circuits.
From a neuroscience perspective, this suggests that underlying neural sub-circuits can vary in density and the number of participating areas and ensembles.
Interestingly, most identified interactions occur between distinct areas (i.e., inter-regional interactions) with either multiple ensembles from the same area affecting together ensembles from a different area (e.g., $\bm{f}_3$), or multiple ensembles from diverse areas converging to the same target area (e.g., $\bm{f}_5$). However, these source or target groups of ensembles vary across sub-circuits, hence the full repertoire of activity cannot be captured by a single one, highlighting the importance of CREIMBO's capacity to disentangle the combined activity of multiple circuits. 
We  also observe, but to a lesser degree, within-area interactions (e.g., black arrows in $\bm{f}_{8}$, $\bm{f}_{11}$). These dynamics emphasize that within-area amplification or regulation do exist in addition to global cross-region computations. 
When exploring the activation patterns of these identified sub-circuits (Fig.~\ref{fig:results_D_y_f_full_letters}D, Fig.~\ref{fig:coefficients_per_subject_and_type}), CREIMBO reveals cross-subject and cross-session variability. Notably, some sub-circuits are consistently used with high variability over time (e.g., $\bm{f}_{8}$ light-green), while others, such as $\bm{f}_{12}$, are not utilized at all in any exemplary session (Fig.~\ref{fig:results_D_y_f_full_letters}C, F, e.g., $\bm{f}_{12}$).
Moreover, when exploring sub-circuits with high activity variability across sessions (Fig.~\ref{fig:results_D_y_f_full_letters}D,F), we observe that some are used within-subjects but across both tasks (e.g., $\bm{f}_2$ in subject 14, Fig.~\ref{fig:results_D_y_f_full_letters}D), while others are shared across both subjects and tasks (e.g., $\bm{f}_{11}$, used by subjects 10 and 14 in both task phases).
The analysis of within-session pairwise correlations between sub-circuit coefficients (Fig.~\ref{fig:example_rate_estimate}I) 
reveals low correlations ($<$0.1) for most pairs. This indicates that the activities of distinct sub-circuits differ and reflect different cognitive processes.

{
Since ground-truth components are unavailable in real data, direct validation of such components is impossible. Hence, we sought a proxy to evaluate CREIMBO’s performance on real-data by testing its 
consistency under increasing observation noise levels. }
{Particularly, we re-ran CREIMBO on the human recordings after adding  increasing variance \textit{i.i.d} Gaussian noise} (with $K=8, p_j = 6$ 
Fig.\ref{fig:fig_robust_noise}).
CREIMBO remains robust to increasing noise, 
and experiences a phase transition at a specific noise level 
 ($\frac{\sigma_\text{noise}}{\sigma_\text{data}} \sim 0.2$,Fig.\ref{fig:fig_robust_noise} A,B, E, F,H,I), which align with the dictionary-learning literature~\citep{studer2012dictionary}.
In this noise experiment, CREIMBO further identifies that within-area ensemble composition correlations (Fig.\ref{fig:fig_robust_noise}C,D) remain stable until a noise level of around $\sigma_\text{noise} \approx 10$, after which these correlations weaken as noise increases, which may imply that high noise in data can obscure meaningful relationships between ensembles.
Additionally, it identified that the dynamics' coefficients remain robust until $\sigma_\text{noise} \approx 1.8$, while exhibiting an increasing internal frequency as the noise increases, 
potentially to account for noise that does not follow dynamical rules (Fig.~\ref{fig:fig_robust_noise}G,  ~\ref{fig:noise_and_c}). This suggests that frequency of fast transitions 
in CREIMBO's ${\bm{c}_{k,t}}$ could indicate noise in the system that does not adhere to the underlying dynamics.

\begin{figure}[t]
    \centering
    \includegraphics[width=\textwidth]{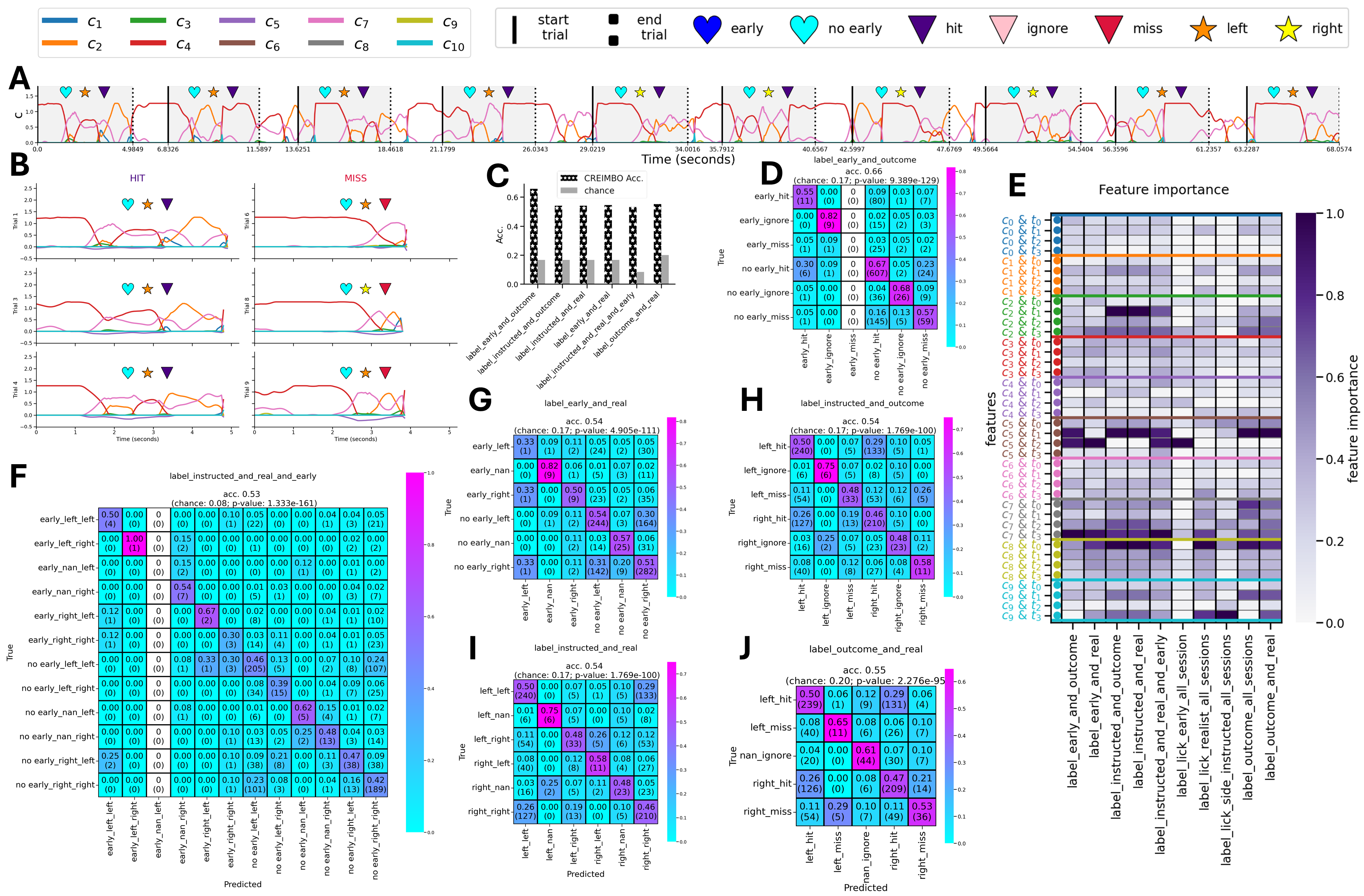}
\caption{{
Results of task-variable prediction using CREIMBO's dynamic coefficients as input. Prediction was done based on coefficients from all used trials and sessions (19 sessions, 40-60 trials each, see Sec.~\ref{sec:meso_experiment}). 
\textbf{A:} Dynamic coefficients from example traces, shade marks within trial, bordered by starting and ending point. 
\textbf{B:} Dynamic coefficients for correct (``hit'') vs. wrong (``miss'') trials. 
\textbf{C:}  Accuracy score of predicting task variables based on coefficients vs. chance. 
\textbf{D, F,G,H,I:} Confusion matrices of predicting varying task variables based on dynamic coefficients. 
\textbf{E:} importance of the different coefficients and time points for prediction. 
    }}
    \label{fig:mesoscale_classifier}
    \vspace{-0.55cm}
\end{figure}

{ \textbf{CREIMBO Discovers Regional Interactions Predictive of Task Variables}}: 
{
We tested CREIMBO's ability to infer task-related variables from mice whole-brain Neuropixels multi-session data during a memory-guided movement (data from~\cite{chen2024brain}, Fig.~\ref{fig:mesoscale_all_areas},~\ref{fig:mesoscale_all_areas_b}).
CREIMBO identifies intra- and inter-area brain interactions via the sub-circuits (Sec.~\ref{sec:meso_experiment}, Fig.~\ref{fig:mesoscale_identified_components}C,D), including cross-regional flows into or from key areas responsible for memory (from $\bm{f}_5$ and to $\bm{f}_8$ the hippocampus), planning (from frontal cortex, e.g. $\bm{f}_3$), and movement (from primary motor cortex, $\bm{f}_4$), which are needed for the task. 
Within-area sub-circuits (Fig.~\ref{fig:mesoscale_identified_components}D) further showed within-area ensemble interactions (e.g., the basal ganglia in $\bm{f}_1$ and the secondary motor cortex in $\bm{f}_2, \bm{f}_7$, along with self-activation/inhibition of ensembles in other sub-circuits (e.g. $\bm{f}_4$).  
Moreover, CREIMBO's dynamic coefficients (${\bm{c}_{kt}}$) captured task-related patterns across trials and sessions (Fig.~\ref{fig:mesoscale_classifier}A,B). When training a regularized logistic-regression model, we found that we could predict from CREIMBO's circuits activations ($\bm{c}_{kt}$) various task variables including outcome, early lick, and lick side far above chance levels (Fig.~\ref{fig:mesoscale_classifier}C-J, p-values < \(1 \times 10^{-10}\)).
Furthermore,  CREIMBO's sub-circuit activations across different task periods highlight how specific multi-regional interactions capture different aspects of the task. For instance, towards the trial end ($t_3$ window), which includes the lick movement, $c_7$ shows increased importance (Fig.~\ref{fig:mesoscale_classifier}E). Notably, $c_7$ corresponds to the activity of the $\bm{f}_7$ sub-circuit that captures flows to the secondary motor cortex (Fig.~\ref{fig:mesoscale_identified_components}C). 
Another example is $c_8$ (activity of sub-circuit $\bm{f}_8$ that includes flows into hippocampus~\ref{fig:mesoscale_identified_components}C), which shows increased feature importance in the first time window $t_0$.  This aligns with the potential involvement of memory-related processes required with the stimulus appearance. 
Thus CREIMBO demonstrates the ability to capture interpretable neural dynamics and predict task variables that reveal complex regional interactions.
(see Sec.~\ref{sec:meso_experiment} for more detailed results).
}

\section{Discussion, Limitations, and Future Work}
\vspace{-10pt}
{Here,} we introduced CREIMBO---a 
novel approach for uncovering multi-regional dynamics in neural data collected across multiple sessions. CREIMBO addresses the challenge of integrating non-simultaneous neural recordings by joint dynamical inference and sparsity regularization to capture the underlying neural sub-circuits governing brain activity. We further demonstrated the efficacy of CREIMBO through multiple synthetic and neural data,  and found that CREIMBO recovered ground truth components and is robust to noise in identifying cross-regional motifs that span cross-session interactions.
CREIMBO offers several advantages over existing methods. 
Chiefly, it identifies a cross-session shared latent space where non-stationary ensemble interactions are governed by a time-varying decomposition of universal basis dynamics. 
By structuring these 
dynamics in terms of global sub-circuits, CREIMBO allows the discovery of sub-circuits meaningful for various cognitive processes, enabling the identification of variability in neural activity across subjects and sessions.
\\
\textbf{Limitations \& Future Directions:}
An important feature of CREIMBO is its ability to unify  
sessions with different neuron subsets through the universal  dictionary of dynamical interactions prior 
 that aligns ensembles in terms of functionality.
This can enable the inference of their activity even if they are not observed in some sessions. This ability, however, depends on the overlap between the used basis interactions across sessions, the distinctiveness of different $\bm{f}_k$s, and the  premise that the basis interactions 
capturing interactions with ensembles missing from a session include dynamic values for ensembles observed in that same session
(Sec.~\ref{sec:assumptions}). 
Another limitation is CREIMBO's reliance on linear projections from ensemble to session observation space, restricting flexibility. Extending to non-linear projections offers potential for development, though it introduces computational and interpretability challenges. 
Finally, CREIMBO uses dictionary learning, which is computationally demanding, and future iterations will include parallel processing.
Extending CREIMBO to additional data types and applications, including 
non-neural data (e.g., immune-cell counts), or for identifying pathologies, is an exciting future step.
{Moreover, integrating CREIMBO with mDLAG~\citep{gokcen2024uncovering}, with the latter identifying optimal communication delays and the former leveraging these to further identify the underlying set of co-active LDSs, presents an exciting future work.}



\newpage
\bibliography{main}
\bibliographystyle{iclr2025_conference}
\newpage

\appendix

\begin{figure}[ht]
    \centering
    \includegraphics[width=0.99\textwidth]{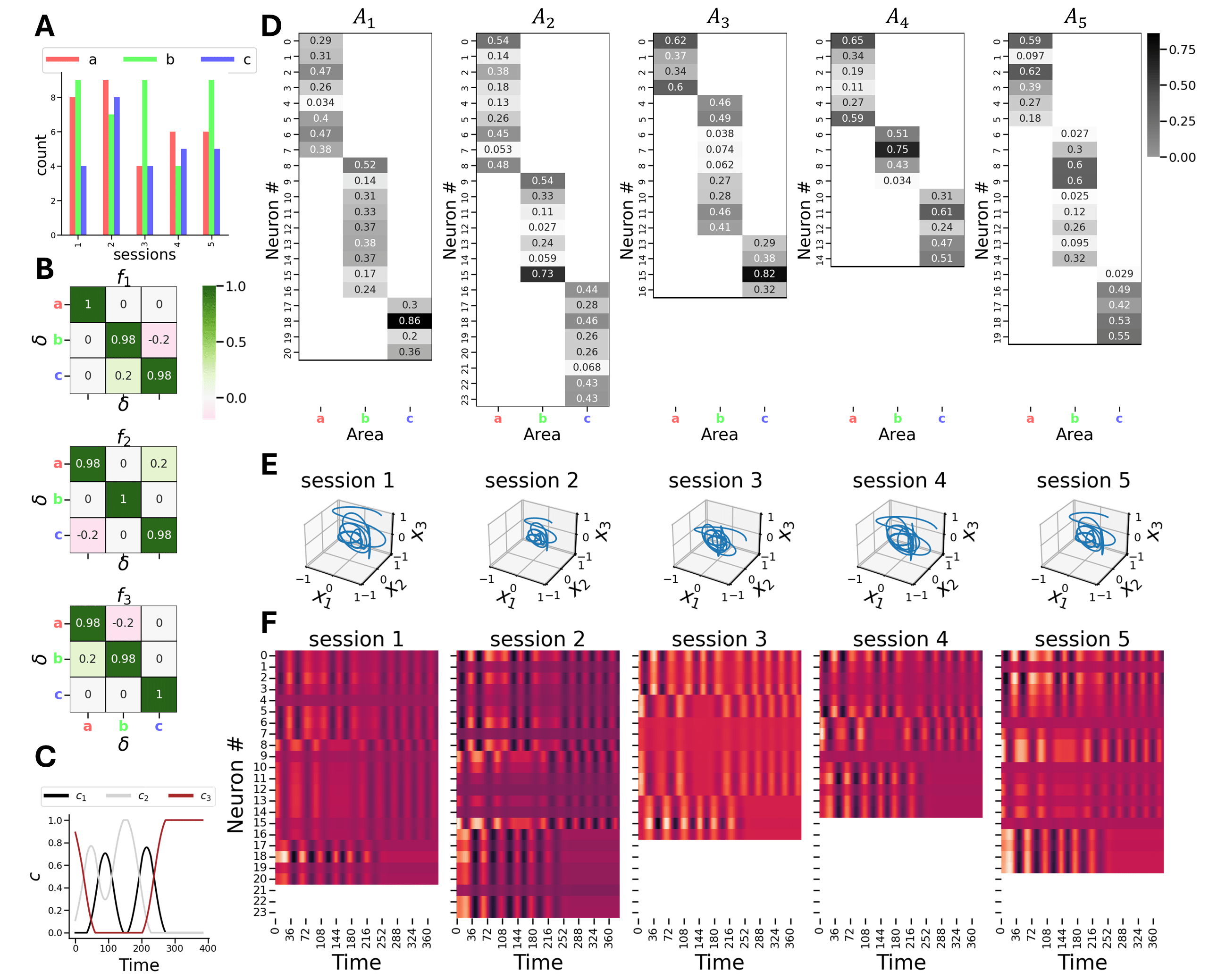}
\caption{
\textbf{A:} Distribution of neurons across areas (areas $a$, $b$, and $c$) over the $5$ different synthetic sessions. 
\textbf{B:} Ground  truth sub-circuits are $K = 3$ rotational matrices, each captures a different rotational direction. 
\textbf{C:} Time-coefficients of synthetic sub-circuits. 
\textbf{D:} Synthetic ensembles ($\{ \bm{A}^d \}^5_{d=1}$
\textbf{E:} Latent (ensemble) dynamics. Different trajectories in the latent space emerge as distinct cross-session time-changing decomposition of the ensemble sub-circuits. 
\textbf{F:} Synthetic Ground Truth observations ($\{\bm{Y}_d\}$)
    }
    \label{fig:synth_simple_description}
\end{figure}

\begin{figure}[ht]
    \centering
    \includegraphics[width=0.99\textwidth]{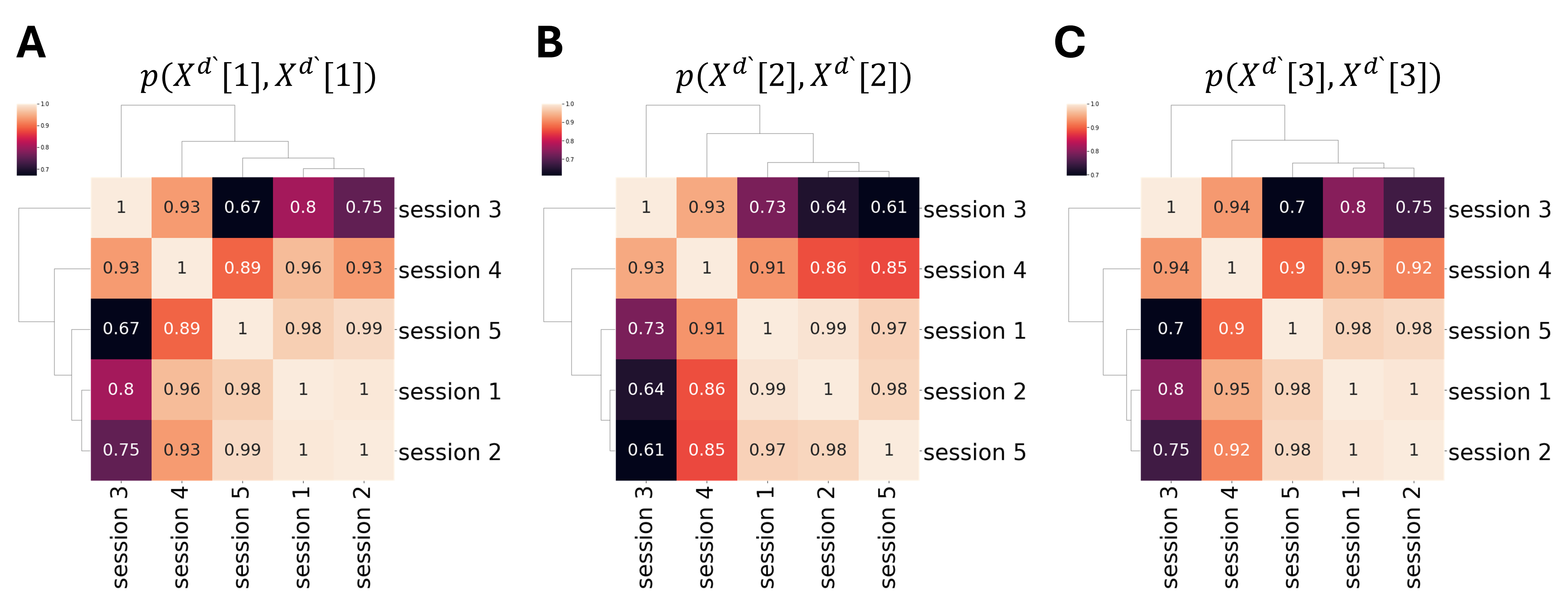}
\caption{{Clustermaps of correlations of cross-session latent dynamics. Each subplot captures one dimension of the latent dynamics.}}
    \label{fig:cross_session_synth_similarity}
\end{figure}

\begin{figure}
    \centering
    \includegraphics[width=0.99\textwidth]{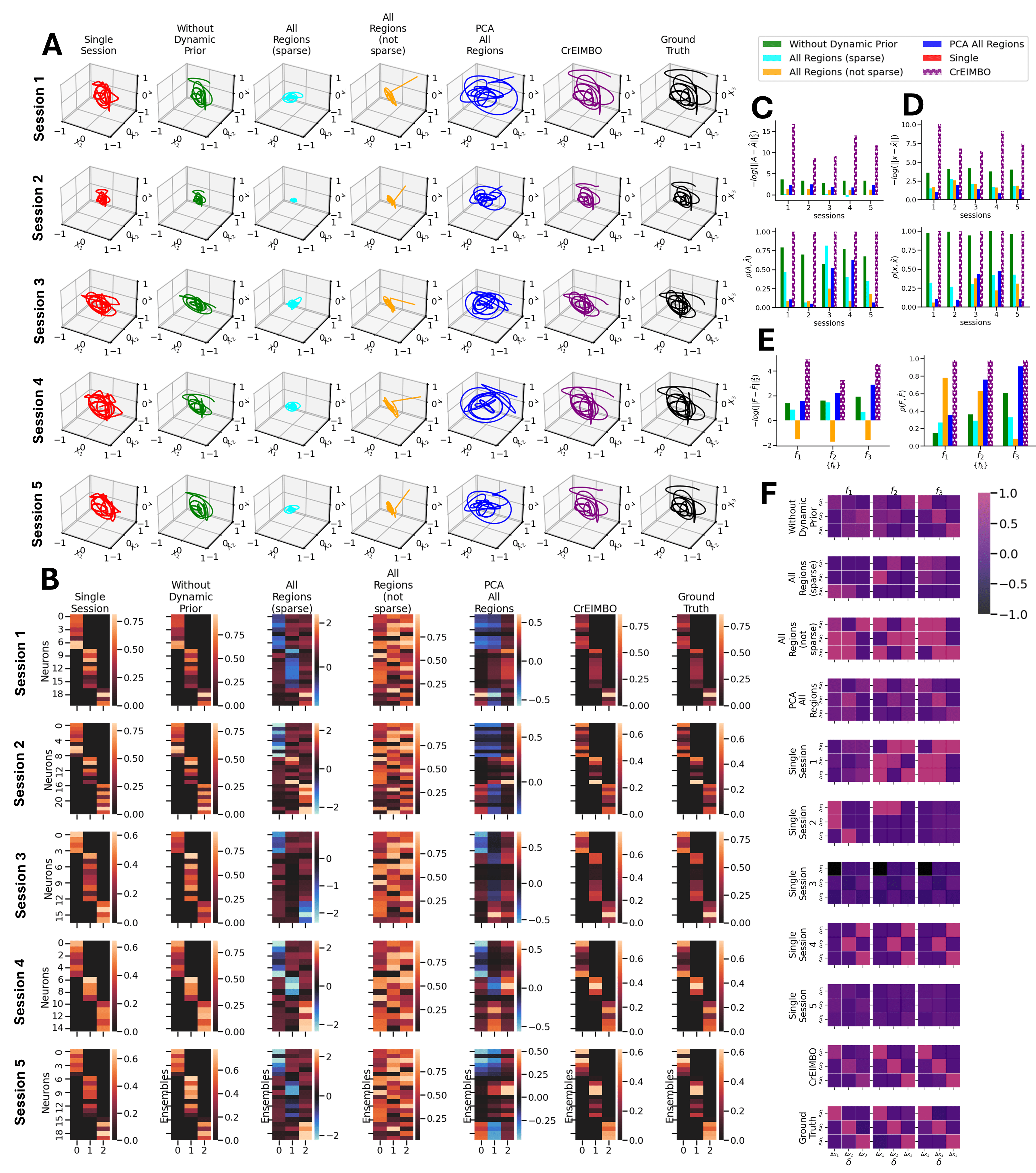}
\caption{
Comparison between CREIMBO to the baselines reveal that CREIMBO outperforms other approaches (see baselines details in Sec.~\ref{sec:baslines}). 
\textbf{A:} Learned latent dynamics ($\bm{X}$) for each of  the $5$ sessions (rows) for the different methods (columns). 
\textbf{B:} Learnt ensemble compositions ($\bm{A}$) for each of  the $5$ sessions (rows) for the different methods (columns). 
\textbf{C-E:} $-log(MSE)$ and correlation between the ground truth and the the identified ensemble compositions (in \textbf{C}), latent dynamics (in \textbf{D}), and sub-circuits (in \textbf{E}).
\textbf{F:} Heatmaps of the identified sub-circuits for the different methods. 
}
    \label{fig:synth_baselines_comp}
\end{figure}

\begin{figure}[ht]
    \centering
    \includegraphics[width=0.99\textwidth]{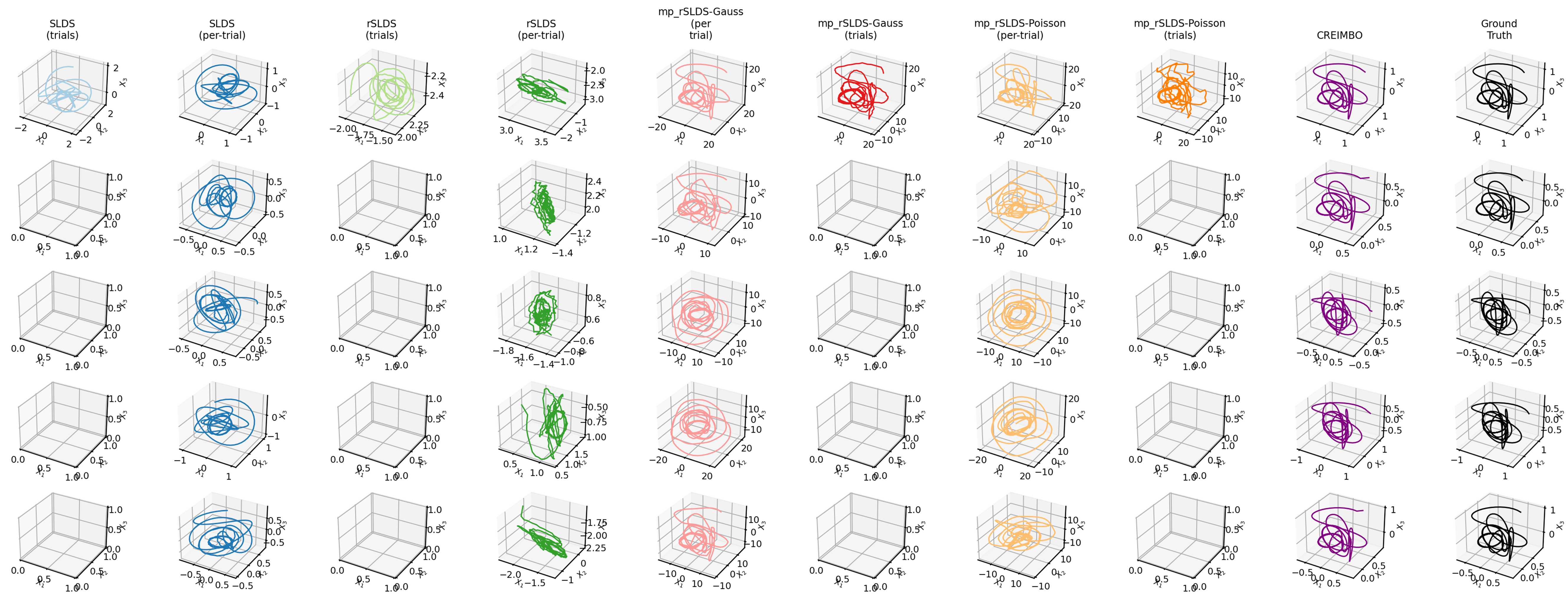}
\caption{Further comparisons of the learned latent dynamics to these learned by SLDS, rSLDS, and multi-region rSLDS, with their variations (see Sec.~\ref{sec:baslines} for details). Different columns represent different methods. Different rows capture different sessions. 
    }
    \label{fig:slds_compare}
\end{figure}

\begin{figure}
    \centering
    \includegraphics[width=0.99\textwidth]{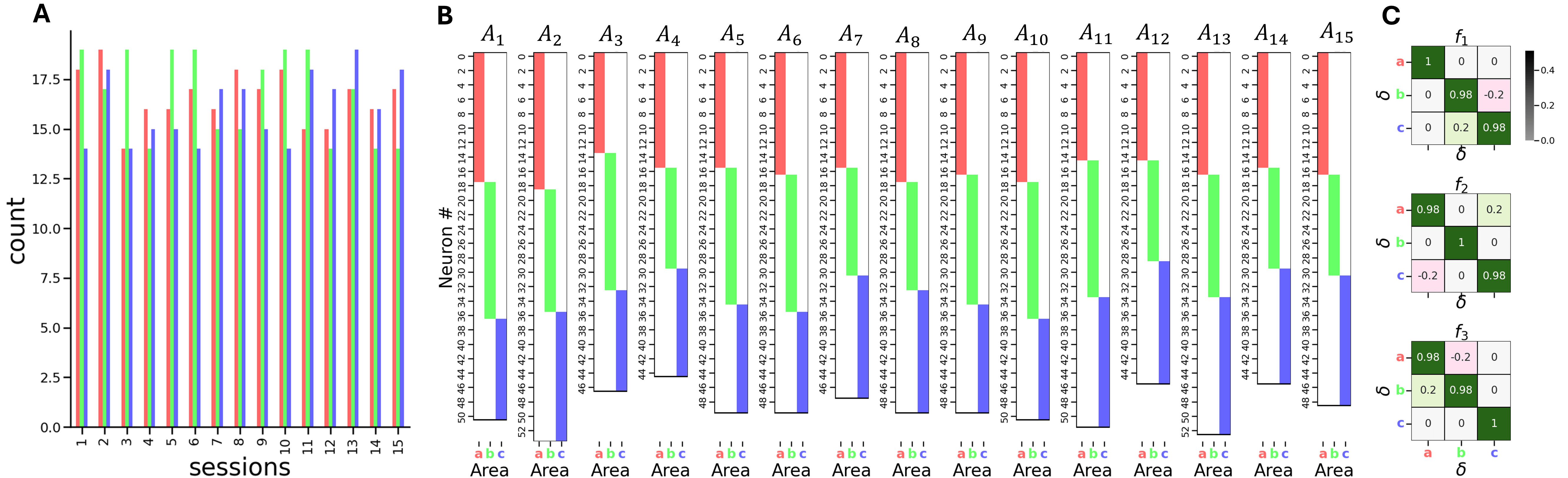}
\caption{Distribution of areas in second synthetic dataset.  
\textbf{A:} Histogram of neural counts per area and session. 
\textbf{B:} Ensemble masks ($\bm{A}_\textrm{mask}$) across sessions. 
\textbf{C:} Ground truth sub-circuits. 
    }
    \label{fig:second_synth_area_dist}
\end{figure}

\begin{figure}[ht]
    \centering
    \includegraphics[width=\textwidth]{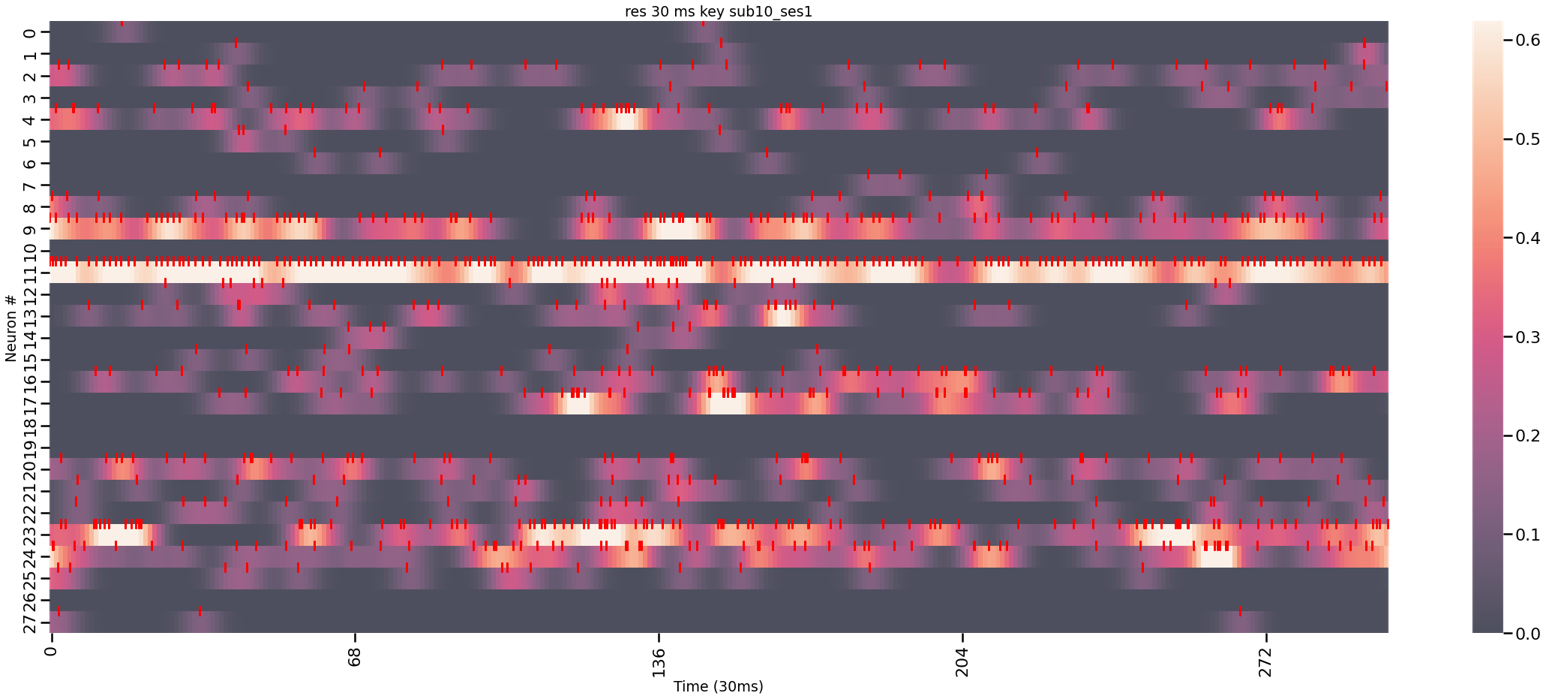}
    \caption{Rate estimation example for a specific session using a 30 ms Gaussian kernel.}
    \label{fig:example_rate_estimate}
\end{figure}


\begin{figure}[ht]
    \centering
    \includegraphics[width=0.99\textwidth]{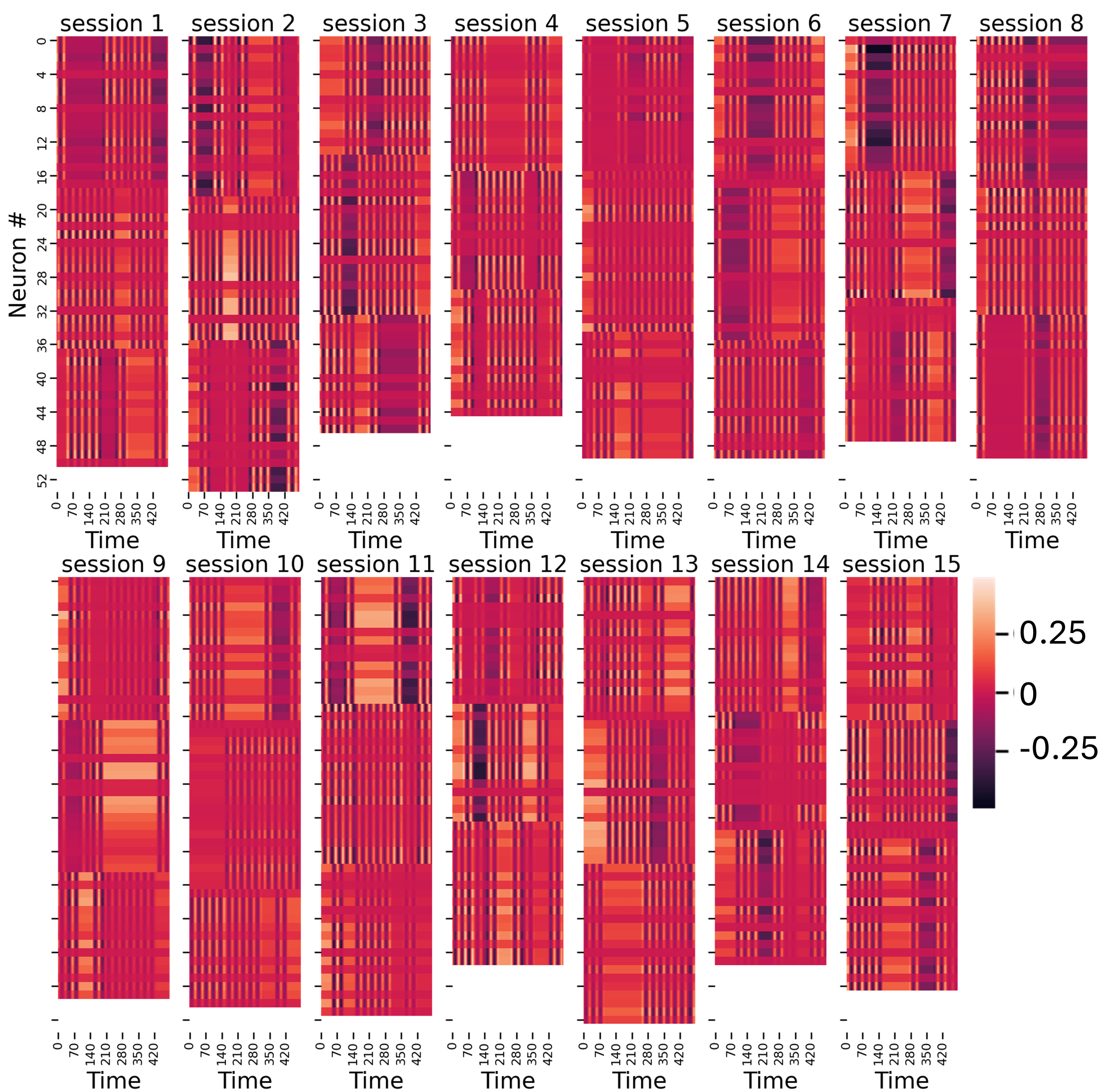}
    %
\caption{ Data and results for the second synthetic experiment (overall $D = 15$ sessions, $J = 3$ regions and $p_j = 1$ ensembles for each region $j = 1\dots J$). 
\textbf{A:} Ground truth synthetic observations. 
\textbf{C:} Ground truth ensemble 
    }
    \label{fig:results_for_2nd_synthetic_experiment}
\end{figure}



\begin{figure}[t]
    \centering
    \includegraphics[width=\textwidth]{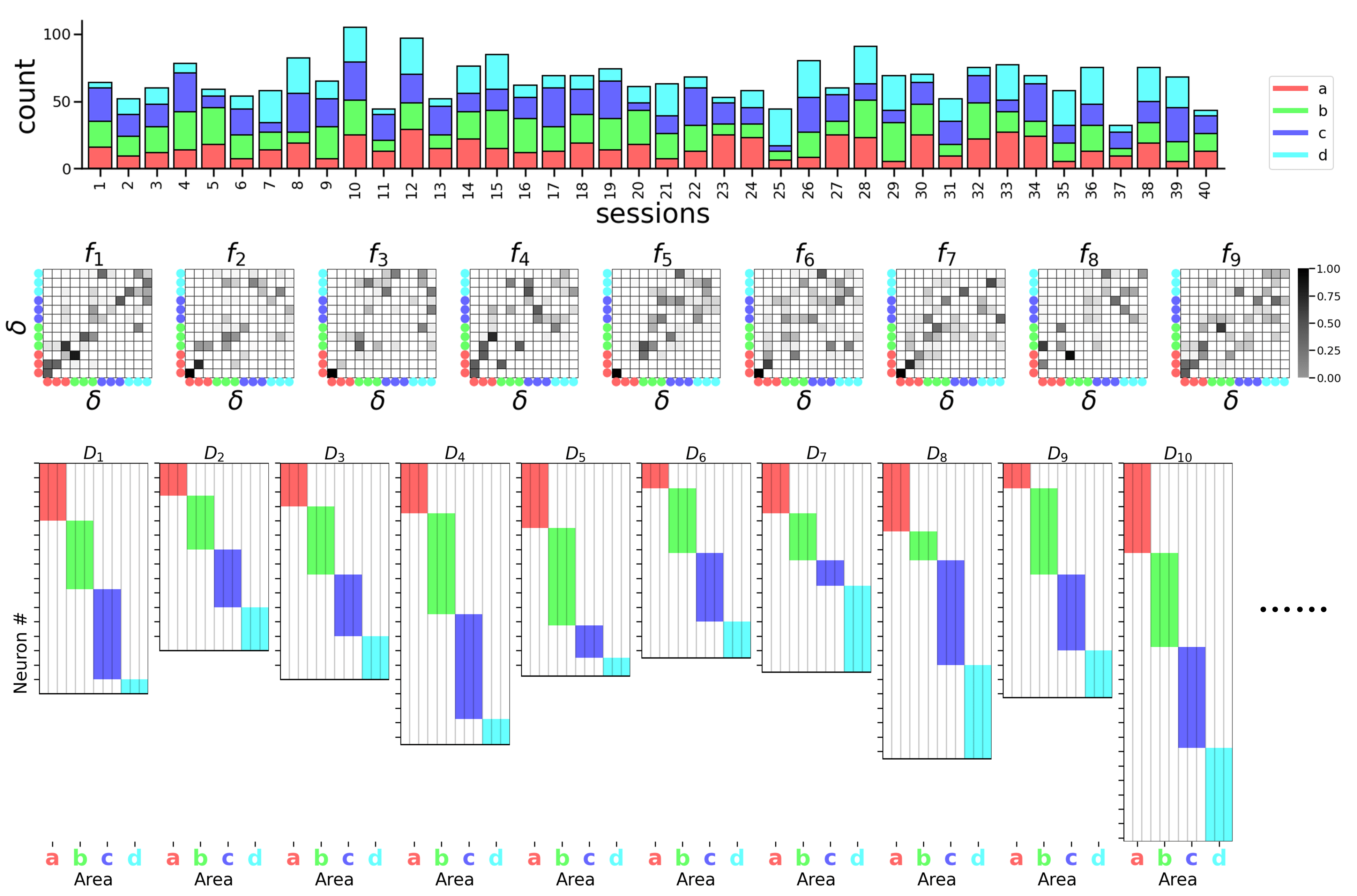}
    \caption{Multi-ensemble synthetic data.
    Top: Distribution of areas across sessions.
    Middle: Ground truth sub-circuits.
    Bottom: Masks for ensembles identification. }
    \label{fig:synth_fig_multi_ens}
\end{figure}

\begin{figure}[t]
    \centering
    \includegraphics[width=\textwidth]{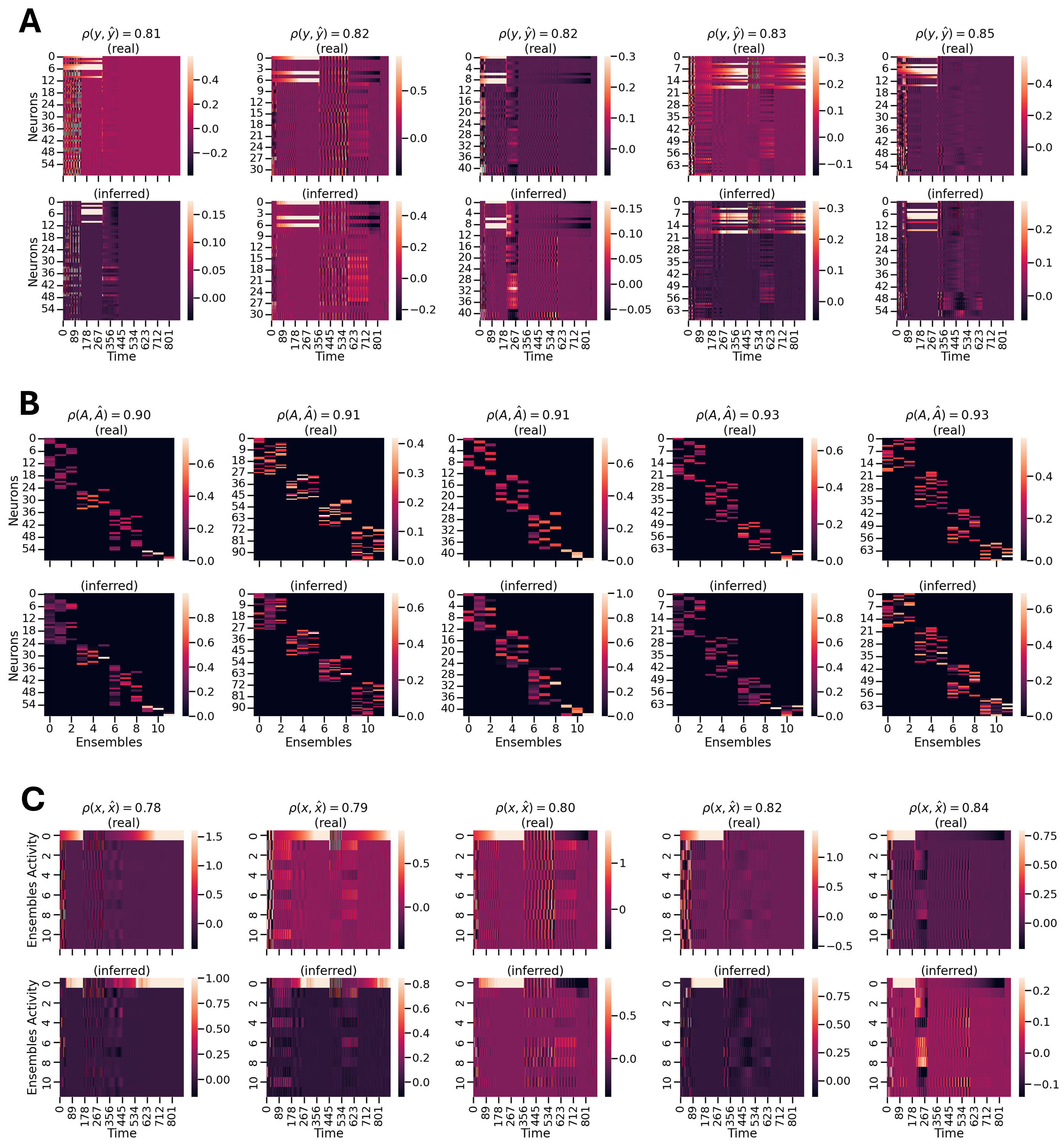}
    \caption{Multi-ensemble synthetic data results.
    Top: Recovering the observations $\bm{Y}$ (five exemplary sessions).
    Middle: Identified Ensembles (five exemplary sessions).
    Bottom: Identified latent dynamics (five exemplary sessions).}
    \label{fig:synth_fig_multi_ens_res}
\end{figure}

\begin{figure}[ht]
    \centering
    \includegraphics[width=\textwidth]{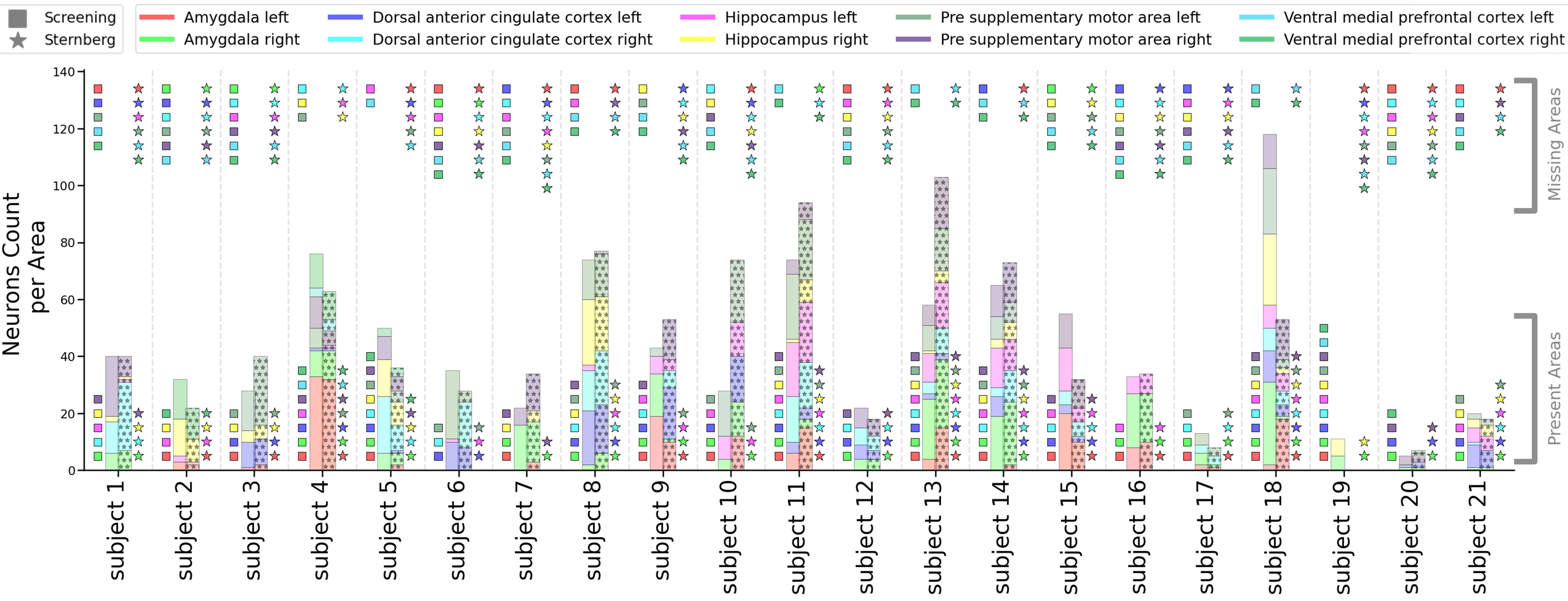}
    \caption{Areas distribution in real world human neural data.}
    \label{fig:sars_regions_figure}
\end{figure}

\begin{figure}[ht]
    \centering
    \includegraphics[width=\textwidth]{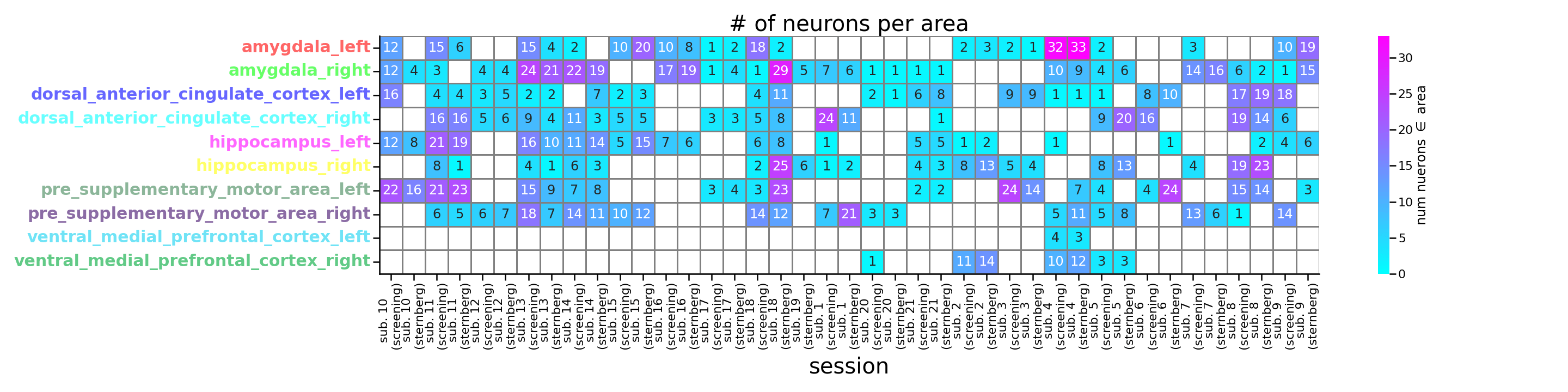}
    %
\caption{   Number of neurons per-area and per-session in the real-world human data taken from~\cite{kyzar2024dataset}. 
    }
    \label{fig:num_neurons_per_area_per_session}
\end{figure}

\begin{figure}[ht]
    \centering
    \includegraphics[width=\textwidth]{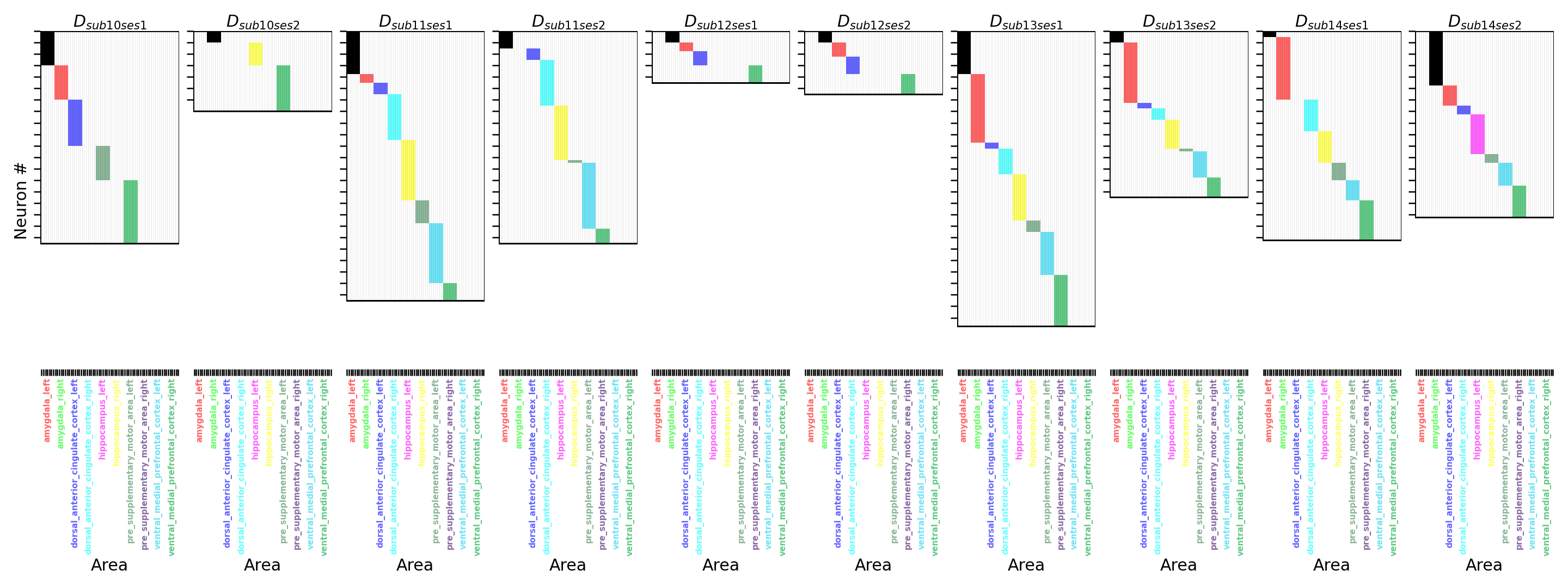}
    \caption{Multi-regional Masks for real-world human data.}
    \label{fig:D_masks_real_world}
\end{figure}

\begin{figure}[ht]
    \centering
    \includegraphics[width=\textwidth]{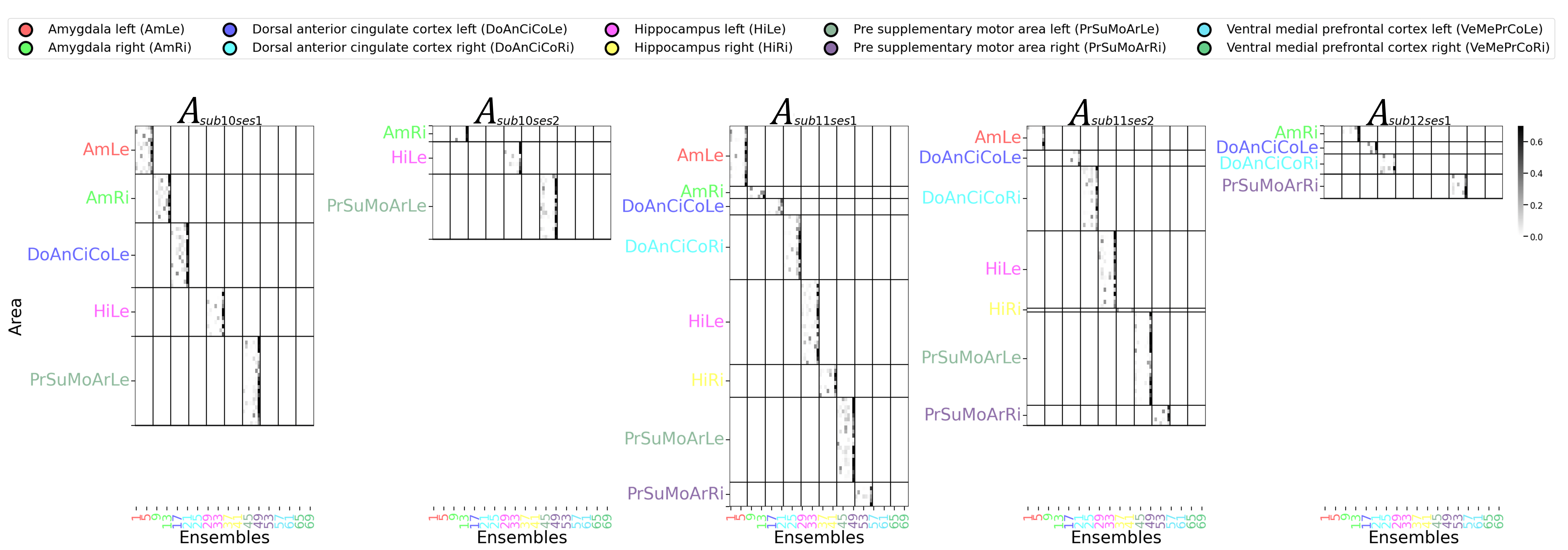}
    \caption{Five Exemplary Ensemble Matrices of the Real World Data, as Identified by CREIMBO}
    \label{fig:Ds_real}
\end{figure}

\begin{figure}[ht]
    \centering
    \includegraphics[width=\textwidth]{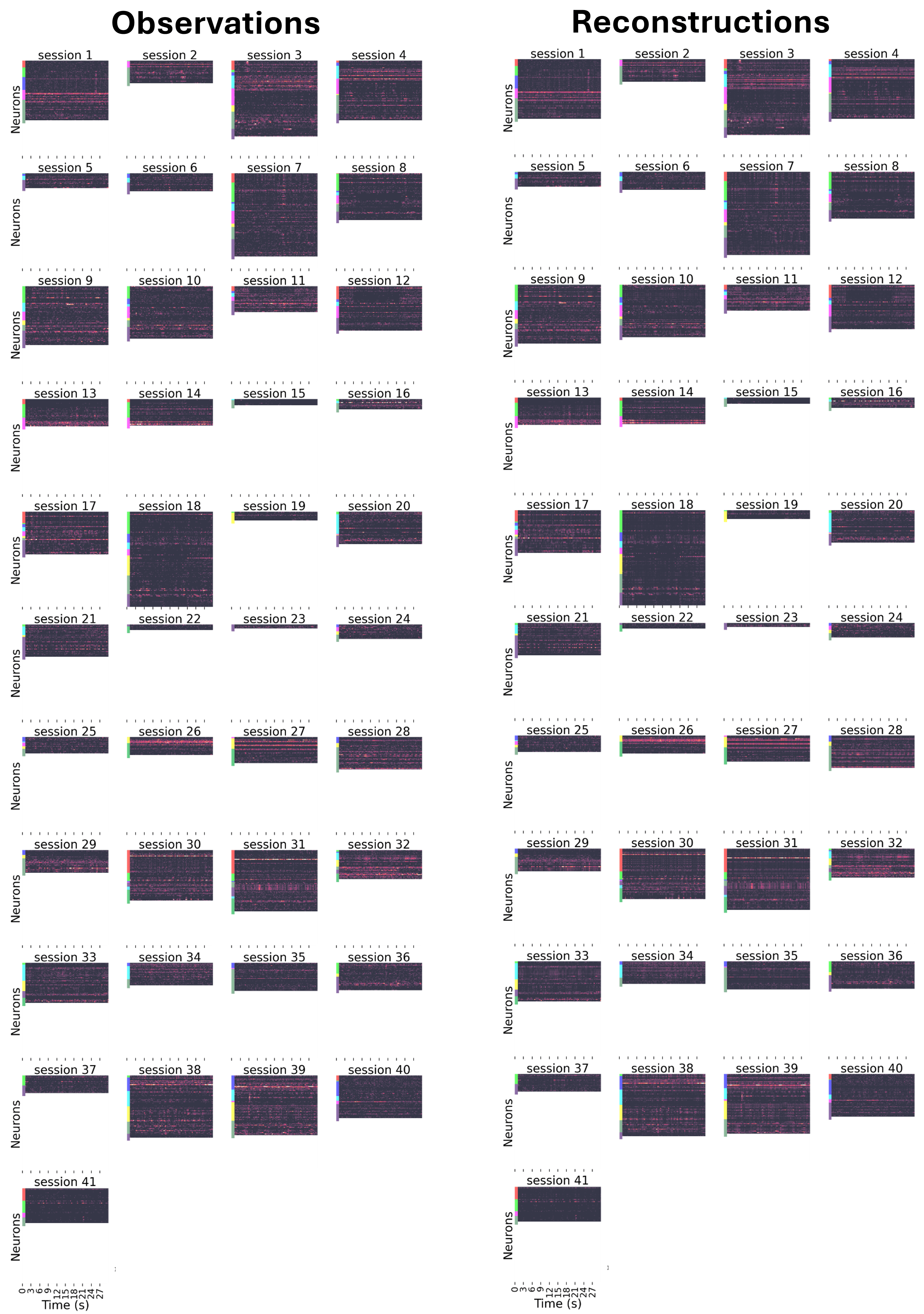}
    \caption{Observations vs. CREIMBO's reconstructions for the real-world human data experiment.}
    \label{fig:res_real}
\end{figure}

\begin{figure}[ht]
    \centering
    \includegraphics[width=\textwidth]{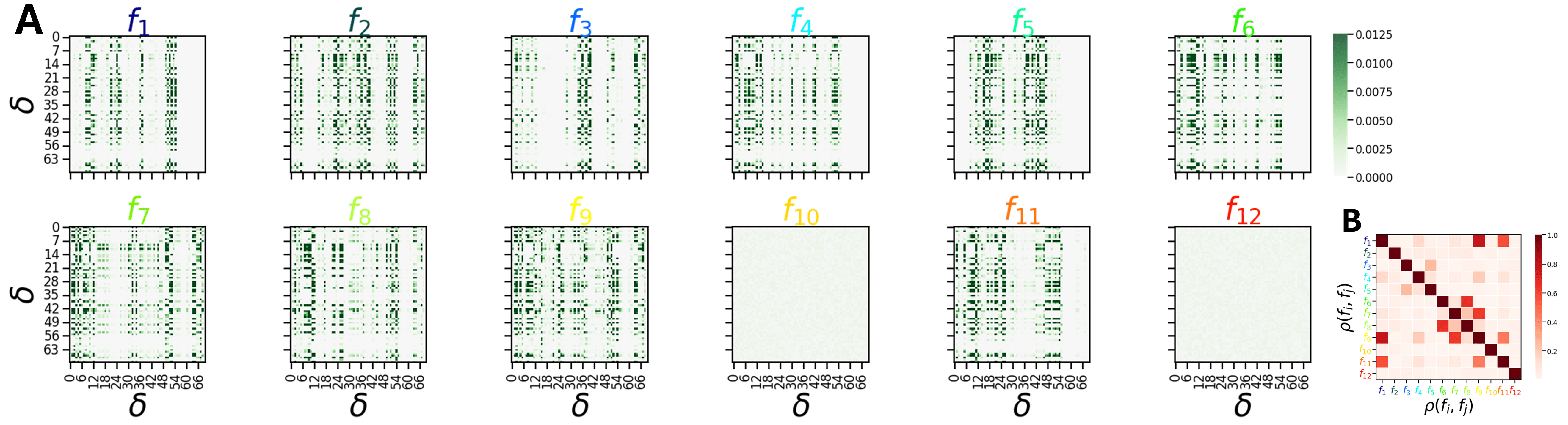}
    \caption{The sub-circuits underlying human data as identified by CREIMBO. \textbf{A}: The identified $12$ sub-circuits.
    \textbf{B}: Pairwise correlations between the identified sub-circuits. }
    \label{fig:fs_heatmaps}
\end{figure}

\begin{figure}[ht]
    \centering
    \includegraphics[width=1\textwidth]{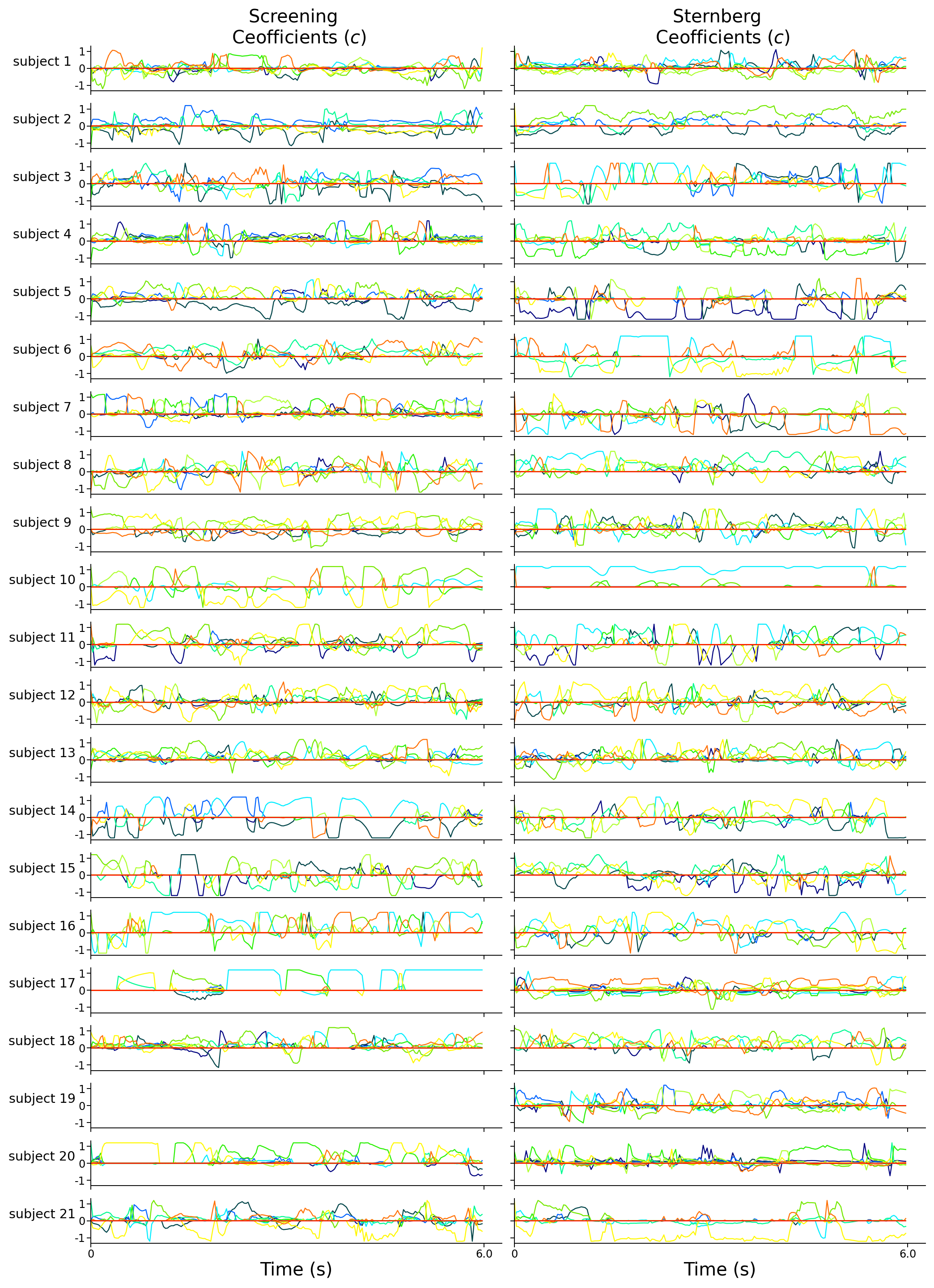}
\caption{  Sub-circuits  Coefficients separated by task type (``Screening'' vs ``Sternberg'') and subject. 
    }
    \label{fig:coefficients_per_subject_and_type}
\end{figure}

\begin{figure}[ht]
    \centering
    \includegraphics[width=1\textwidth]{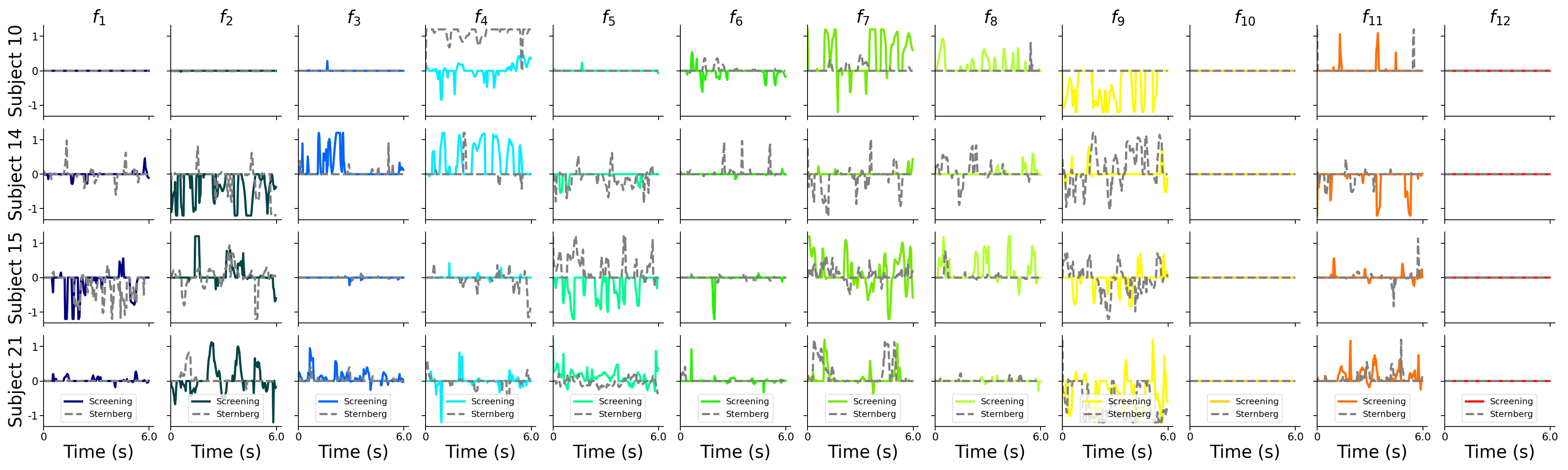}
\caption{  Comparing  the use  of the same coefficient under 'Sternberg' vs 'Screening' within four exemplary subject, reveals usage of similar sub-circuits within subjects. 
    }
    \label{fig:coefficients_per_subject}
\end{figure}

\begin{figure}[ht]
    \centering
    \includegraphics[width=1\textwidth]{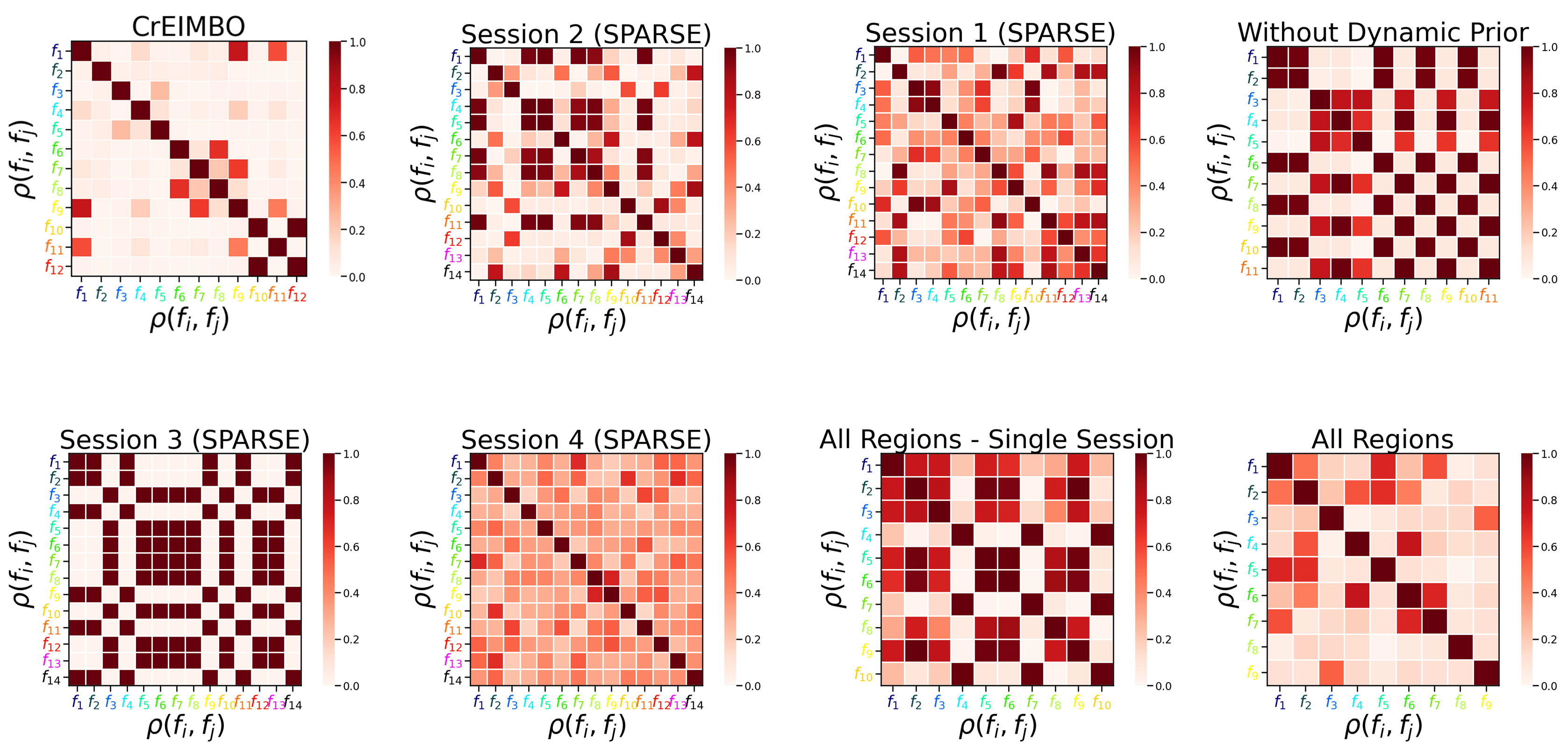}
\caption{ Correlations between the brain sub-circuits  identified by CREIMBO compared to the other methods. These results highlight the CREIMBO discovers more distinct sub-circuits than these idetified by the other methods. }
    \label{fig:f_corrs_baselines_human}
\end{figure}

\begin{figure}[t]
    \centering
    \includegraphics[width=1\textwidth]{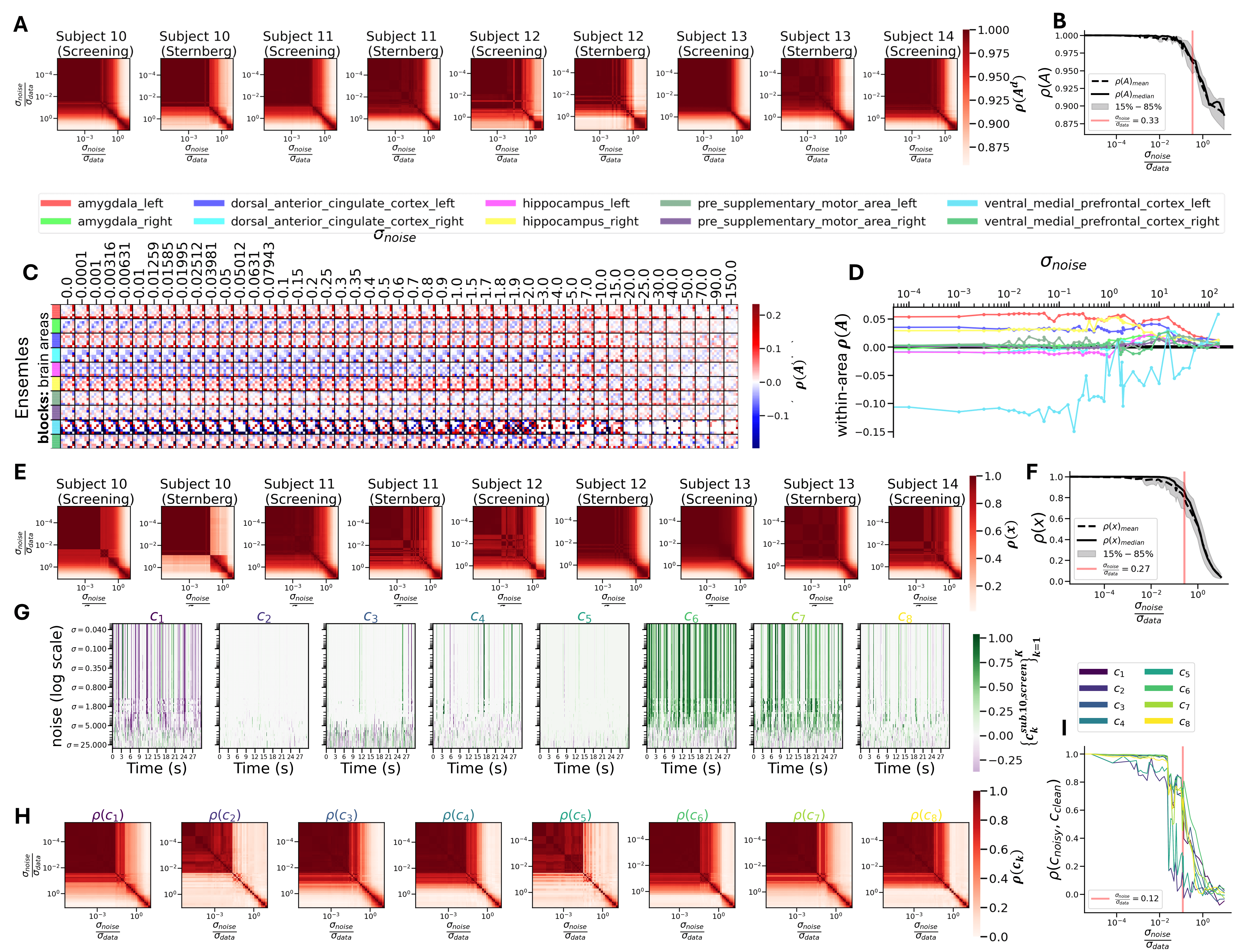}
    \caption{
\textbf{CREIMBO's Robustness to Increasing Random Normal Noise Levels.}
\textbf{A~\&~B:}~Correlations of identified ensemble compositions (\(\bm{A}^d\)) under increasing noise, for individual sessions (\textbf{A}) and all sessions combined (\textbf{B}). Robustness decreases only when \(\frac{\sigma_\textrm{noise}}{\sigma_\textrm{data}} = \frac{1}{3}\).
\textbf{C~\&~D:}~Correlations of ensemble compositions (from data concatenated across all conditions) displayed in a heatmap, with areas represented by rows of blocks (vertical) and noise levels by columns (horizontal). Block \((i,j)\) captures the correlation of ensembles from area \(i\) at the \(j\)-th noise level. A sharp drop in within-area correlations (\textbf{D}) occurs at \(\sigma_\textrm{noise} = 1\) (\(\frac{\sigma_\textrm{noise}}{\sigma_\textrm{data}} \approx \frac{2}{3}\)).
\textbf{E~\&~F:}~Correlations of ensembles' trajectories (\(\{\bm{x}^d\}_{d=1}^D\)) for each condition (\textbf{E}) or combined (mean/median) (\textbf{F}) show robustness until \(\frac{\sigma_\textrm{noise}}{\sigma_\textrm{data}} = 0.27\).
\textbf{G:} Time-varying interaction coefficients for the 1st condition ``(sub. 10, Screening)'' under increasing noise reveal similar but more frequent changes, with a sharp frequency change at \(\sigma_\textrm{noise} = 5\) (\(\frac{\sigma_\textrm{noise}}{\sigma_\textrm{data}} = 0.33\)).
\textbf{H~\&~I:}~Correlations between corresponding \(\{\bm{c}_k\}_{k=1}^K\) ``(sub. 10, Screening)'' over increasing noise (\textbf{H}), and compared to the coefficients identified from the original data (\textbf{H}). 
\vspace{-10pt}
    }
    \label{fig:fig_robust_noise}
    \vspace{-2pt}
\end{figure}

\begin{figure}[ht]
    \centering
    \includegraphics[width=1\textwidth]{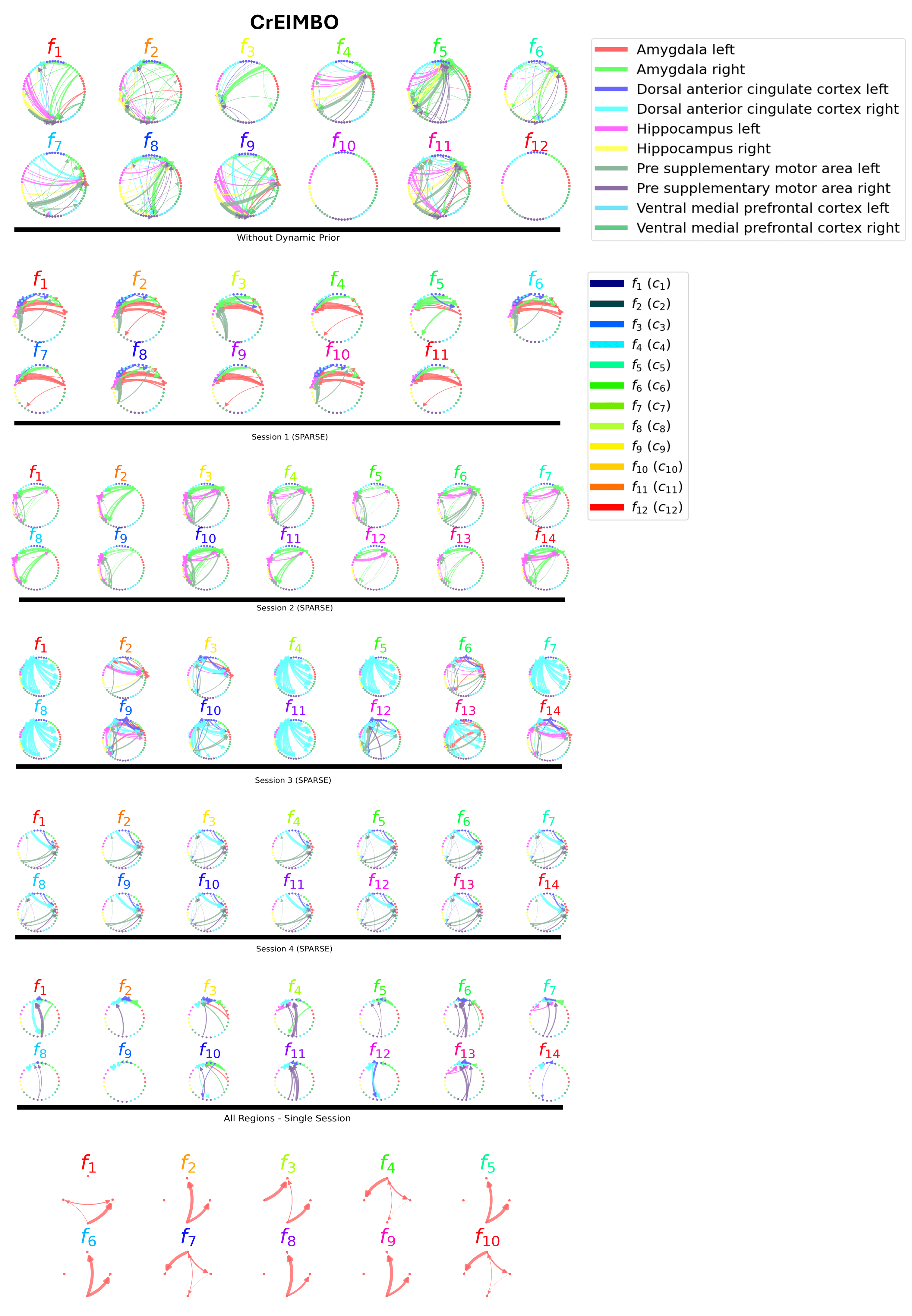}
\caption{Comparison of sub-circuits identified by CREIMBO with those identified by other approaches using the real-world human neural recordings. CREIMBO reveals distinct motifs capturing multi-regional flows to specific areas, including both cross-regional and multi-regional interactions. In contrast, other approaches identify sub-circuits with overlapping flows that fail to capture the full spectrum of multi-regional interactions.}
    \label{fig:human_baselines_f}
\end{figure}

\begin{figure}[ht]
    \centering
    \includegraphics[width=1\textwidth]{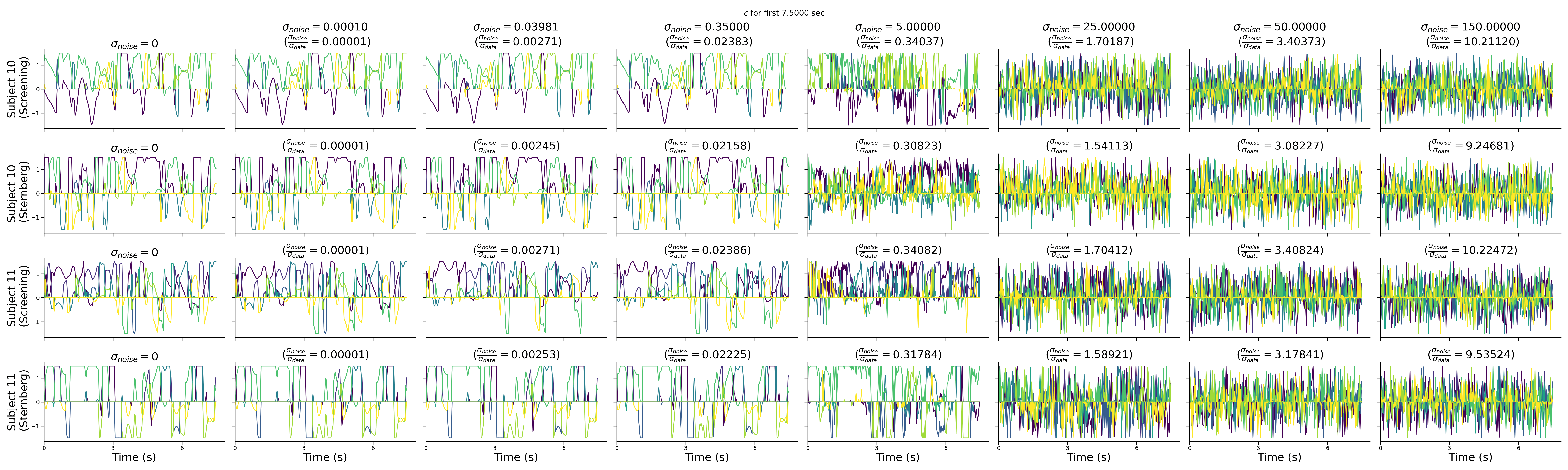}
\caption{Identified dynamics' coefficients (${\bm{c}}_{k=1}^K$) for the human data experiment under increasing noise levels demonstrate robustness, with a rapid increase in internal pattern frequency at higher noise levels.}
    \label{fig:noise_and_c}
\end{figure}

\newpage

\begin{algorithm}
\caption{CREIMBO training}
\label{algo:optimization}
\begin{algorithmic}[1]
    \State \textbf{Input:} $p, K, \sigma_h, \beta_1, \beta_2, \beta_3, \lambda_x, \lambda_c, \lambda_f, \lambda_\rho, \lambda_{\text{obs}}, \text{Batch Size}$
    \State \textbf{Initialize:} $\{f_k\}_{k=1}^K$, $\{c_d\}_{d=1}^D$, $\{x_d\}_{d=1}^D$
    \State \textbf{Pre-calculate:} $\{h_j\}_{j=1}^J$
    \Repeat
        \For{each session $d$ in a random batch of $D$}
            \State Update ensembles
            \State Update hidden dynamics and coefficients
        \EndFor
        \State Select a random batch of sessions from $D$
        \State Update networks $\{f_k\}$
        \If{stuck in local minimum}
            \State Perturb $\{f_k, c_d\}$
        \EndIf
    \Until{convergence}
\end{algorithmic}
\end{algorithm}

\section{Notations}

\noindent
\begin{tabular}{|c|>{\raggedright\arraybackslash}p{11.5cm}|}
    \hline 
    \textbf{Symbol} & \textbf{Description} \\
    \hline
    $D$ & Number of sessions (encompassing both same and different subjects) \\
    \hline
    $M_d$ & Number of trials for each session $d$ \\
    \hline
    $\bm{Y}_m^d$ & Neural recordings firing rate estimation for trial $m$ and session $d$ \\
    \hline
    $N_m^d$ & Number of neurons observed in trial $m$ of session $d$ \\
    \hline
    $\Tau_m^d$ & Number of time points observed in trial $m$ of session $d$ \\
    \hline
    $\{\bm{Y}_m^d\}_{m,d}$ & Overall set of observations \\
    \hline
    $\bm{A}_m^d$ & The neuronal composition of each sparse ensemble in trial $m$ and session $d$ \\
    \hline
    $\bm{A}_{\text{block}}$ & Block of $\bm{A}$ representing neural dynamics in a specific brain area \\
    \hline
    $n_j$ & Total neurons in area $j$ \\
    \hline
    $p_j$ & Maximum number of ensembles discernible within area $j$ \\
    \hline
    $p$ & Total number of ensembles across all areas \\
    \hline
    $J$ & Total number of distinct brain areas \\
    \hline
    $n$ & Total neurons involved across all trials or sessions \\
    \hline
    $\bm{x}_t$ & State vector at time $t$ \\
    \hline
    $\bm{F}_t$ & Transition matrix at time $t$, representing ensemble interactions \\
    \hline
    $K$ & Number of global interacting units \\
    \hline
    $\bm{c}_{kt}$ & Influence of the $k$-th global interaction at time $t$ \\
    \hline
    $\bm{f}_k$ & Basic linear system representing the $k$-th global interaction \\
    \hline
    $\tau = \Tau_m^d$ & Duration of the trial or session \\
    \hline
\end{tabular}

\section{Synthetic Data Experiment Details}\label{sec:params_exp_simple}
{
We simulated the synthetic data to represent the non-stationary dynamics of brain activity including multi-region interactions. Current assumptions about brain activity suggest~\cite{zeki2015massively, nelson1990brain} it functions through distributed, parallel processing across multiple regions, potentially involving co-active circuits that can process a range of variables (e.g., task demands, feedback signals, and sensory information), simultaneously. This biological reality thus motivated  our data generation approach, to reflects the multi-process nature of neural activity, where different processes can interact and contribute to brain dynamics.}
{
We simulated \( D = 5 \) sessions, each sharing a common set of basis rotational dynamics, denoted as \( \{\bm{f}_k\}_{k = 1}^{K=3} \), with \( p_j = 3 \) regions of interest. Different regions were associated with different ensembles, and the dynamics for each session were modeled using sparse decomposition of the basis dynamics. For each session, the dynamics' coefficients, \( \bm{c}_t^k \), were generated over \( T = 500 \) time points. These coefficients were generated by initially giving them binary values (active vs. inactive states), with random switches occurring every \( x \) time points, where \( x \) was uniformly drawn between 10 and 30. The sparsity constraint was maintained at each time point, ensuring that only one dynamic was active at any time (\( \|\bm{c}_t^k\|_0 = 1 \forall t \)).}

{
Then, to introduce smooth transitions between switching states, a Gaussian convolution with a time window of 6 points (standard deviation = 0.7) was applied to the coefficients, to promote a  smoother / more gradual changes between active states. Different sessions maintained the same sequence of active dynamics but allowed for varying durations of each state, introducing temporal variability across sessions.}

{Each session $d$ was augmented with a unique projection matrix  \( \bm{A}^d \) that captures the mapping between the latent space and the observations (neural) space in the respective regions. These projection matrices \( \bm{A}^k \) consisted of \( P = 3 \) columns, each corresponding to one of the three regions. The number of neurons in each region was randomly chosen between 4 and 9, with the count varying independently across regions and sessions. This variability ensured that different sessions were not identical in their neural configurations.
Latent dynamics for each session were generated using the recursive equation \( \bm{x}_t = \sum_{k=1}^K \bm{c}_{kt} \bm{f}_k \bm{x}_{t-1} \), starting from an initial state \( \bm{x}_0 = [1, -1, 1]^\top \). }
{CREIMBO was trained based on the observations only, 
which were defined for each session as $\bm{y}^d = \bm{A}^d \bm{x}^d + \epsilon$ where $\epsilon$ is an addition of \textit{i.i.d} Gaussian noise with standard-deviation of $\sigma_\text{noise} = 0.1$}

The full set of parameters for the synthetic experiment are available in Table~\ref{tab:parameters_synth}.

\begin{table}[ht]\label{tab:params_exp_simple_params_table}
\caption{Parameter values for synthetic experiments}
    \label{tab:parameters_synth}
\centering
    \begin{tabular}{|>{\raggedright}p{3cm}|>{\raggedright}p{2cm}|>{\raggedright\arraybackslash}p{7cm}|}
        \hline
          \textbf{Parameter} & \textbf{Value (range) }  & \textbf{Explanation} \\  \hline
same\_c   & False            &  Indicates if to use shared sub-circuits coefficients across sessions. \\  \hline
step\_f   & 0.1  - 0.5      &  Gradient descent step size for updating $\bm{f}$ \\  \hline
GD\_decay & 0.99 - 0.999  &  Decay over iterations of gradient-descent step size. \\  \hline
max\_error& 1e-09       & The error threshold to stop training the model. \\  \hline
max\_iter & 500         & Threshold on the maximum number of iterations. \\  \hline
include\_D& True         & Indicates the inclusion of a projection to a latent space. \\  \hline
step\_D   & $10^{-5}-10^{-2}$      & Range of Gradient Descent Step size for $\bm{A}$. \\  \hline
seed      & 0 - 4000       & Random seed to choose from\\  \hline
normalize\_eig         & True      , False       & Indicates if to normalize the sub-circuits by \\  \hline
start\_sparse\_c       & False , True    & Whether to initialize $c$ to be sparse.\\  \hline
sparse\_f & True            & Whether to apply sparsity on the sub-circuits \\  \hline
num\_gradient\_steps   & 1 -3       & Number of gradient steps in an iteration. \\  \hline
add\_avg     & True , False      & Whether to add a moving average to an iteration. \\  \hline
sparsity\_on\_f\_max   & 40-60            & Percentile of sparsity applied on each $\bm{f}$\\  \hline
take\_multiple\_gd     & False, True        & Indicates if to take multiple steps of gradient descent. \\  \hline
D\_graph\_driven       & True      & Indicates that D is inferred with graph-driven way. \\  \hline
infer\_x\_c\_together  & False , True      & Indicates if the inference of $\bm{x}$ and $\bm{c}$ occurred simultaneously. \\  \hline
include\_mask          & True         & Indicates the inclusion of a mask on $\bm{A}$\\  \hline
norm\_D\_cols          & True       & Indicates the normalization of columns of $\bm{A}$   \\   \hline
lambda\_D & 0.1 - 0.3        & Regularization weight on updating $\bm{A}$\\   \hline
step\_D\_decay         & 0.99 - 0.9999    & Decay rate for updating $\bm{A}$ parameters\\  \hline
num\_regions           & 3                & Number of brain regions. \\  \hline
lambda\_x & 0  - 0.1           & Sparsity term on the latent dynamics. \\  \hline
latent\_dim& 3               & Latent dimension.\\  \hline
noise\_level           & 0.2-0.6 & Standard deviation of noise added to the data. \\  \hline
num\_subdyns           & 3              & Number of sub-circuits. \\  \hline
lasso\_solver & 'spgl1' & Pylops solver used for $\ell_1$ regularization. \\   \hline
    \end{tabular}
\end{table}

\section{Parameters for human multi-regional experiment}\label{sec:params_human}
The full set of parameters for the human data experiment are available in Table~\ref{tab:parameters_human}.

\begin{table}[ht]
    \caption{Parameter values for real-world human experiment}
    \label{tab:parameters_human}
    \centering
    \begin{tabular}{|>{\raggedright}p{3cm}|>{\raggedright}p{2cm}|>{\raggedright\arraybackslash}p{7cm}|}
        \hline
        \textbf{Parameter} & \textbf{Value} & \textbf{Explanation} \\
        \hline
        same\_c & False & Indicates if sub-circuit coefficients are shared across sessions. \\
        \hline
        step\_f & 0.2 & Step size for gradient descent in updating $\bm{f}$. \\
        \hline
        GD\_decay & 0.992 & Gradient descent step-size decay rate over iterations. \\
        \hline
        max\_error & $1 \times 10^{-9}$ & Maximum allowable error. \\
        \hline
        max\_iter & 100 & Maximum number of iterations. \\
        \hline
        seed & 0 & Random seed. \\
        \hline
        normalize\_eig & True & Normalize eigenvalues. \\
        \hline
        start\_sparse\_c & False & Start with sparse $\bm{c}$. \\
        \hline
        include\_D & True & Whether to include $\bm{A}$. \\
        \hline
        step\_D & 0.0001 & Step size for updating $\bm{A}$. \\
        \hline
        num\_gradient\_steps & 4 & Number of gradient steps. \\
        \hline
        add\_avg & True & Add average. \\
        \hline
        sparsity\_on\_f\_max & 45 & Maximum sparsity on $\bm{f}$. \\
        \hline
        take\_multiple\_gd & False & Take multiple gradient descents. \\
        \hline
        D\_graph\_driven & True & Graph-driven $\bm{A}$. \\
        \hline
        infer\_x\_c\_together & False & Infer $\bm{x}$ and $\bm{c}$ together. \\
        \hline
        include\_mask & True & Include mask. \\
        \hline
        norm\_D\_cols & False & Normalize columns of $\bm{A}$. \\
        \hline
        lambda\_D & 0.3 & Regularization parameter for $\bm{A}$. \\
        \hline
        step\_D\_decay & 0.9999 & Decay rate for step size of $\bm{A}$. \\
        \hline
        l1\_D & 2.74 & $\ell_1$ regularization for $\bm{A}$. \\
        \hline
    \end{tabular}
\end{table}

\section{Information about the baselines:}\label{sec:baslines}

We distinguish between two types of baselines: (1) ablation experiments and (2) comparison to the performance of entirely different methods. It is important to note that the latter cannot fully capture the unique capabilities of CREIMBO, particularly in identifying basis ensemble interactions shared across sessions through sparse decomposition. Therefore, we only compare certain aspects that can be extracted from other methods (e.g., multi-regional SLDS, which requires the same number of neurons per session). However, these comparisons do not imply that other models outperform or underperform CREIMBO; they simply offer different insights, with varying strengths and limitations depending on the application.

Importantly, since the sub-circuits $\{ \bm{f}_k \}$ is invariant to its ordering, we used SciPy's implementation of the `linear\_sum\_assignment` problem~\cite{crouse2016implementing} to re-order the identified sub-circuits before comparing them with the ground truth, matching the \(\bm{f}\)s values by minimizing the \(\ell_2\) error.
\textbf{Ablation Experiments:}
\begin{enumerate}
    \item ``\textbf{Without Dynamic Prior}'': Running CREIMBO while removing the prior over the dynamics from the inference. Instead, inferring the temporal traces of the ensembles using regularized least-squares with the addition of smoothness, de-correlation, and Frobenius norm regularization terms on the dynamics. 
    \item ``\textbf{All Regions (sparse)}'': Running CREIMBO without applying the multi-regional diagonal mask over the ensemble compositions ($\bm{A}$), thus supporting the finding of non-localized ensembles. 
    \item ``\textbf{All Regions (non-sparse)}'': Similar to the previous one but without applying sparsity on the ensemble matrix.
    \item ``\textbf{PCA All Regions}'': Running CREIMBO while replacing the dimensionality reduction step with PCA.
    \item ``\textbf{Single Session \textit{\#}}'': Running CREIMBO on a single session (view) of the data rather than leveraging cross-session information.

\end{enumerate}

\textbf{Other methods:}
\begin{enumerate}
    
    \item ``\textbf{SLDS}'': We used the SSM Python package described by Linderman et al. at~\cite{Linderman_SSM_Bayesian_Learning_2020}. For the (non-recurrent) SLDS option, we used the ``gaussian\_orthog`` emission parameter, the ``bbvi'' as the fitting method, ``variational\_posterior'' was set to ``mf'', and the number of iterations (``num\_iters'') was set to 500. The rest of the fitting parameters were left as default. We were inspired by this notebook ~\url{https://github.com/lindermanlab/ssm/blob/master/notebooks/3-Switching-Linear-Dynamical-System.ipynb} by ~\cite{Linderman_SSM_Bayesian_Learning_2020} for our comparison.
    This includes:
    \\
    1) ``\textbf{SLDS (trials)}'' which captures the training of SLDS with ``bbvi''  fitting method from across all trials (by stacking trials information vertically), 
    \\ 2)
    ``\textbf{SLDS (per-trial)}'' which refers to training SLDS individually per-trial. 
    
    \item ``\textbf{rSLDS}'': Similarly to SLDS, for the recurrent version (rSLDS,~\citep{linderman2016recurrent}), we used the same SSM package provided by~\cite{Linderman_SSM_Bayesian_Learning_2020}. Here, we used the recurrent option for the transitions (i.e., the ``transitions'' parameter was set to``recurrent\_only''), using ``diagonal\_gaussian'' dynamics and ``gaussian\_orthog'' emissions. This includes:
    \\ 3)  ``\textbf{rSLDS (trials)}'' which captures training rSLDS with the ``bbvi''  fitting method from across all trials (by stacking trials information vertically), 
    \\ 4)
    ``\textbf{rSLDS (per-trial)}'' which refers to training rSLDS individually per-trial. 

        \item \textbf{``mp\_rSLDS''}: Multi-Regional rSLDS, as described in~\cite{glaser2020recurrent}. To run our comparison, we were inspired by the colab-notebook for this ``mp\_rslds'' method, provided by the SSM Python Package~\cite{Linderman_SSM_Bayesian_Learning_2020}   (at~\url{https://colab.research.google.com/github/lindermanlab/ssm/blob/master/notebooks/Multi-Population-rSLDS.ipynb}).
        This includes:
        \\
        5)  ``\textbf{mp\_rSLDS-Gauss (trials)}'' which captures training ``mp\_rSLDS'' from across all trials (by stacking trials information vertically) under Gaussian statistics.
        \\
        6)
    ``\textbf{mp\_rSLDS-Gauss (per-trial)}'' which refers to training ``mp\_rSLDS'' individually per-trial  under Gaussian statistics.
    \\
        7)  ``\textbf{mp\_rSLDS-Poisson (trials)}'' which captures training ``mp\_rSLDS''  with the ``bbvi''  fitting method from across all trials (by stacking trials information vertically)  under Poisson statistics.
        \\
        8)
    ``\textbf{mp\_rSLDS-Poisson (per trial)}'' which refers to training  ``mp\_rSLDS'' individually per-trial  under Poisson statistics.


\end{enumerate}


\newpage
\section*{Assumptions}\label{sec:assumptions}
The CREIMBO model is based on a core set of assumptions over the nature of the data. These assumptions draw on both known properties of neural processes and general well-known statistical models used widely in data science. We categorize these assumptions into two types:
\begin{enumerate}
    \item Priors over the underlying neural processes.
    \item Priors over Observational Constraints in order for CREIMBO to properly infer the underlying system by leveraging multi-session information.
\end{enumerate}


\subsection{Priors over the underlying neural processes}

\begin{enumerate}
    \item The neural dynamics in each session $d$ are assumed to lie on a low-dimensional manifold, embedded in a $P << N^d$ low-dimensional space that is defined by $P$ functional groups of neurons (referred to as ``ensembles'').
    
    \item We assume that these functional ensembles consist of neurons with co-activation patterns, where neurons can belong to more than one, but only \textit{a few} (a sparse number) of ensembles, with varying degrees of membership.
    
    \item We further propose that the interactions between these ensembles drive the evolution of the latent manifold over time and are key to encoding changes in conditions and behavior.
    
    \item For CREIMBO to be effective, we assume these interactions arise from the joint synchronous activity of multiple co-occurring processes, captured by a limited-size set of ``basis-ensemble-interactions''. The time-varying, sparse decomposition of these basis-ensemble-interactions, weighted by their time-local contributions in every time point, can adequately describe the manifold's evolution over time.
    
    \item We assume that different dynamics basis elements capture distinct processes or behaviors. Some  processes are required to be globally related to the cognitive task, while others capture session- or subject-specific processes (see Sec.~\ref{sec:data_priors}).
    
    \item Each of these ensemble interactions may capture between-area interactions, within-area interactions, or both. 
\end{enumerate}

\subsection{Priors over Observational Constraints}\label{sec:data_priors} 

To effectively learn a unified representation by leveraging information across sessions, we make the following assumptions on the statistics of the data. 

\begin{enumerate}
    \item 
Our ability to infer the underlying latent state (see Fig.~\ref{fig:demofig}B) is contingent upon the overall observability of the dynamical system. In traditional linear systems, the observability matrix can guarantee that any state is visible in a finite time horizon output. Non-stationary systems, as in CREIMBO, do not have succinct guarantees, the closest of which come in the form of observability conditions on switched linear systems~\citep{tanwani2012observability}. 

As the heart of the observability condition is that the dynamics ``well mixes'' the state such that any state element in the null space of the readout matrix will eventually be rotated into the its span, and thus visible. We thus assume that the spectral radius of each dynamical system (a measure of mixing in linear systems~\citep{simchowitz2018learning}) in the basis is close to one. This assumption is loose, due to the lack of theoretical guarantees on general nonstationary systems, and further analysis should identify tighter, and less stringent, assumptions.




 \item Another prior we want to consider is which $\bm{f}$s capture ensembles unobserved in certain sessions, allowing us to assert that CREIMBO can recover the activity of these unobserved ensembles during those sessions. To achieve this, we must ensure that in the sessions where these ensembles are unobserved, at least some $\bm{f}$s representing them also capture the activity of observed ensembles. For example, if certain missing ensembles are represented solely by $\bm{f}$s that do not capture any observed ensembles, we will not be able to infer the activity of these unobserved ensembles in the sessions where they are absent, because there are no observed ensembles linked through the dynamic prior.

  \item 
To ensure uniqueness of the sparse decomposition of the dynamical systems model at each session and time point, we assume that the effective spark of the dynamics basis is large ($S^*\geq 2S)$. Essentially, there is no $\bm{f}_k$ that can be linearly composed of $2S^*$ other dynamical systems. This ensures that there cannot exist two (or more) equivalent decompositions $\sum_{k=1}^K \bm{f}_k c_{kt}$ and $\sum_{k=1}^K \bm{f}_k c_{kt}^{*}$ for distinct coefficients that both compose the same dynamical matrix.

 \item 
Lastly, to ensure accurate cross-session alignment, we propose that each dynamical system projects uniquely onto the obseverved subset of ensembles in each session. This assumption prevents two dynamical systems (e.g., ``ensemble 1 $\rightarrow$ ensemble 2 $\rightarrow$ ensemble 3'' and ``ensemble 1 $\rightarrow$ ensemble 4 $\rightarrow$ ensemble 3'') from being indistinguishable from the data available in that session (e.g., a session with only ensembles 1 and 3 recorded). 


\end{enumerate}

\section{Computational Complexity}\label{sec:model_complexity}

CREIMBO's learning process involves learning both the ensemble compositions per session (\( \{ \bm{A}^d \}_{d=1}^D \)), the ensembles' temporal activity (\( \{\bm{x}_t^d \}_{d=1}^D \)), and its underlying temporal evolution, which requires identifying the global (session-invariant) dynamic operators (\( \{\bm{f}_{j}\}_{j=1}^J \)) and their per-session temporal coefficients (\( \{ \bm{c}_t \}_{t=1}^T \)). 

\subsection{Complexity of ensemble-matrix update}
The loading matrix update relies on 4 main computational steps: 

\textbf{1) Channel Graph Construction:} This operation, performed once for all $N$ channels of every state $d = 1 \dots D$, generates a channel graph $\bm{H}^d \in \mathbb{R}^{N \times N}$ for each state $d \in[1, D]$ by concatenating within-state trials $1 \dots M_d$ horizontally, resulting in a $N \times \sum_{m=1}^{M_d} T_m^d$ matrix. 
For simplicity, let $\widetilde{T} = \sum_{m=1}^{M_d} T_m^d$.
The computational complexity of calculating the pairwise similarities of this concatenated matrix for all $D$ states is thus $\mathcal{O}\left(D \widetilde{T}^2 N(N-1)\right)$. 

\textbf{2) The k-threshold step:} involves keeping only the $k$ largest values in each row while setting the other values to zero\textemdash the complexity will be $\mathcal{O}\left(\widetilde{T} \log k\right)$ per row for a total computational complexity of  $\mathcal{O}\left(DN \widetilde{T} \log k\right)$   for $N$ rows and $D$ states.

\textbf{3) State Graph Construction:} This is a one-time operation that involves calculating the pairwise similarities between each pair of states. For simplicity, if we assume the case of user-defined scalar labels, and as in this case there are $D$ states (and accordingly $D$ labels), the computation includes $D(D-1)) / 2$ pairwise distances for $\mathcal{O}\left(D^2\right)$.

\textbf{4) Ensemble Inference (Eq.~\eqref{eqn:updateLambda}):} 
This iterative step involves per-channel re-weighted $\ell_1$ optimization. If the computational complexity of a weighted $\ell_1$ is denoted as $\mathcal{C}$, then the computational complexity of the re-Weighted $\ell_1$ Graph Filtering is $N L \mathcal{C}+L N k$, where $N$ is the number of channels, $L$ is the number of iterations for the RWLF procedure, and $k$ is the number of nearest neighbors in the graph. 
For the last term in Eq.~\eqref{eqn:updateLambda}, there are $p^2$ multiplicative operations involving the vector $\nu$ and the difference in ensembles, arising from the $\ell_2^2$ norm. Additionally, there is an additional multiplication step involving  $\bm{P}_{dd'}$. For each state $d$, this calculation repeats itself $D - 1$ times (for all $d' \neq d$). This process is carried out for every $d = 1 \dots D$. In total, these multiplicative operations sum up to $\left(p^2+1\right) D(D-1)$, resulting in a computational complexity of $\mathcal{O}\left(D^2 p^2\right)$.

\subsection{Complexity of inferring the ensembles activity and interactions coefficients (Eq.~\eqref{eqn:update_x_c})}
The inference of $\bm{c}_t^d$ for each $t = 1 \dots T$ amd each condition $d=1\dots D$, involves:
$\widehat{\bm{c}}_t^d = \arg \min_{\bm{c}_t,\bm{x}_t}
 \| \bm{y}_t^d -  \bm{a}\bm{x}_t \|_F^2 
 +
\| \widetilde{{\bm{x}}_{t+1}}^d -  \bm{FX}_{K}^d \bm{c}_t^d \|_F^2 + \lambda_c \|\bm{c}_t\|_1$
where  
$\bm{FX} \in \mathbb{R}^{p \times K}$ is the horizontal concatenation of $\{\bm{f}_k \bm{x_t} \}_{k=1}^K$.

This simplifies to a Lasso problem of the form  

$$
\min_\theta \frac{1}{2} \|\xi - \mathcal{M}\theta\|^2 \quad \text{s.t.} \quad \|\theta\|_1 \leq \tau, 
$$ 

with dimensions \( \xi: (N + p) \times 1 \), \( \mathcal{M}: (N + p) \times (p + K) \), and \( \theta: (p + K) \times 1 \), where 

$$
\xi = \begin{bmatrix} y_t \\ 0_{p \times 1} \end{bmatrix} \in \mathbb{R}^{(N+p) \times 1}, \quad 
\theta = \begin{bmatrix} x_t \\ c_t \end{bmatrix} \in \mathbb{R}^{(p+K) \times 1}, 
$$ 

and 

$$
\mathcal{M} = \begin{bmatrix} A & 0_{N \times K} \\ I_{p} & \bm{FX} \end{bmatrix} \in \mathbb{R}^{(N+p) \times (p+K)}.
$$ 

This can be solved with SPGL1 with up to \(\text{num\_iters}\) iterations, resulting in a computational complexity of  

$$
O(\text{num\_iters} \cdot (N + p)(p + K)).
$$

per session and time point and overall (across all sessions and time points):

$$
O(\text{num\_iters} \cdot (N + p)(p + K) \sum_{d=1}^D T^d).
$$
%
\subsection{Complexity of inferring the dynamics' dictionary} 
The inference of the basis-interactions dictionary $\{ \bm{f}_k\}_{k=1}^K$ (via Eq.~\eqref{eq:F_infer})
involves solving
\begin{align}
{\widehat{\bm{F}}_{all}} = \arg \min_{{\bm{F}}_{all}} {\| {\bm{x}}^{+} - \bm{F}_{all} \bm{({CX})} \|_F^2}, \nonumber
\end{align}
where
$\bm{x}^{+} \in \mathbb{R}^{p\times \sum_{d \in \textrm{batch} \Tau^d}}$
and
 $\bm{cx} \in \mathbb{R}^{Kp \times \sum_{d \in \text{batch}} \Tau^d }$.
 For simplicity, assuming that $\lambda_{\rho} = 0$, The problem simplifies to a LASSO formulation, which is solved using SPGL1 for \( F^{\text{all}} \), with an overall computational complexity per iteration of \( O(\text{num\_iter} \cdot Kp^2 \tau) \), where \(\text{num\_iter}\) represents the number of iterations required for convergence.

\section{Application to multi-Regional cross-Session Mice Neural Activity}\label{sec:meso_experiment}
\begin{figure}[ht]
    \centering
    \includegraphics[width=0.95\textwidth]{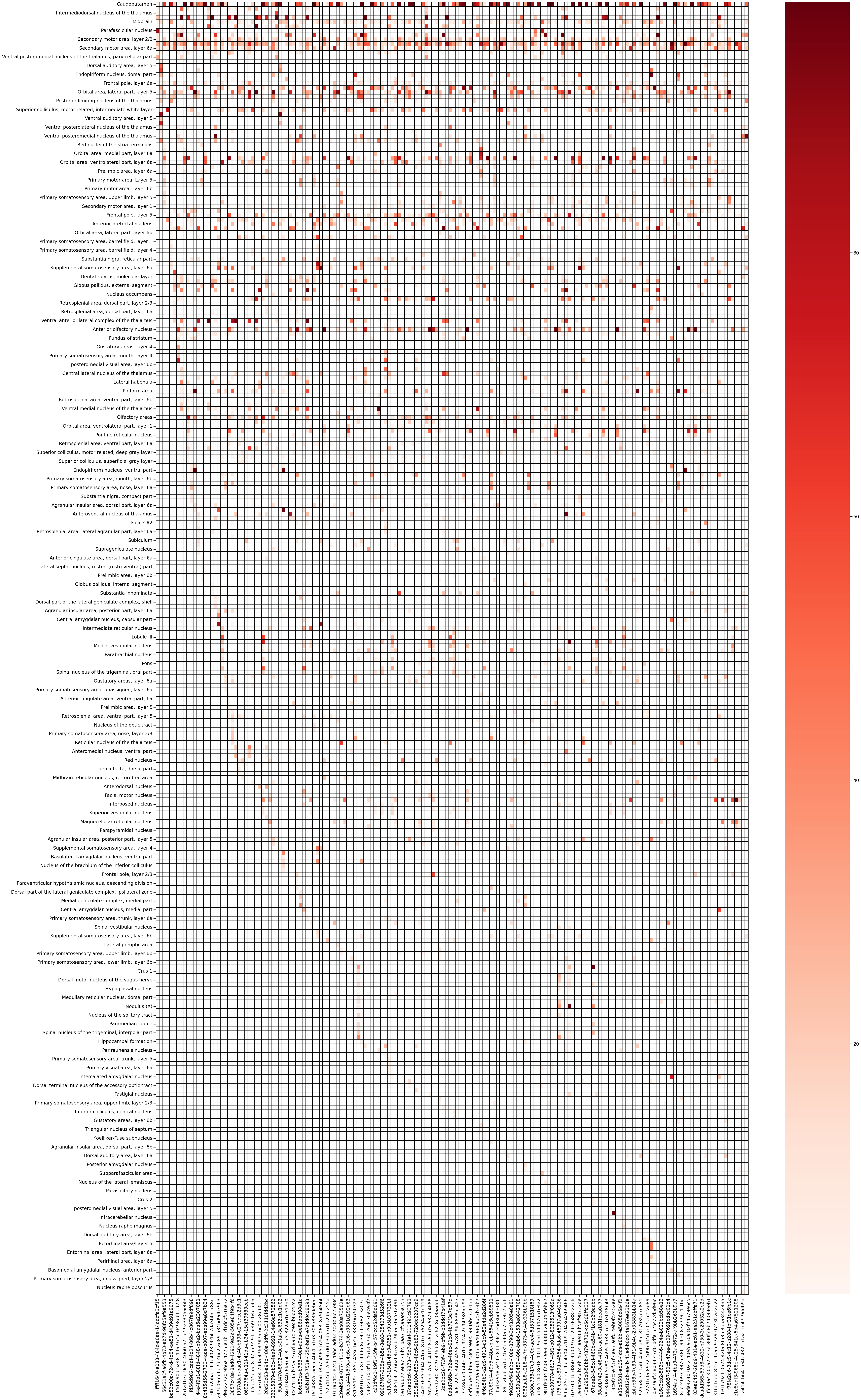}
\caption{
number of neurons per area across all files
    }
    \label{fig:mesoscale_all_areas}
\end{figure}

\begin{figure}[ht]
    \centering
    \includegraphics[width=0.99\textwidth]{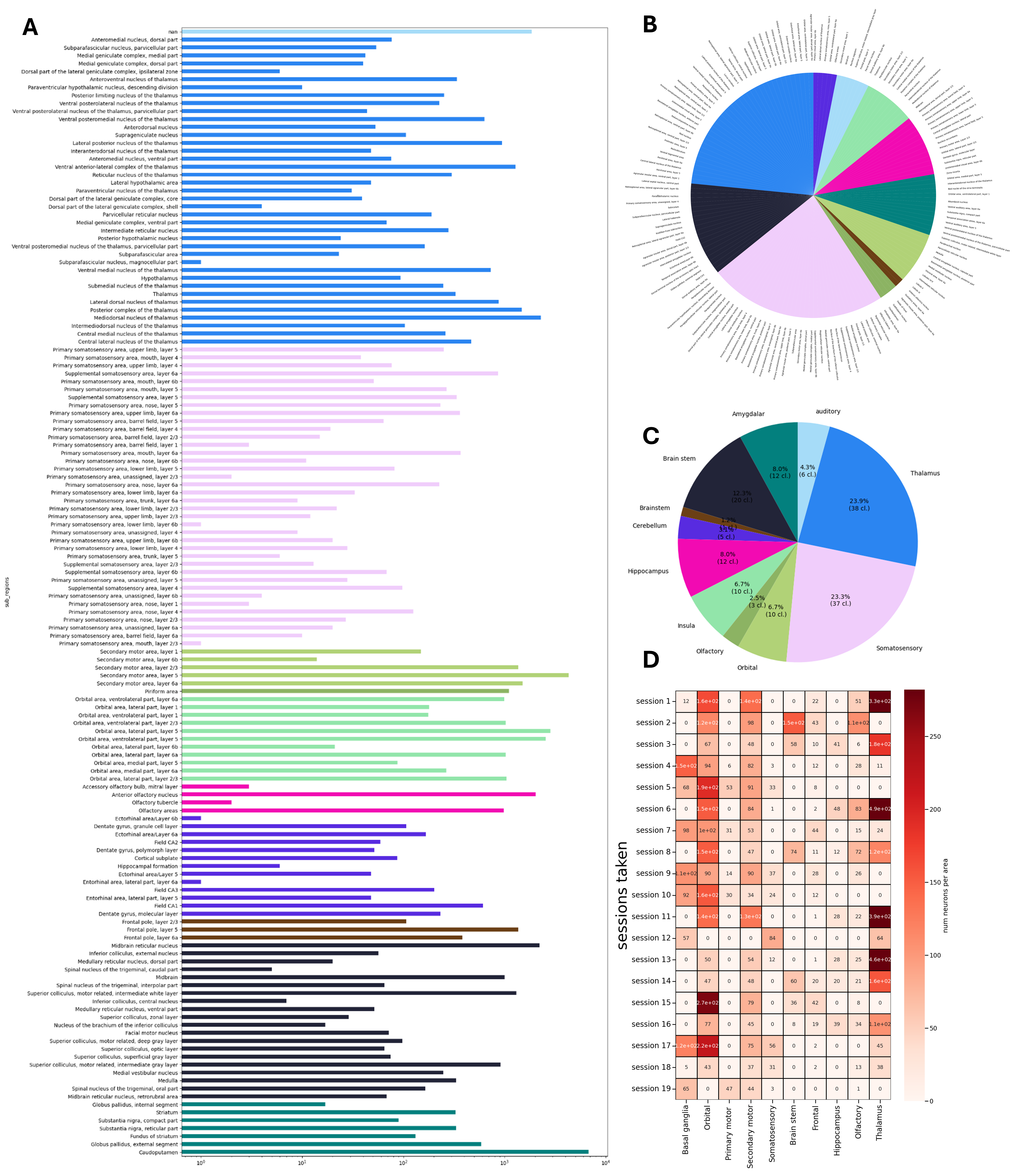}
\caption{
number of neurons per area as a motivation for area-focusing. 
\textbf{A:} Overall number of neurons per sub-area (colored by parent area) across all sessions.
\textbf{B:} Distribution of areas with sub-area name across all sessions.
\textbf{C:} Distribution of areas with parent-area name across all sessions.
\textbf{D:} Number of neurons per parent-area (across the 10 selected areas) in the 19 used sessions. 
    }
    \label{fig:mesoscale_all_areas_b}
\end{figure}


\begin{figure}[ht]
    \centering
    \includegraphics[width=0.99\textwidth]{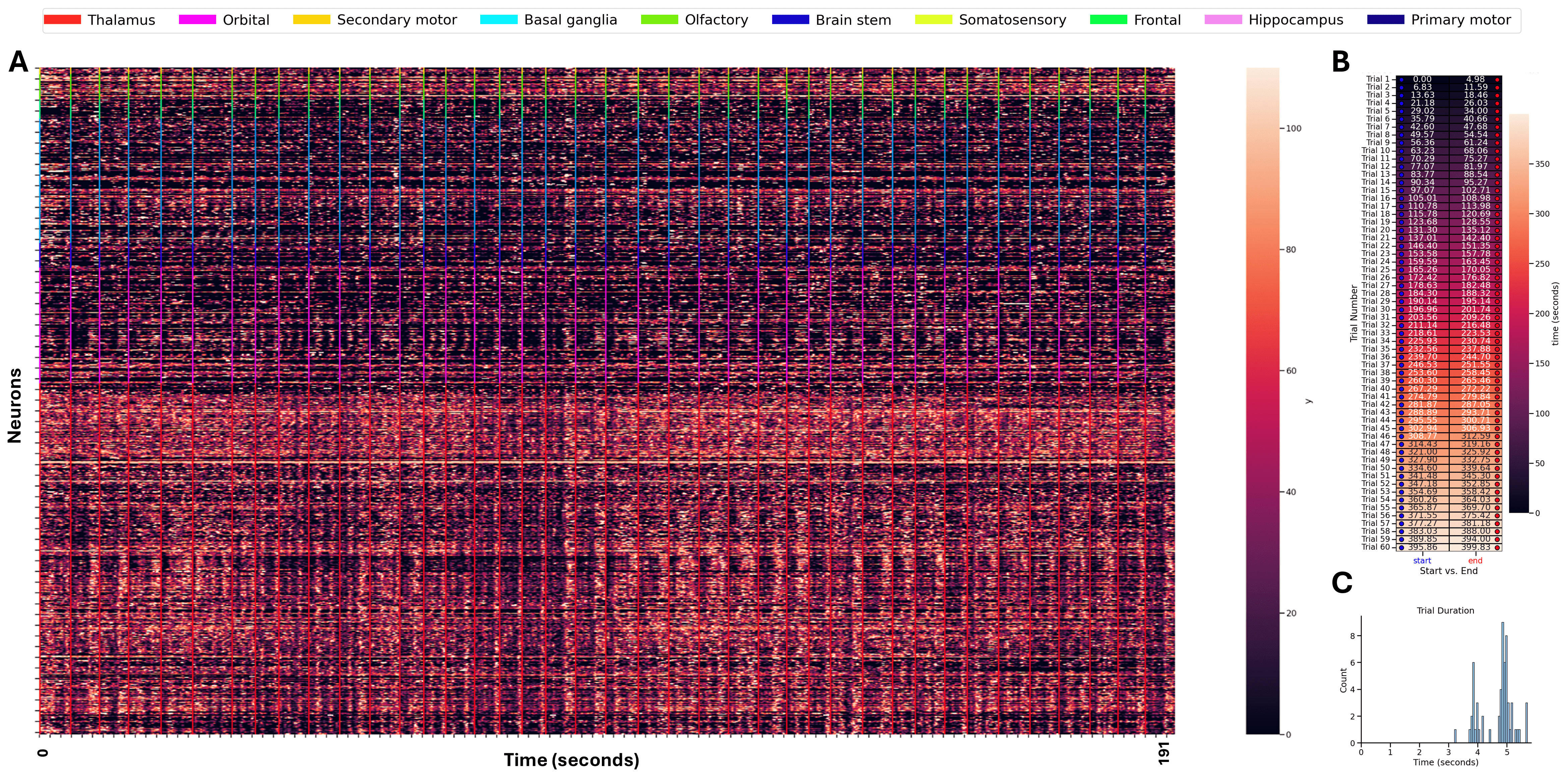}
\caption{{
Example Session ~\cite{chen2024brain}.
\textbf{A:} Data firing rate over time separated by trials in vertical lines. the color of the vertical line captures the relevant brain area (example random session).  
\textbf{B:} Trials start and end times (in seconds) for examples session. 
\textbf{C:} Trial duration distribution for example random session. 
}
    }
    \label{fig:mesoscale_example_session_results}
\end{figure}

\begin{figure}[ht]
    \centering
    \includegraphics[width=0.99\textwidth]{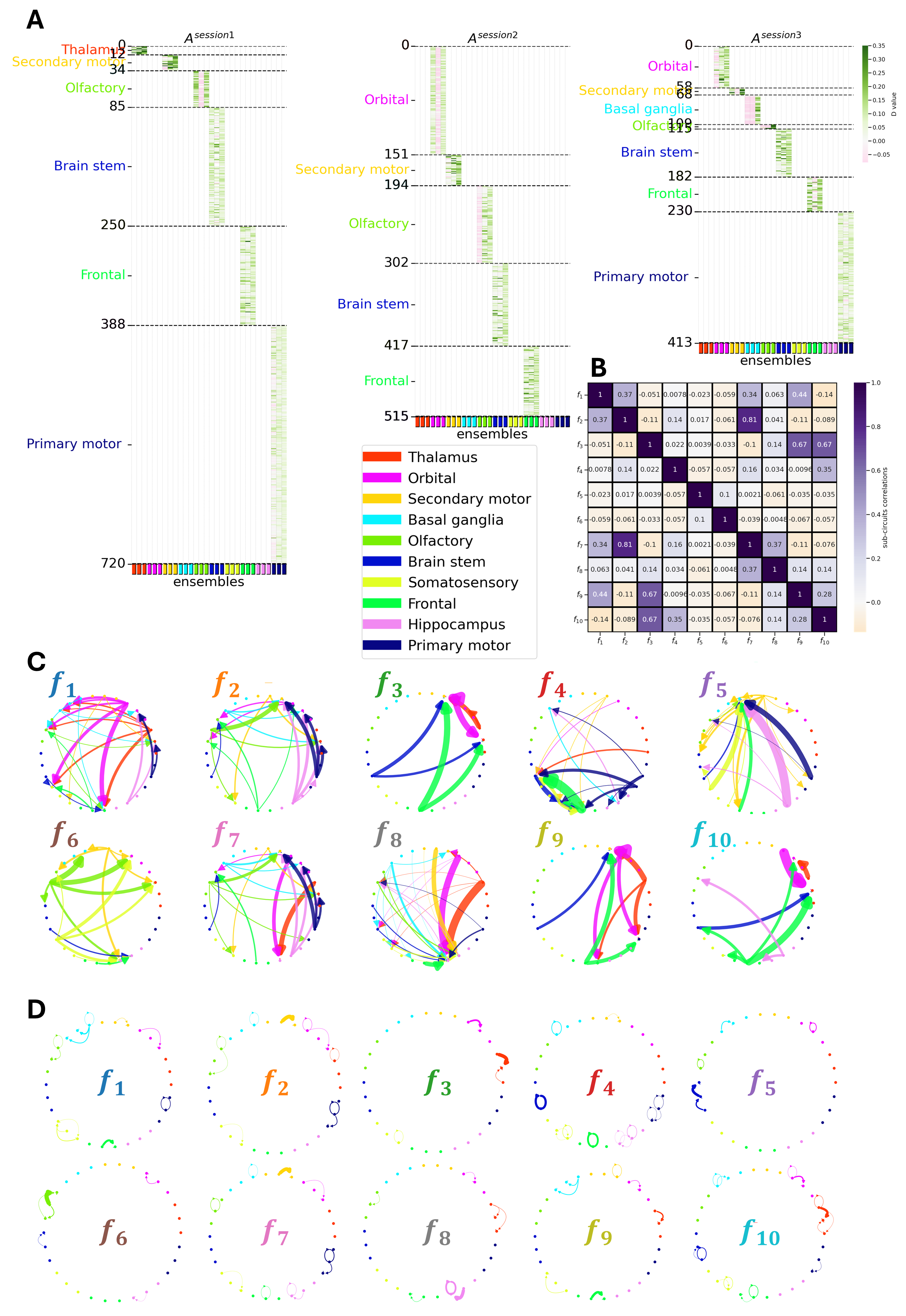}
\caption{{
Identified components of the Mesoscale data~\citep{chen2024brain} using 19 random sessions with 40-60 trials within each.  
\textbf{A:} Identified ensemble matrices for three random sessions.
\textbf{B:} Correlations between circuits. 
\textbf{C:} Between-areas interactions (extracted from $\{\bm{f}_k\}_{k=1}^K$ via a block off-diagonal mask)
\textbf{D:} Within-areas interactions (extracted from $\{\bm{f}_k\}_{k=1}^K$ via a block diagonal mask)
    }}
    \label{fig:mesoscale_identified_components}
\end{figure}

\begin{figure}[ht]
    \centering
    \includegraphics[width=0.99\textwidth]{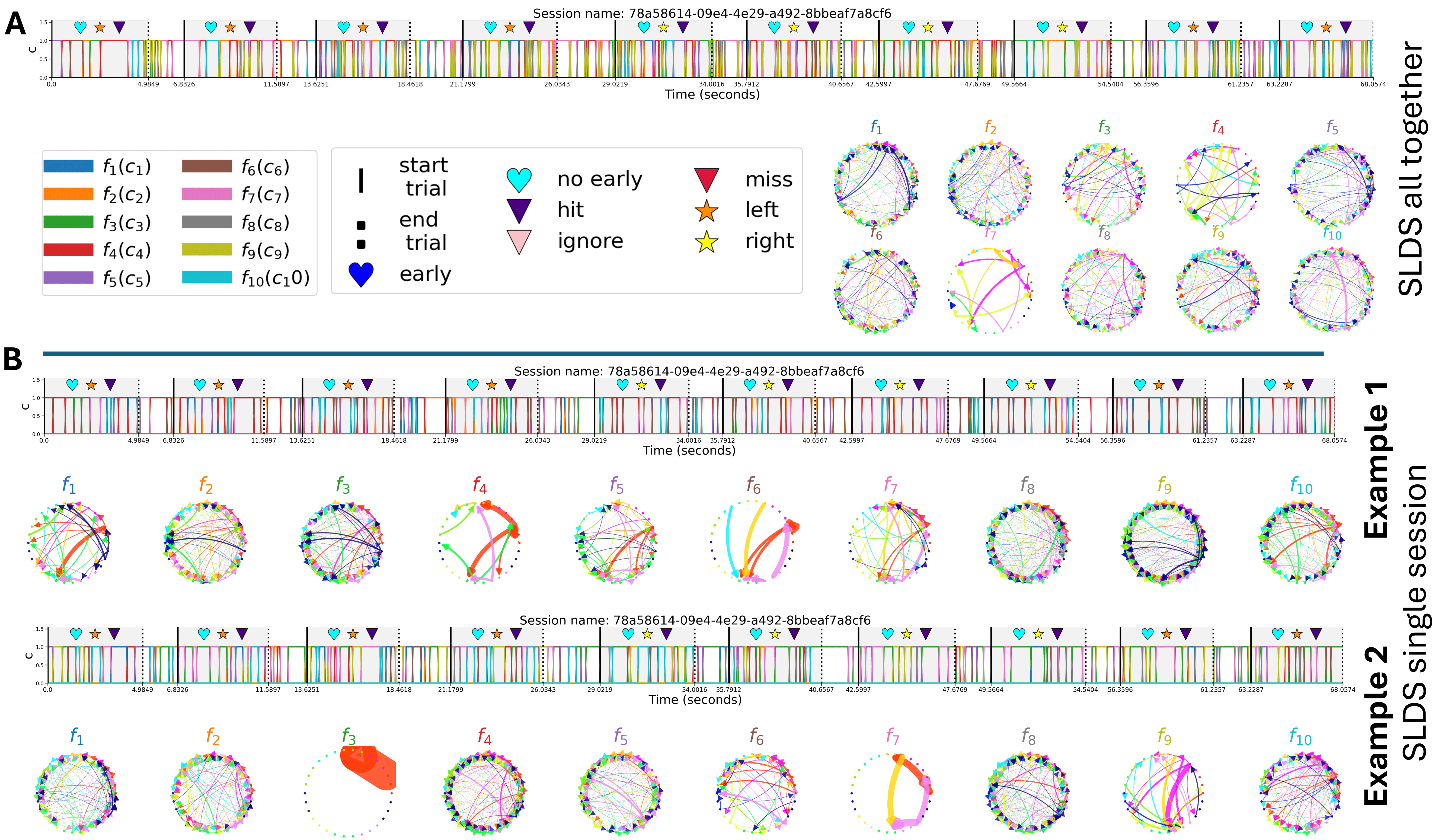}
\caption{{
The components identified by the SLDS (Fig.~\ref{fig:mesoscale_identified_components_by_slds}) and rSDLS (Fig.~\ref{fig:mesoscale_identified_components_by_rslds}) baselines~\citep{linderman2016recurrent, Linderman_SSM_Bayesian_Learning_2020}, when we applied them to the mouse mesoscale data from all sessions together (\textbf{A}) or individual sessions (\textbf{B}). For each model, we showed both the sub-circuits and their active coefficients for the first few trials. Compared to CREIMBO, the identified sub-circuits are more distributed and dense, and the dynamic coefficients capture only sharp transitions, i.e., unable to represent multiple co-occurring processes, in contrast to CREIMBO. 
Checking individual sessions with SLDS (\textbf{B}) or rSLDS (\textbf{C}), resulted in per-session set of sub-circuits (\textbf{F}s) that cannot enable cross-session sub-circuits comparison. 
All models ran with $K=10$ sub-circuits and $p_j = 3$ ensembles per-region, to keep maximum consistency with our CREIMBO results from Figure~\ref{fig:mesoscale_identified_components}.
    }}
    \label{fig:mesoscale_identified_components_by_slds}
\end{figure}

\begin{figure}[ht]
    \centering
    \includegraphics[width=0.99\textwidth]{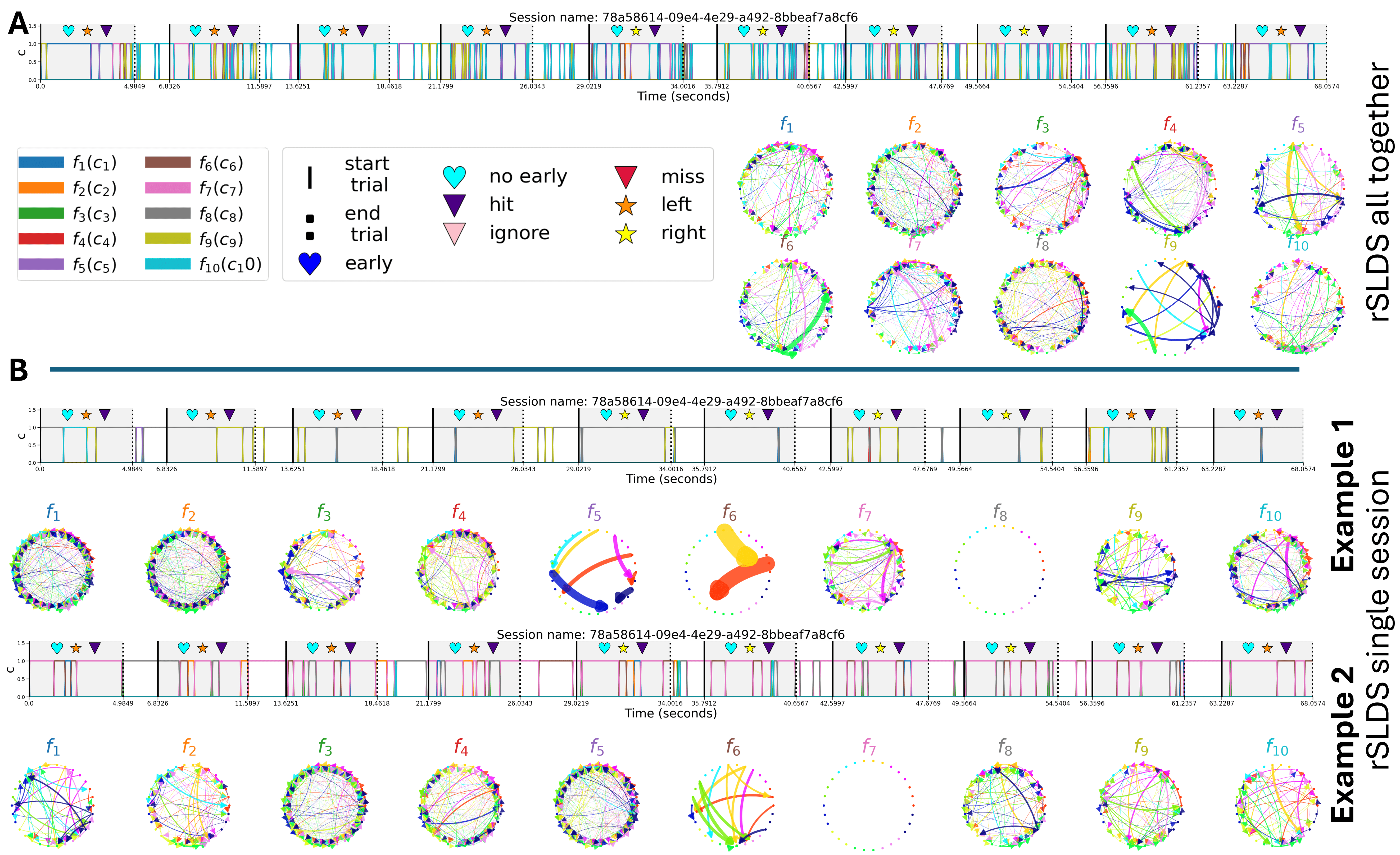}
\caption{{
Same as Figure~\ref{fig:mesoscale_identified_components_by_slds}, but for the rSLDS baseline~\cite{linderman2016recurrent}.
    }}
    \label{fig:mesoscale_identified_components_by_rslds}
\end{figure}


\begin{table}[ht]
\centering
\caption{Parameter table for the CREIMBO on Mice Mesoscale.}\label{tab:parameters_meso}
\begin{tabular}{|l|l|}
\hline
\textbf{Parameter} & \textbf{Value} \\
\hline
max\_error & 1e-09 \\
seed & 0 \\
normalize Fs & True \\
sparse\_f & True \\
sparsity\_on\_f\_max & 40 \\
increase\_in\_sparsity\_f & 1.4318632048860938 \\
norm\_D\_cols & True \\
D\_graph\_params & \{'with\_kNN': True, 'with\_norm': True, 'k': 15\} \\
lambda\_D & 0.4392235894758376 \\
update\_type\_D & spgl1 \\
latent\_dyns\_initialization & random \\
D\_with\_lasso & True \\
l1\_D & 2.7440675196366238 \\
params\_D & \{'update\_c\_type': 'spgl1'\} \\
num\_regions & 10 \\
reg\_type\_on\_c & spgl1 \\
lambda\_x & 1.0291322176748077 \\
latent\_dim & 30 \\
num ensembles per region & 30 \\
number of sub-dynamics & 10 \\
seed\_f & 0 \\
sigma\_mix\_f & 0.1 \\
sparse\_f\_params & \{'axis': '1', 'percent0': 20\} \\
\hline
\end{tabular}
\end{table}

We further tested CREIMBO's ability to infer meaningful task variables underlying  whole-brain multi-session data from mice~\cite{chen2023mesoscale,chen2024brain} while performing a memory-guided movement task. 
The measurements were obtained using multi-electrode extracellular electrophysiology, with the overall dataset stored in DANDI Archive under the NWB format~\cite{chen2023mesoscale}.

Particularly, the data include electrophysiology recordings of neural activity from multiple brain areas per session (Figs.~\ref{fig:mesoscale_all_areas}, ~\ref{fig:mesoscale_all_areas_b}), across overall $175$ sessions, with numerous trials in each.
The task the mice performed during the recordings involved selecting one of two ``lick ports'' based on auditory cues, with the left or right port leading to a reward depending on whether they heard a high or low tone. Neural activity was recorded across several brain regions in each session, with some regions overlapping between sessions, though no two sessions had identical coverage (refer to Fig. 1., \textbf{A-B}, in~\cite{chen2024brain}). 

We first loaded data from all sessions and selected areas for CREIMBO that featured the highest number of sessions including them. We identified the parent area of each sub-area labeled in the dataset and removed neurons with unlocated or NaN areas. Inspired by our exploratory analysis of parent-area distributions across sessions and neurons (Fig.\ref{fig:mesoscale_all_areas_b}B,C), we selected the 10 parent areas with the highest number of neurons across sessions\footnote{Used Parent Areas: Thalamus, Orbital Cortex, Secondary Motor Cortex, Basal Ganglia, Olfactory Cortex, Brain Stem, Somatosensory Cortex, Frontal Cortex, Hippocampus, Primary Motor Cortex}. We then selected 20 random sessions. One of these 20 sessions contained only `NaN`-located neurons, leaving 19 sessions for this analysis (Fig.~\ref{fig:mesoscale_all_areas_b}D)\footnote{Sessions DANDI names: `78a58614-09e4-4e29-a492-8bbeaf7a8cf6`, `ab103954-d99f-46b1-98ac-e6f91a1e9313`, `32701cfa-8932-4d9b-9f4d-cc05cb22ae89`, `b080c738-17c9-4e0d-aee4-6d5371518f69`, `886c4302-846a-4ef5-996a-6f02d6a81a5f`, `ba5b1ff3-753e-425c-baf0-e7cdd0c08093`, `ea856208-4240-404a-9034-d729ad6f4cda`, `ed759efa-dceb-4472-a2a3-4f7357d67665`, `8a2ce9b2-2e98-4c37-8f2a-3b9c8b542086`, `22791d80-26dc-4495-b4c0-651fe10e3298`, `a007fac1-96e7-4028-ad84-31a1b80db089`, `b39eb52a-0774-411b-b374-6eb08a73562e`, `8dbc25ee-cc17-4504-a1b9-7d43643b9466`, `981feb84-f209-4994-8838-02a0048feb87`, `585dc9d4-be9c-4244-8fa7-e64309019afc`, `aa74e4d7-a79b-4179-adf3-497fb1237edc`, `ebfab58b-7c80-4f31-9b4e-2b33883bb14a`, `8c724097-3876-48fc-94e0-832779e4f1be`, `2e9cb8a1-457d-49ea-97fd-f7023434d231`}\label{fn:sessions}.

We pre-processed the data using a 30 ms Gaussian kernel for firing rate estimation across all the sessions used (see ~\ref{fn:sessions}). We then selected only the neurons from the 10 parent areas and limited the analysis to the first 60 trials. Some sessions exhibited no activity or disconnection in certain trials, meaning some sessions included fewer than 60 trials (but no fewer than 40).

Next, we ran CREIMBO with 10 sub-circuits ($K=10$) and $p_j=3$ ensembles per area. The full list of parameters is shown in \ref{tab:parameters_meso}, and an example exploration of a random session is shown in (Fig.\ref{fig:mesoscale_example_session_results},\textbf{A-C} show the firing rate, trial start-end times, and distribution of trial durations, respectively).

We identified ensembles per region (exemplary in~\ref{fig:mesoscale_identified_components}A) with sub-circuits that capture both between-area (Fig.\ref{fig:mesoscale_identified_components}C) and within-area interactions (Fig.\ref{fig:mesoscale_identified_components}D), showing diverse motifs with varying degrees of cross-circuit correlations (Fig.~\ref{fig:mesoscale_identified_components}B). Some cross-area interactions reveal effects originating from specific regions involved in the task. For example, $\bm{f}_1$ demonstrates hippocampal effects, which align with the task's memory demands, while, e.g., $\bm{f}_3$ shows interactions involving the frontal cortex, including follow-up activations, corresponding to the task's need for planning and execution. Other sub-circuits reveal inputs converging on specific areas. For instance, $\bm{f}_8$ shows motor cortex inputs, relevant for moving the tongue during the lick response, and $\bm{f}_2$ reflects inputs into the secondary motor cortex, which is involved in coordinating motor actions.

When extracting only within-area interactions, we observe unique motifs as well. For example, $\bm{f}_1$ shows further internal processing between basal ganglia ensembles, which may reflect the internal coordination and integration of motor planning and execution signals. $\bm{f}_7$ and $\bm{f}_2$ show strong effects of one ensemble on another within the secondary motor cortex, which could represent the modulation and refinement of motor commands for more precise movements. $\bm{f}_5$ demonstrates brainstem effects, with two ensembles influencing two other ensembles, potentially supporting basic motor functions such as muscle activation and coordination. $\bm{f}_4$ shows multiple self-activations of ensembles, which may capture self-regulatory processes like feedback inhibition or facilitation that modulate motor output and prevent over-activation.

We further examined CREIMBO's dynamics coefficients (${\bm{c}_t}$) and identified task-related patterns with similar coefficient profiles across trials for the same task variable (Fig~\ref{fig:mesoscale_classifier}A for an example session, and B for hit vs. miss outcomes). Similar differences were observed across other task variables. To quantify the predictive power of the dynamic coefficients, we used them as input for training a simple one-vs-rest logistic regression classifier. Each trial was split into four equal-duration time windows, and the 10 dynamic coefficients were averaged within each window, resulting in 40 features (4 windows × 10 coefficients). These features were then used to predict task variables, including outcome (hit/miss/ignore), early lick (whether a lick was early), lick side instructed, and lick side performed in practice. Additionally, we tested the model to predict 2-3 variables simultaneously, which was more challenging due to the need to distinguish between multiple options. The resulting accuracy levels (Fig~\ref{fig:mesoscale_classifier}C, black stars) were significantly higher than chance levels (gray bars). We further analyzed the confusion matrix for each variable’s prediction and applied the $\chi^2$ test to assess whether the distribution of predictions significantly deviated from the expected distribution under chance. We found p-values well below \(1 \times 10^{-10}\) (see subtitles of Fig~\ref{fig:mesoscale_classifier}D, F, G, H, I, J), strongly supporting CREIMBO's predictive power.

Furthermore, we assessed the importance of subcircuits' activations across the four time windows (Fig.\ref{fig:mesoscale_identified_components}E), with each block of rows representing one subcircuit and the blocks corresponding to different time windows. CREIMBO revealed patterns where certain time windows and dynamic coefficients were pivotal for different task variables. For instance, $c_7$ at $t_3$ was critical, while other coefficients, like $c_9$ at $t_3$, were specific to encoding the lick side. This specificity aligns with the fact that this time point likely reflects the final stages of processing during the task, including aspects of learning. When we examined $c_9$ (Fig.\ref{fig:mesoscale_identified_components}C), we observed multiple inputs into the secondary motor cortex, related to the execution of the lick towards the trial's end, highlighting CREIMBO's ability to pinpoint region-specific interactions. Additionally, $c_8$ showed increased importance in the initial window $t_0$, contributing across multiple variables and capturing inputs into the hippocampus—an area integral to memory processing—underscoring its role in early trial stages. These findings further validate CREIMBO’s ability to capture and predict task-related neural dynamics, offering insights into circuit-level brain activity across time.


\begin{figure}[ht]
    \centering
    \includegraphics[width=0.89\textwidth]{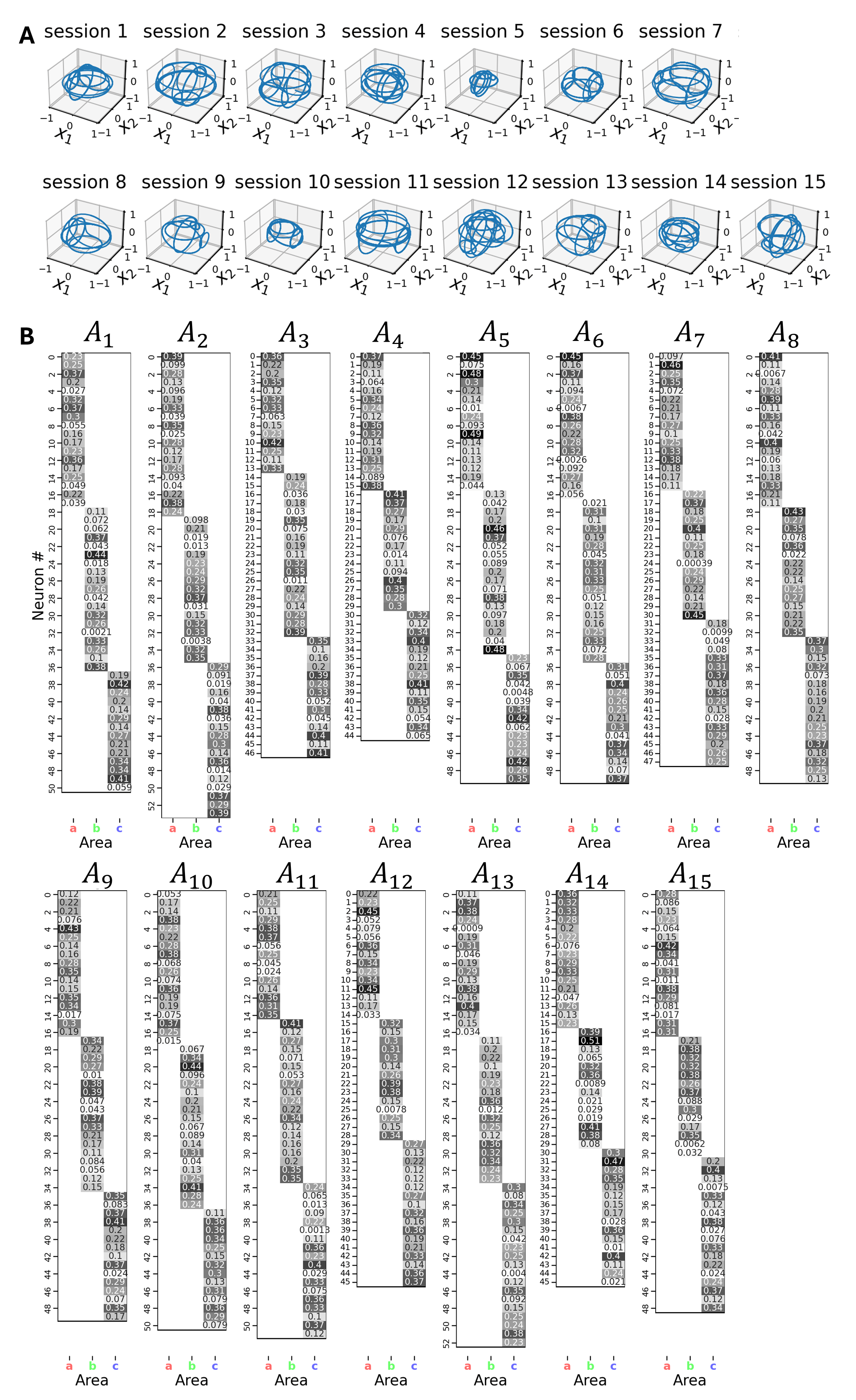}
    %
\caption{Cont. of Figure~\ref{fig:results_for_2nd_synthetic_experiment}.
\textbf{A:} Ground truth latent dynamics ($\{\bm{X}^d \}^{D = 15}_{d=1}$). 
\textbf{B:} Ground truth ensemble compositions  ($\{\bm{A}^d \}^{D = 15}_{d=1}$). 
    }
    \label{fig:results_for_2nd_synthetic_experiment_b}
\end{figure}

\begin{figure}[ht]
    \centering
    \includegraphics[width=0.99\textwidth]{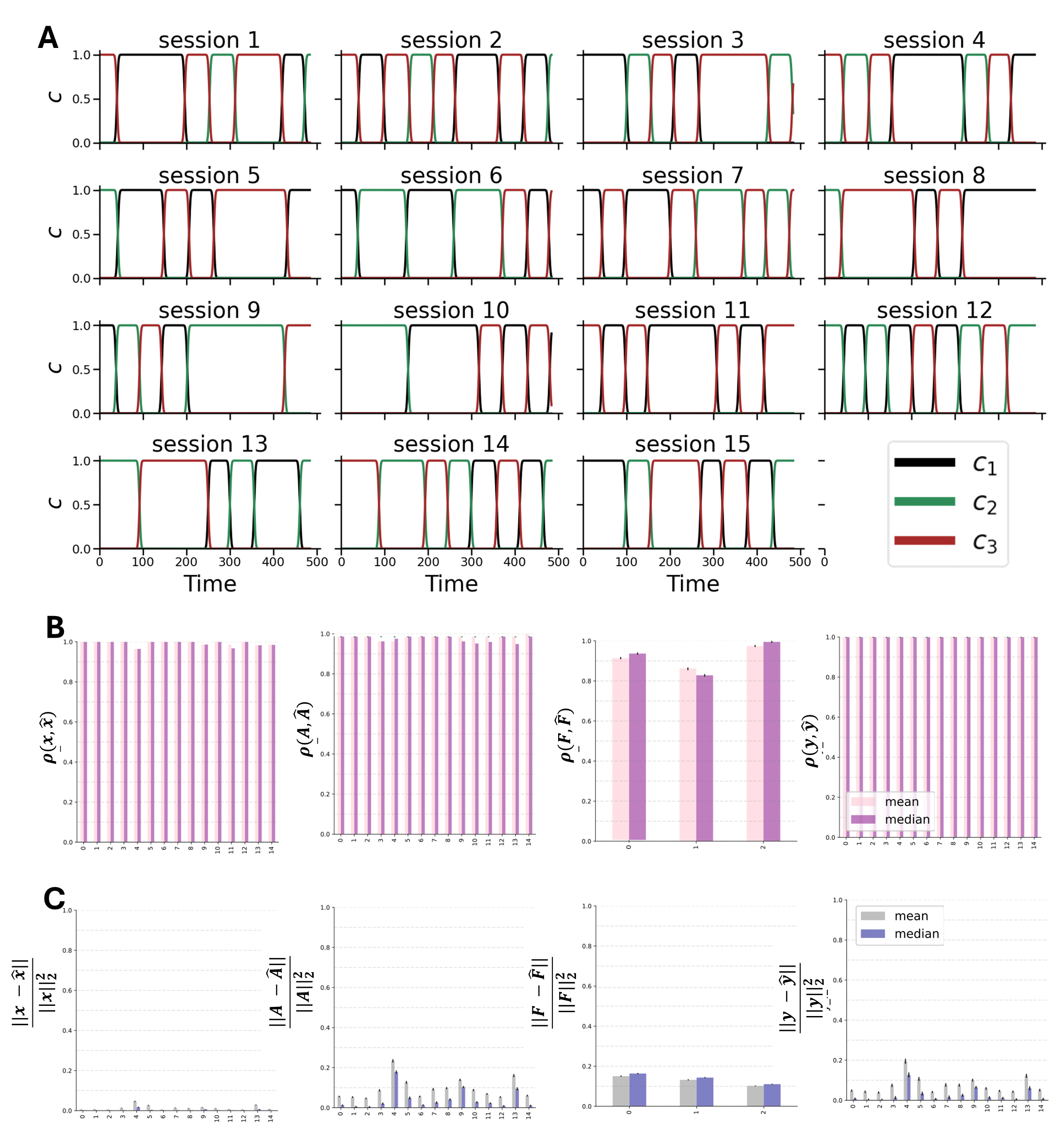}
    %
\caption{Cont. of Figure~\ref{fig:results_for_2nd_synthetic_experiment}.
\textbf{A:} Ground truth sub-circuits coefficients ($c$) vary between sessions.
\textbf{B:} Correlation between ground truth and identified components for \textbf{A} (left), \textbf{x} (middle left),  \textbf{F} (middle right), \textbf{Y} (right). 
\textbf{C:}  $\ell_2$ distance between the components learned by CREIMBO and the ground-truth components for \textbf{A} (left), \textbf{x} (middle left),  \textbf{F} (middle right), \textbf{Y} (right). 
    }
    \label{fig:results_for_2nd_synthetic_experiment_c}
\end{figure}

\section{Inference Assuming Poisson Statistics}
\label{sec:poisson_dev}
CREIMBO, as outlined in the main text, assumes Gaussian-distributed  \textit{i.i.d}  noise. This assumption is reflected in the inference procedure, which minimizes the $\ell_2$ and Frobenius norms (e.g.,~\eqref{eqn:updateLambda}, \eqref{eqn:update_x_c}). Particularly, these norms arise from minimizing the negative log-likelihood of Gaussian-distributed observations noise ($p(\bm{Y}^d_{n,t}) = \frac{1}{\sqrt{2\pi \sigma^2}} \exp\left( -\frac{\bm{Y}^d_{n,t} - [A^d X^d]_{n,t})^2}{2\sigma^2} \right),$
with some noise standard-deviation $\sigma$.

Electrophysiology, however, records trains of action potentials (spikes) that are often modeled as a Poisson processes. Binned spike counts are thus typically approximated as a Poisson random variable. However, when firing rates (FRs) are high enough, the Gaussian assumption is an adequate approximation, 
and thus
the Gaussian assumption remains common in data-driven neuroscience models (e.g.,~\cite{linderman2016recurrent,aoi2018model}). 
However, some species (e.g., bats \citep{allen2021effect})  exhibit extremely low FRs, which may require Poisson likelihood functions. Thus we propose an extension to CREIMBO that with a Poisson likelihood function for ensemble and traces inference, that can replace the Gaussian model in low spiking-rates scenarios. 


Our Poisson extension first involves transforming the spike-timing data to binned spike counts for all sessions $d=1\dots D$ ($\{\bm{Y}^d\}_{d=1}^D$). 
Next, we consider the same goal of finding the ensembles $\bm{A}^d$ and their traces $(\bm{X}^T)^d$, under the assumption that the data follows a Poisson distribution.

To simplify, we focus on a single session $d$, with $\bm{y} = (\bm{Y}^d)^T$ ($T$ for transpose), and note that the extension below applies to all sessions.
In the Poisson model, for each neuron $n=1\dots N$, the  
likelihood of a certain spike count at time $t$ given a latent rate $\lambda_{t,n}$
is:
$$
p(\bm{y_{t,n}}|\lambda_{t,n}) = \frac{\lambda_{t,n}^{y_{t,n}} e^{-\lambda_{t,n}}}{y_{t,n}!}
$$
where $y_{t,n}!$ denotes the factorial of $y_{t,n}$. 

Here we model the unknown rate $\bm{\lambda}\in\mathbb{R}^{T\times N}$ as the low-dimensional representation  captured via the ensembles $\bm{A}^d\in\mathbb{R}^{N\times p}$ and their traces $ \widetilde{\bm{x}} := (\bm{X}^d)^T \in \mathbb{R}^{T \times p}$ ($T$ for transpose), i.e., $\bm{\lambda}=\bm{A}^d\bm{X}^d$, such that the likelihood becomes
$$
p(\bm{y}|\bm{\widetilde{x}}, \bm{A}) = \prod_{n=1}^N \prod_{t=1}^T p(y_{t,n}|\lambda_{t,n})
$$
$$
= \prod_{n=1}^N \prod_{t=1}^T p(y_{t,n}|[\bm{\widetilde{x}} \bm{A}^T]_{t,n})
$$
$$
= \prod_{n=1}^N \prod_{t=1}^T \frac{([\bm{\widetilde{x}} \bm{A}^T]_{t,n})^{y_{t,n}} e^{-([\bm{\widetilde{x}} \bm{A}^T]_{t,n})}}{y_{t,n}!} .
$$


This can be trained by iteratively updating $\bm{\widetilde{x}}$ and $\bm{A}$ to minimize the negative log of the above likelihood, which can be represented mathematically as:

\begin{align}
   \{ \widehat{\bm{\widetilde{x}}},\widehat{\bm{A}}\} &= 
   \arg\min_{\widetilde{\bm{x}}, \bm{A}} \left[-log \left( \prod_{n=1}^N \prod_{t=1}^T \frac{([\bm{\widetilde{x}} \bm{A}^T]_{t,n})^{y_{t,n}} e^{-([\bm{\widetilde{x}} \bm{A}^T]_{t,n})}}{y_{t,n}!} \right)\right] \nonumber
   \\ &=
   \arg\min_{\bm{\widetilde{x}}, \bm{A}} \sum_{n=1}^N \sum_{t=1}^T \left[[\bm{\widetilde{x}} \bm{A}^T]_{t,n} - y_{t,n} \log([\bm{\widetilde{x}} \bm{A}^T]_{t,n}) \right] \label{eq:phi_mle}
\end{align}
where the logarithm is taken using the natural exponential base. Notably, in the above $\arg\min$, we  chose to omit the constant term that emerge from taking the logarithm (i.e., $log(y_{t,n}!)$) as it does not affect the argument minimization. 

In this Poisson case, in contrast to the Gaussian, we can no longer solve for $\bm{x}$ using e.g., least squares or LASSO; instead, we will update $\widetilde{\bm{x}}$ via Gradient Descent. The first step is to compute the gradient of the cost function in~\eqref{eq:phi_mle} with respect to $\bm{\widetilde{x}}$.

To simplify the calculation, we first notate the two components obtained in~\eqref{eq:phi_mle} by the auxiliary terms 
$$
g_1(\bm{\widetilde{x}}, \bm{A}) = \sum_{t,n} [\bm{\widetilde{x}} \bm{A}^T]_{t,n}
$$
and
$$
g_2(\bm{y}, \bm{\widetilde{x}}, \bm{A}) = \sum_{t,n} -y_{t,n} \cdot \log([\bm{\widetilde{x}} \bm{A}^T]_{t,n}).
$$


The cost function from~\eqref{eq:phi_mle} thus can be written in terms of these functions as:

\begin{align}
\{ \widehat{\bm{\widetilde{x}}}, \widehat{\bm{A}}\} = \arg\min_{\bm{\widetilde{x}}, \bm{A}} \left[ g_1(\bm{\widetilde{x}}, \bm{A}) + g_2(\bm{y}, \bm{\widetilde{x}}, \bm{A}) \right].
\label{eq:g1_g2}
\end{align}

Hence, for updating $\bm{\widetilde{x}}$ via Gradient Descent, we need to find the gradients of $g_1$ and $g_2$. 

We first establish two notations to make subsequent steps clearer:
\begin{itemize}
    \item let $[1]_{(m,k)}$ denote a matrix of ones of size $m \times k$.
    \item let $\delta_{(M,P)_{m,p}}$ denote a matrix of shape $M \times P$ whose entries are all zeros except for the entry at index $(m,p)$, which is set to 1.
\end{itemize}

\textbf{Calculate $\frac{\partial g_1 (\bm{\widetilde{x}}, \bm{A})}{\partial \bm{\widetilde{x}}}$:}

To begin, we can rewrite $g_1(\bm{\widetilde{x}}, \bm{A})$ as
$$
g_1 (\bm{\widetilde{x}}, \bm{A}) = \sum_{t,n} [\bm{\widetilde{x}} \bm{A}^T]_{t,n} = [1]_{(1,T)} \bm{\widetilde{x}} \bm{A}^T [1]_{(N,1)}
$$
and follow the identity  (taken from~\cite{petersen2008matrix} Eq. 20):
\begin{align} 
\frac{\partial [\bm{a}^T \bm{M} \bm{b}]}{\partial \bm{M}} = \bm{a} \bm{b}^T  
 \hspace{5pt} \text{ where } \bm{a},\bm{b}  \text{ vectors, and  } \bm{M} \text{ matrix}
        .\label{eq:identity}
\end{align}

Building on this, let $[1]_{(1,T)}$ correspond to the $\bm{a}^T$ vector from~\eqref{eq:identity},  $\bm{A}^T [1]_{(N,1)}~\in~\mathbb{R}^{p\times 1}$ correspond to the $\bm{b}$ vector of~\eqref{eq:identity}, and $ \bm{\widetilde{x}}$ correspond to the matrix $\bm{M}$ . Then, the gradient of $ g_1(\bm{\widetilde{x}}, \bm{A}) $  with respect to $ \bm{\widetilde{x}} $  is:

\[
\frac{\partial g_1 (\bm{\widetilde{x}}, \bm{A})}{\partial \bm{\widetilde{x}}} = \frac{\partial \left( [1]_{(1,T)} \bm{\widetilde{x}} \bm{A}^T [1]_{(N,1)} \right)}{\partial \bm{\widetilde{x}}}
= [1]_{(T,1)} [1]_{(1,N)} \bm{A}
= [1]_{(T,N)} \bm{A}
\]

\textbf{Calculate $\frac{\partial g_2 (\bm{y},\bm{\widetilde{x}}, \bm{A})}{\partial \bm{\widetilde{x}}}$:}


First, we will rewrite the term $ [\bm{\widetilde{x}} \bm{A}^T]_{t,n} $  as:
\begin{align}  
[\bm{\widetilde{x}} \bm{A}^T]_{t,n} = \sum_{j=1}^p \bm{\widetilde{x}}_{t,j} (\bm{A}^T)_{j,n}
= \delta_{{(1,T)}_{1,t}} \bm{\widetilde{x}} \bm{A}^T \delta_{{(N,1)}_{n,1}}\label{eq:g2_der}
\end{align}

$ \delta_{{(1,T)}_{1,t}} $  is a row vector of length $ T $  with a value of 1 at index $ t $  and 0 elsewhere, and $ \delta_{{(N,1)}_{n,1}} $  is a column vector of length $ N $  with a value of 1 at index $ n $  and 0 elsewhere.

We will use again the identity from~\cite{petersen2008matrix} presented in~\eqref{eq:identity},  and now mark $\delta_{{(1,T)}_{1,n}} $ with the~\eqref{eq:identity}'s vector $\bm{a}^T$, $ \bm{\widetilde{x}}$ is again $\bm{M}$, and $b := \bm{A}^T \delta_{{(N,1)}_{j,1}}$. 
When applying the identity in~\eqref{eq:identity} to~\eqref{eq:g2_der}, we thus obtain
$$ 
\frac{\partial [\bm{\widetilde{x}} \bm{A}^T]_{t,n}}{\partial \bm{\widetilde{x}}} = \delta_{{(T,1)}_{t,1}} \delta_{{(1,N)}_{1,n}} \bm{A} = \delta_{(T,N)_{t,n}} \bm{A}
$$ 

Given the above, we receive
\begin{align}
\frac{\partial g_2(\bm{y},\bm{\widetilde{x}},\bm{A})}{\partial \bm{\widetilde{x}}}
&= 
\frac{\partial \left(\sum_{n=1}^N \sum_{t=1}^T -y_{t,n} \log \left( [\bm{\widetilde{x}} {\bm{A}}^T]_{t,n} \right) \right)}{\partial \bm{\widetilde{x}}} \\
&= -\sum_{n=1}^N \sum_{t=1}^T \frac{y_{tn}}{[\bm{\widetilde{x}} \bm{A}^T]_{t,n}} \delta_{(T,N)_{t,n}} \bm{A} \nonumber \\
&= -\sum_{t,n} \psi_{t,n} \delta_{(T,N)_{t,n}} \bm{A} \nonumber \\
&= - \left (  \sum_{t,n} \psi_{t,n} \delta_{(T,N)_{t,n}} \right ) \bm{A} \nonumber \\
&= - \bm{\Psi} \bm{A} \nonumber
\end{align}
where $\psi_{t,n} = \frac{y_{t,n}}{[\bm{\widetilde{x}} \bm{A}^T]_{t,n}}$, and $\bm{\Psi} \in \mathbb{R}^{T \times N}$ is a matrix whose entry at index $(t,n)$ is $\psi_{t,n}$.

Going back to the expression in~\eqref{eq:g1_g2},
the full gradient of the Poisson loss with respect to $\bm{\widetilde{x}}$ is:
\[
\frac{\partial [g_1(\bm{\widetilde{x}}, \bm{A}) + g_2(\bm{y}, \bm{\widetilde{x}}, \bm{A})]}{\partial \bm{\widetilde{x}}} = [1]_{(T,N)} \bm{A} - \bm{\Psi} \bm{A} = ([1]_{(T,N)}  - \bm{\Psi}) \bm{A}, 
\]
and 
the projected gradient step in $\bm{\widetilde{x}}$
at iteration $m$ thus follows:
$$\bm{\widetilde{x}}^{m+1} \leftarrow \Pi \left( \bm{\widetilde{x}}^{m} - \eta^{m} \left( [1]_{(T,N)}  -  \bm{\Psi} \right) \bm{A} \right)$$
where $\Pi$ can project the columns of $\bm{\widetilde{x}}$ onto some constraint set, with the identity transform used if no projection is desired, and $\eta^{m}$ is the gradient descent step size at iteration $m$.

For inferring $\bm{A}$,  at each iteration $m$, we can then use the Spiral-Tap package~\cite{harmany2012spiral} to infer the sparse ensembles.

The above ensemble model (including $ \bm{A} $ and $ \bm{x} $) is the only part that directly connects to the Poisson-distributed spike-counts ($\bm{Y}^d$), while the latent evolution of $ \bm{x} $  via $x_{t} = \bm{F}_t \bm{x}_{t-1}$ 
is independent of the Poisson assumption on the spike-counts $\bm{Y}^d$.
Particularly, as the transition from $ \bm{x_t} $ to $ \bm{x_{t+1}} $ 
is dictated by the ensembles' dynamic interactions ($\bm{F_t} = \sum_{k=1}^K c_{kt}\bm{f}_k$), which we continue to assume follows a Gaussian distribution in the latent space, 
the dynamic inference part of CREIMBO 
remains unchanged from the main text.
\clearpage
\newpage
\end{document}